\documentclass{aa}  
\usepackage{graphicx}
\usepackage{txfonts,rotate,psfig,subfigure,supertabular,lscape,times}
\begin{document}
   \title{The soft X-ray properties of AGN from the CJF sample \thanks{Tables 3--6 are only available in the online edition of the Journal.} }

   \subtitle{A correlation analysis between soft X-ray and VLBI properties}

   \author{S. Britzen\inst{1,2,3} \and W. Brinkmann\inst{4} \and R.M. Campbell\inst{5}
     \and M. Gliozzi\inst{4, 6} \and A.C.S. Readhead\inst{7} \and I.W.A. Browne\inst{8} \and P. Wilkinson\inst{8}
          }

   \offprints{S. Britzen}

   \institute{Max-Planck-Institut f\"ur Radioastronomie, Auf dem H\"ugel 69, D-53121 Bonn, Germany\\ \email{sbritzen@mpifr-bonn.mpg.de}
   \and Landessternwarte, K\"onigstuhl, 69117 Heidelberg, Germany
   \and ASTRON, Oude Hoogeveensedijk 4, 7991 PD Dwingeloo, The Netherlands
   \and Max--Planck--Institut f\"ur extraterrestrische Physik, Giessenbachstrasse~1, D-85748 Garching, Germany
   \and Joint Institute for VLBI in Europe, Oude Hoogeveensedijk 4, NL-7991 PD Dwingeloo, The Netherlands
   \and Department of Physics and Astronomy, George Mason University, 4400 University Drive, MS 3F3, Fairfax, VA 22030, USA
   \and California Institute of Technology, Department of Astronomy, 105-24, Pasadena, CA 91125, USA
   \and University of Manchester, Nuffield Radio Astronomy Laboratories, Jodrell Bank, Macclesfield, Cheshire SK11 9 DL, England UK
             }

   \date{Received ; accepted }

% \abstract{}{}{}{}{} 
% 5 {} token are mandatory
 
  \abstract
   {We present the soft X-ray properties obtained in the ROSAT All-Sky survey
   and from pointed PSPC observations for the AGN in the complete flux-density limited Caltech-Jodrell Bank flat spectrum sample (hereafter CJF). CJF is a VLBI survey (VLBA observations at 5 GHz) of 293 AGN with detailed information on jet component motion.}
   {We investigate and discuss the soft X-ray properties of this AGN sample and examine the correlations between X-ray and VLBI properties, test beaming scenarios, and search for the discriminating properties between the sub-samples detected and not detected by ROSAT.} 
   {Comparing the observed and the predicted X-ray fluxes by assuming an Inverse Compton (IC) origin for the observed X-rays, we compute the beaming or Doppler factor, $\delta_{\rm IC}$, for the CJF sources and compare it with  the equipartition Doppler factor, $\delta_{\rm EQ}$. We further contrast the Doppler factors with other beaming indicators derived from the VLBI observations, such as the value of the expansion velocity, and the observed and intrinsic brightness temperature. 
   We calculate two different core dominance parameters ($R$): the ratio of total VLBI flux to single-dish flux, $R_{\rm V}$, and the ratio of the VLBI core-component flux to single-dish flux, $R_{\rm C}$. In addition, we investigate the large-scale radio structure of the AGN and the difference between the pc- and kpc-scale structure (misalignment) with regard to the X-ray observations.} 
   {We find a nearly linear relation between X-ray and radio luminosities, and a similar but less stringent behaviour for the relation between optical and X-ray luminosities.
   The CJF-quasars show faster apparent motions and larger values of
$\delta_{\rm IC}$ than the radio galaxies do. The quasars detected by ROSAT have a different $\beta_{\rm app}$-redshift relationship compared to the non-detected ones. We find no significant difference in $R$ between the quasars detected and not detected by ROSAT. We find evidence that $R$ is smaller for quasars and BL Lac objects than it is for radio galaxies, in accordance with unification scenarios. ROSAT-detected sources tend to reveal extended large-scale radio structures more often.}
   {We conclude that beaming alone cannot explain the observed dichotomy of ROSAT detection or non-detection and assume that the large-scale jet structure plays a decisive role.}
   \keywords{Radio sources: general; Galaxies: active; X--rays: general}
   \maketitle

\section{Introduction}
The investigation of the origin and nature of radiation processes in AGN requires the combination of multifrequency flux-density observations and morphological information to pinpoint the emission regions. 
Only recently have {\it CHANDRA} observations been able to clarify the dominant X-ray emission mechanism for a number of individual sources. 78 radio galaxies, quasars, and BL Lac objects with known X-ray emission from jets or hotspots are now known (http://hea-www.harvard.edu/XJET/). The nature of the X-ray emission processes in most of the remaining AGN is a matter of debate and a statistical treatment of this phenomenon for a larger sample of AGN is still lacking.\\
The CJF survey has recently provided a kinematic database suited for an improved statistical treatment of such questions related to AGN. All CJF sources have been monitored in at least three epochs at 5 GHz with the VLBA. Detailed information concerning the observations, the data reduction, and the source parameters resulting from the model-fitting procedure can be found in Britzen et al. 2007a (hereafter Paper I). The kinematic analysis of the sources is described in Britzen et al.  2007b (hereafter Paper II). Most (57\%) of these sources have been detected in the {\it ROSAT} All-Sky Survey. Although we lack simultaneous observations when comparing the radio and the {\it ROSAT} data, we do have the possibility to check the basic concepts of beaming with this largest uniform sample so far. \\
The discovery of well-collimated, one-sided, apparently superluminal jets
 on parsec scales by VLBI has revealed the dominant
 effects of relativistic beaming on the appearance of these objects
 (e.g., Witzel et al. 1988; Readhead 1993; Zensus \& Pearson 1987),
 and has motivated the development of the so-called ``unified theories'' for
 quasars and radio galaxies (Orr \& Browne 1982; Barthel 1989; Urry \& Padovani 1995).
 The viewing angles are expected to differ among the source classes, with quasars and BL Lacertae objects being radio galaxies seen with their jets forming a small angle with respect to the line of sight.
 The observed superluminal motions strictly require that some
 ``phase'' or ``pattern'' speed of a wave traveling along the jet is
 relativistic, but there are strong arguments also for the bulk velocity of
 the radiating plasma to be relativistic, with associated forward beaming of
 the emitted radiation (e.g., Witzel et al. 1988; Eckart et al. 1989).
 The comparison of Doppler factors calculated on the basis of velocity
 and X-ray information may
 answer the question whether the pattern and the bulk velocities are different.\\
Since part of the X-ray emission is isotropic and part is beamed Inverse Compton radiation from the radio jet, we can place limits on the IC Doppler factor $\delta_{\rm IC}$. The Doppler factors can be derived via the standard
synchrotron self-Compton (SSC) argument, from equipartition arguments ($\delta_{\rm EQ}$), and
 from the apparent velocities determined from VLBI observations.\\ 
Most models, accounting for the observed broadband
spectra of blazars, attribute the radio
through optical emission to synchrotron radiation, and X-ray through $\gamma$-ray
emission to Compton scattering (e.g., Marscher 1980; K\"onigl 1981).
The models
differ in the location and structure of the acceleration and emission
region(s). However, in the case of the so-called High Peaked BL Lac objects (HBL), even the X-ray emission is thought to be due to synchrotron radiation (e.g., Padovani \& Giommi 1995).  
The X-ray emission from knots in radio jets can mainly be attributed to synchrotron emission. Convincing evidence for this has been found in the optical polarization of M87, suggesting that the optical emission as well as the radio emission are produced via the synchrotron process (Harris et al. 1998; Biretta et al. 1991). On the other hand, X-ray intensities that lie well above the extrapolation of the radio/optical synchrotron spectrum are taken to be strong evidence against the ``simple" synchrotron model. In addition, every synchrotron source must also produce IC emission from at least the CMB and the synchrotron photons themselves.\\
In this paper we combine and correlate information from different parts of the
electromagnetic spectrum, i.e. the radio and the X-ray regime. We test the beaming and unification scenario on the basis of data obtained in the CJF survey and the {\it ROSAT} All-Sky Survey (RASS).\\
The RASS was the first soft X-ray survey  of the whole sky
 using an imaging telescope (Tr\"umper 1983).
It was performed from August 1, 1990 to February 1, 1991 and
yielded $\sim$ 125000 X-ray sources  with a positional accuracy
such that 68\% of the sources are found within 20\arcsec\ from
their corresponding optical counterparts (Voges et al. 2000) 
and a  limiting sensitivity of a few times
 $10^{-13}$ erg/cm$^{2}$/s in the 0.1--2.4 keV energy
band, depending on the spectral form and the amount of galactic
absorption. The survey was followed by a period of pointed observations
lasting until February 12, 1999; the PSPC (Position Sensitive Proportional Counter) detector (Pfeffermann et al. 1986) exhausted its gas supply earlier, in September 1994.\\ 
This paper is organized as follows:  sections 2 and 3 introduce the CJF data and the {\it ROSAT} X-ray data respectively; section 4 presents the results of the correlation analysis with regard to the X-ray properties of the CJF survey, the $\beta_{\rm app}$-relation, the IC Doppler factor, the equipartition Doppler factor, the brightness temperature, the core dominance parameter, the misalignment, and the relation between the large scale structure of AGN and the observed X-ray emission. Finally, section 5 briefly discusses our results.
\section{The CJF data}
The CJF, defined by Taylor et al. (1996), is
a complete flux-limited sample of 293 flat-spectrum radio sources, drawn
from the 6 cm and 20 cm Green Bank Surveys (Gregory \& Condon 1991; White
\& Becker 1992)
with selection criteria as follows: $S$(6~cm)$\ge $350 mJy,
{$\alpha _{20}^{6}$}{$\ge $}{$-0.5$}, $\delta $(1950)$\ge $35$^{\circ}$,
and {$|b^{\rm II}|$}{$\ge $}10$^{\circ}$. This sample is mostly a
superset of the flat-spectrum sources in the
Pearson-Readhead Survey (Pearson \& Readhead 1988)
based on the 6~cm MPI-NRAO 5 GHz surveys (K\"uhr et al. 1981),
the First Caltech-Jodrell Bank Survey (CJ1: Polatidis et al. 1995; Thakkar
 et al. 1995; Xu et al. 1995) and the
 Second Caltech-Jodrell Bank Survey (CJ2: Taylor et al. 1994;
 Henstock et al. 1995).
Continued VLBI observations of the CJF
sources have been performed since 1990. For the unambiguous 
determination of the jet
component position and motion parameters at least
three observing epochs (spread over roughly 4 years) have been obtained for all of the 293 objects. All epochs for all sources have been analyzed in the same systematic way in order to create a homogeneous, statistically valid
database. A reanalysis of ``early" epochs has been done in parallel with
the acquisition and analysis of new observational epochs, which led to a
simultaneous completion of the observation, reduction, and reanalysis parts.
For details on the observations and data reduction see Paper I.\\
The redshifts of the identified quasars in CJF range from
$z$=0.227 to $z$=3.889, with an average of 16
quasars per redshift interval of 0.2 in the range $z$=0.6--2.6. This
provides us with the opportunity to investigate possible correlations over
a broad range of redshifts and to address important cosmological
questions, such as AGN evolution with cosmic epoch.
The CJF is now known to contain some 25 radio galaxies and 11 BL Lac
objects at $z>$0.6, enough to allow a meaningful comparison of the properties of
these source classes at the same redshift and luminosity.\\
Preliminary results have already been discussed in Britzen et al. (2001) and Britzen (2002). The kinematics of the complete sample is investigated and discussed in detail in Paper II.\\
The results presented here are based on a careful identification of the jet components across the epochs and multiple checks of the resulting motions in the xy-plane (see Paper I and II). It turned out that not all jet components can yield proper motion estimates of equal significance. We assigned a quality factor to each component that takes the following properties of each jet component into account:\\
-the jet component has to be clearly separated from other components and the core\\
-the component should be unambiguously identifiable in each epoch in which it is detected (i.e., it should not appear to merge with other components nor appear to split into two components at any epoch)\\
-the component should be visible in at least three epochs\\
Single, bright jet components that are clearly separate in all epochs were given a quality 1, while all sources that merge or split were assigned a quality 3. With these additional quality criteria, we can additionally select only the most reliable jet component proper motions for further consideration.
For a detailed discussion of these quality factors please see Paper II.
For source-based comparisons in this paper, e.g., between $\beta_{\rm app}$ and $\delta_{\rm IC}$, we use a ``representative'' subsample of jet components. In several tests we determined these jet components to be representative for jet component motion in this source. The so-called ``brightest'' subsample -- described in detail in Paper II -- fulfills these criteria and will be used as comparison VLBI-sample for the correlation analysis between {\it ROSAT}- and VLBI-information presented in this paper. Wherever possible, e.g., the calculation of the core dominance parameter $R$ or the misalignment, the complete sample of CJF sources has been used. 
\begin{figure*}[htb]
\begin{center}                                                                                                   \subfigure[]{\includegraphics[clip,width=5.3cm,angle=-90]{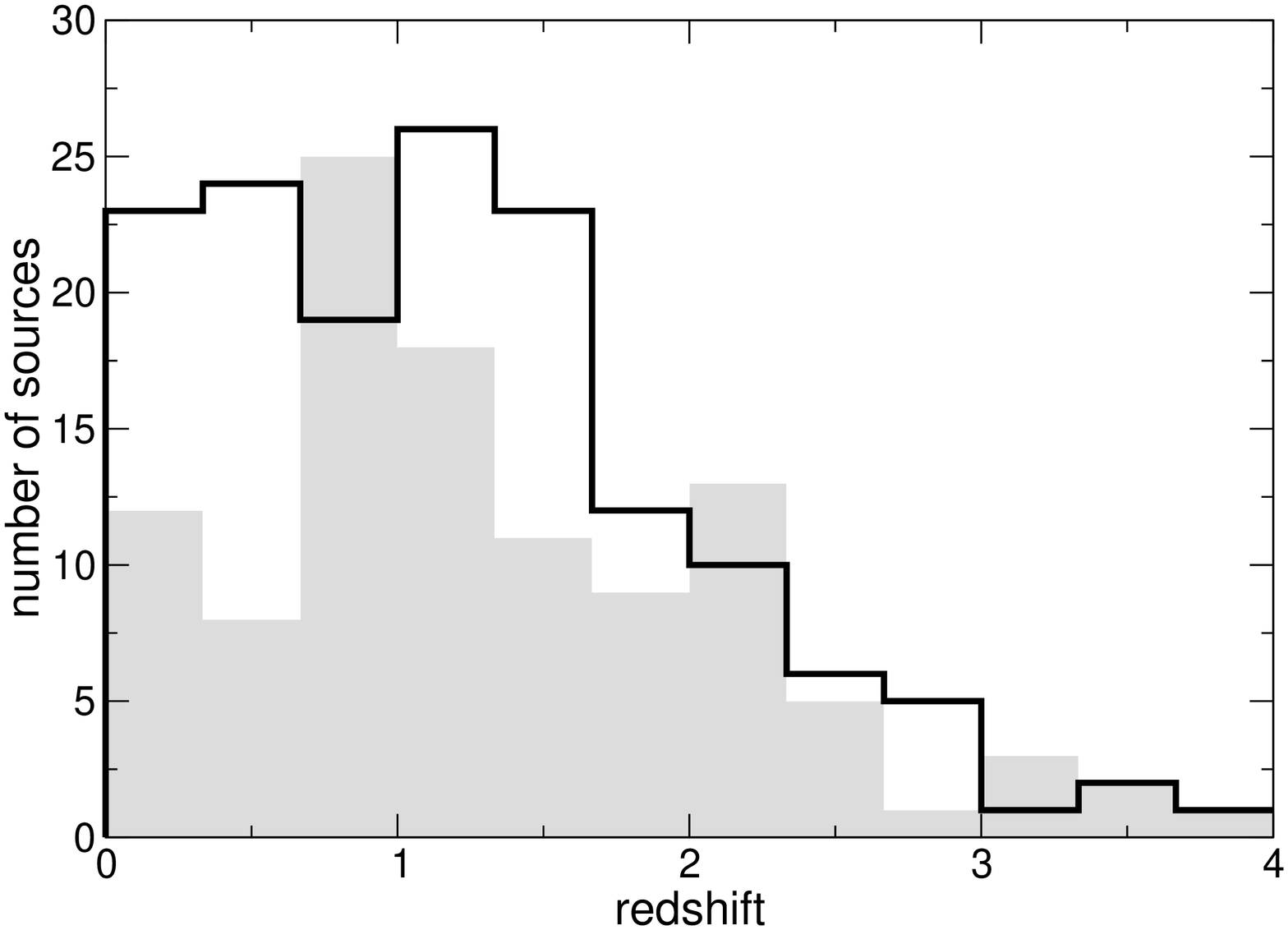}}\\
\end{center}                                                                                                     \hspace*{-1.0cm}\subfigure[]{\includegraphics[width=4.4cm,angle=-90,clip]{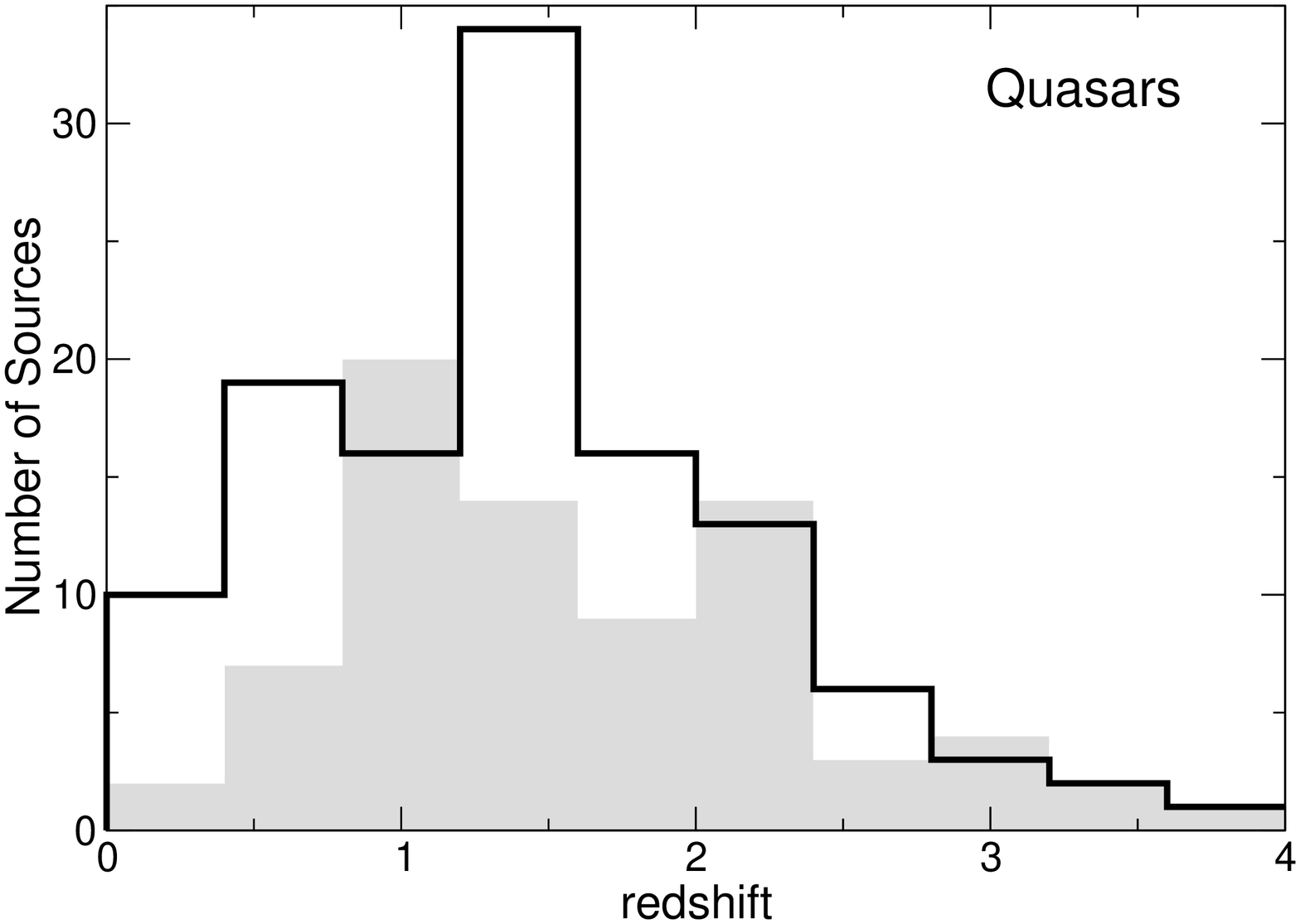}}
\hspace*{0.7cm}\subfigure[]{\includegraphics[clip,width=4.4cm,angle=-90]{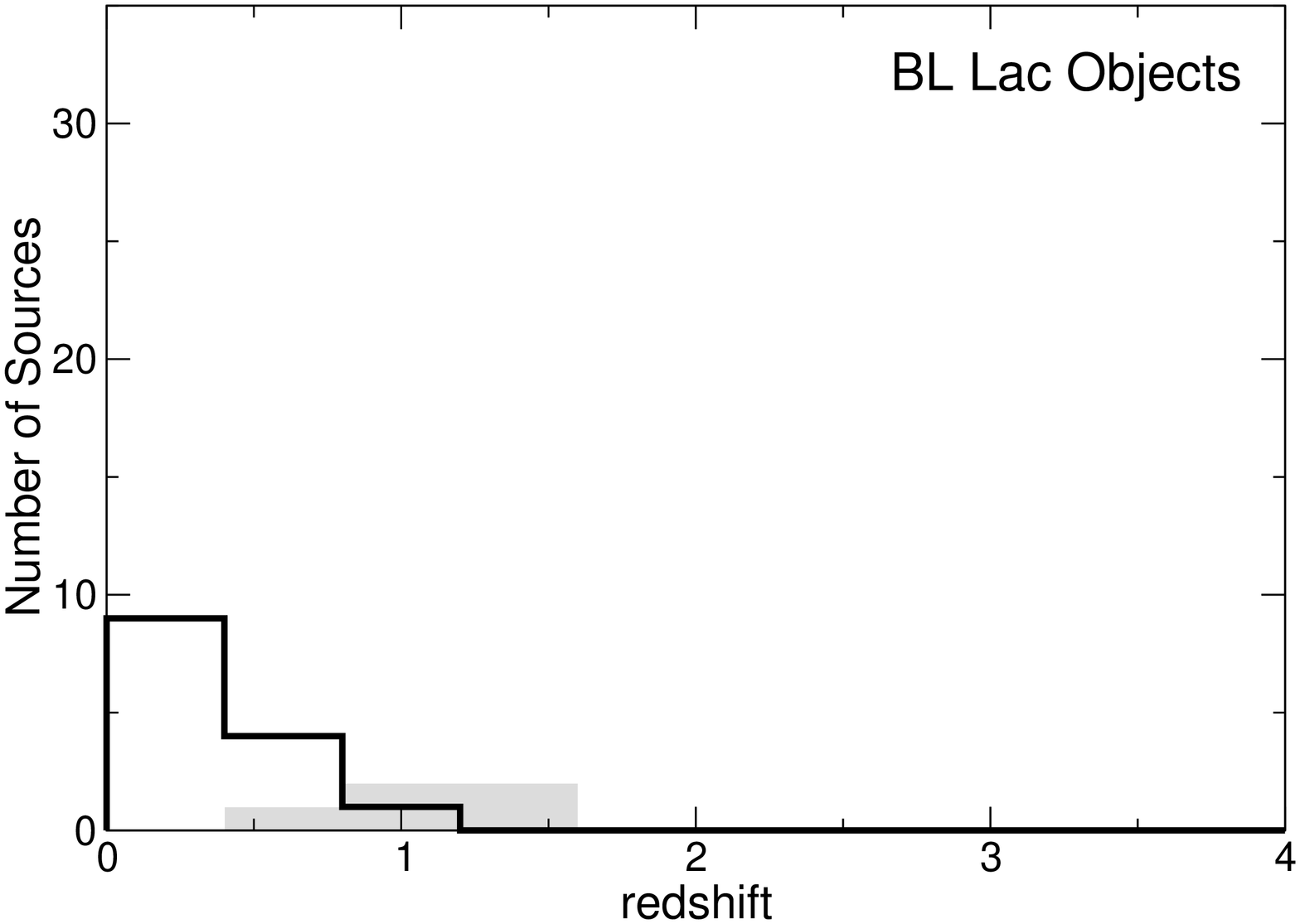}}
\hspace*{1.0cm}\subfigure[]{\includegraphics[clip,width=4.4cm,angle=-90]{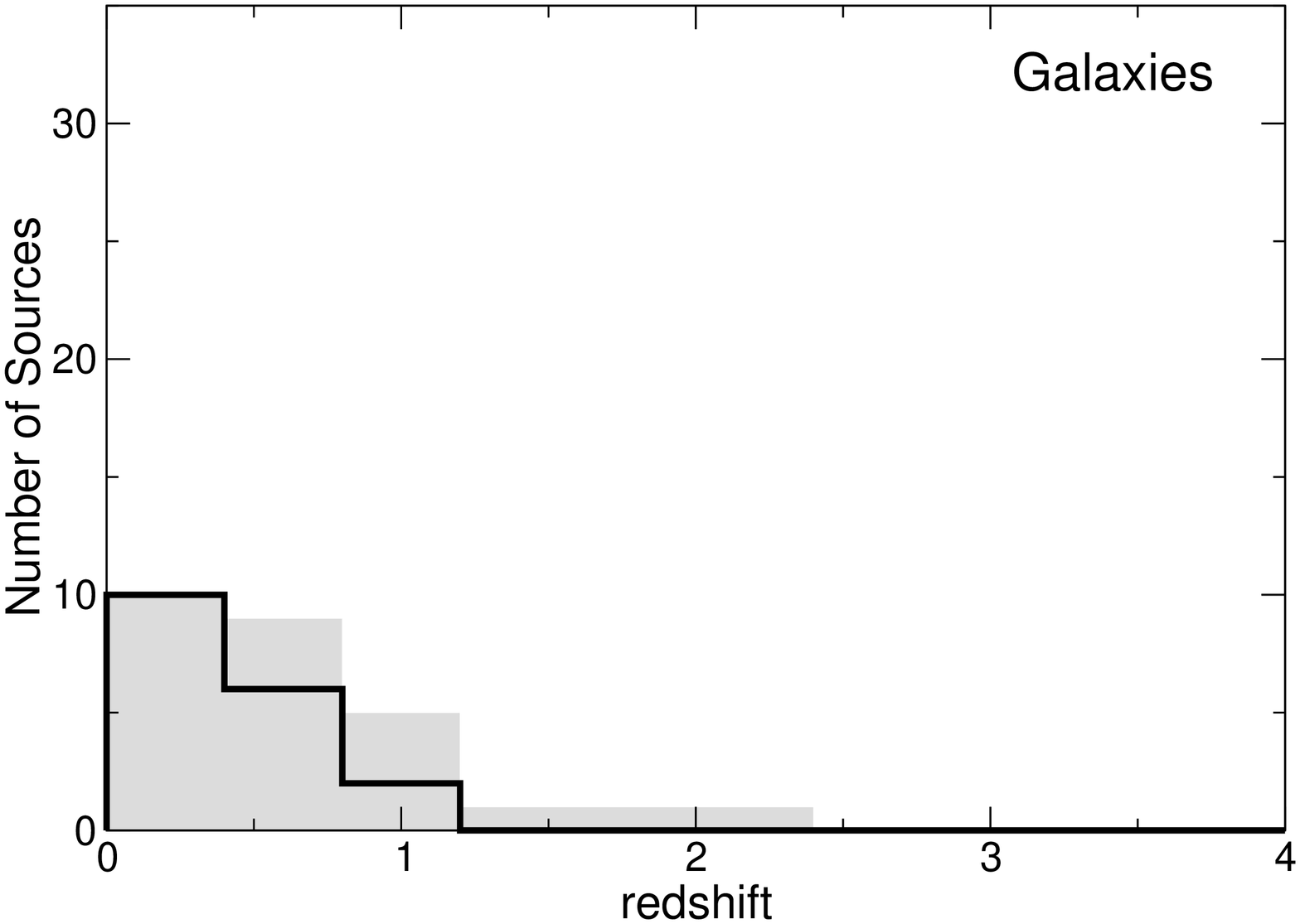}}
\caption{(a) shows the distribution of CJF sources with redshift. The source sample that has been detected by {\it ROSAT} is indicated by the solid line whereas the distribution of the non-detected source sample is shown in grey. The histograms (b-d) show the different $z$-N distributions for the three classes of objects: (b) quasars, (c) BL Lac objects, (d) radio galaxies; the detected objects are shown by the black line, the non-detections are shown in grey.}
\label{redshift}
\end{figure*}
\section{The {\it ROSAT} X-ray data}
\label{rosat}
For each of the sample sources a 1$^{\circ}\times$1$^{\circ}$ field
centered on the radio position was extracted from the RASS and analyzed using a procedure based on standard routines
within the  EXSAS environment (Zimmermann et al. 1994). This
procedure uses a maximum-likelihood source detection algorithm
(Cruddace et al. 1988), which
returns the likelihood of the existence of an X-ray source at the specified
radio position, the number of source photons within 5 times the FWHM of
the PSPC's point spread function, and the error in the number of source
photons (Cruddace et al. 1988). For the RASS, the FWHM of the PSPC point
spread function is estimated to be $\sim 60$ arcsec 
(Zimmermann et al. 1994).
Since it is known that an AGN is present at the position of the radio
source, we considered a radio source to be detected in X-rays if the
likelihood of existence is greater than 5.91, which corresponds
to 3$\sigma$. If no X-ray source is detected above the specified significance
level, we determined the 2$\sigma$ upper limit on the number of X-ray
photons. To calculate the corresponding count rates we used the
vignetting-corrected RASS exposure averaged over a circle with radius
5 arcmin centered on the radio position.\\
Several of the sources, mostly well-studied radio sources of the CJF, had
been targets of pointed observations. We extracted the data for these 
from the {\it ROSAT} archive and used them in the following analysis, as they are in
general of superior statistical significance.
In particular, if a source is not detected in the survey but is a target
of a pointed observation or found serendipitously in a pointed observation, we used the pointed data in the analysis. \\
The unabsorbed X-ray fluxes are calculated  from the measured
count rates  by  assuming power laws for the X-ray spectra 
with average photon indices of  $\Gamma = 1.8$ for radio galaxies,
 $\Gamma = 2.2$ for quasars, and  $\Gamma = 2.1$ 
for all other objects and Galactic absorption (Dickey \& Lockman 1990, Stark et al. 1992;
for details see Brinkmann et al. 1994). For correlations with the recent radio data we used for 
the luminosity 
determination a cosmology with
$h=0.71$, $\Omega_m h^2=0.135$, and $\Omega_{\rm tot}=1.02$ (for details see sect.\ref{tab}).
It should be noted that a slightly 
incorrect value of the power law slope  does not influence the flux
determination dramatically; the main source of uncertainty is the
amount of absorption of the soft X-rays.
The stated errors merely reflect the errors in
the counting statistics  of the survey sources and do not
incorporate deviations from the assumed power law slope, additional
absorption, or other systematic errors.
Therefore, for weaker sources a  total error of the X-ray flux  of
the order of $\la 25\%$ must be regarded as a conservative estimate.
 
\section{Results}
In Table~1 we present the relevant data for all 293 sources.
Column 1 and 2 give the IAU designation of the radio source,
 followed by a common name of the object, if available, and the JVAS J2000 name (Wilkinson et al. 1998, Browne 1998). In case the source is not observed within JVAS, we used the V\'{e}ron-Cetty \& V\'{e}ron (2001) edition of {\it A Catalogue of Quasars and Active Nuclei}, or the CLASS source name (Myers et al. 2003). Column 3 lists the type of the object: Quasars (Q), BL Lac objects (B), radio galaxies (G), High Polarized Quasars (HPQ), Low Polarized Quasars (LPQ), Compact Symmetric Objects (CSO), Seyfert galaxies Type 1 (Sy 1), Seyfert galaxies Type 2 (Sy 2), sources with peculiar properties (pec) or sources belonging to a cluster of objects (cluster). The next three columns give the redshift, its optical magnitude (mostly obtained from NED) and the 5~GHz Green Bank flux.
 It should be noted that in several cases
different data exist for one source, mainly for the type and the magnitude,
and we were not always able to resolve the discrepancies. 
The next column (7, EXML) gives the likelihood of the existence of an X-ray source
at the radio position in the RASS, as described in section \ref{rosat}, followed in
column (8) by the 0.1--2.4~keV X-ray flux density with its
statistical error in units of 10$^{-12}$erg~cm$^{-2}$s$^{-1}$.
If the source is not detected at a $3\sigma$ level and no pointed observations
are available, we give the $2\sigma$ upper limit to the flux density and no uncertainty in column (8).
We further want
to caution that for a couple of sources the given X-ray flux may not
originate exclusively from the radio/optical object: there are sources
with another prominent object nearby (which is not the radio source)
or which reside in a cluster of radio galaxies (like NGC~1275) where
the X-ray emission must be primarily attributed to the cluster.\\
Column 9 indicates whether a source has been found by the standard analysis of the 
{\it ROSAT} Survey data (``s''), and thus the data have been published
already elsewhere, or whether it was found in the field of a pointed PSPC or HRI observation (``p'').\\ 
Further radio data follow in columns 10 to 13. Column 10 introduces a system of comments to describe the VLA structure of the sources. In addition, references for the VLA maps are given.
The abbreviations are explained explicitly in Table~\ref{comments}. In addition, this column lists in boldface a value describing jet structure on large (VLA) scales in order to quantify the complexity of the extended emission. This is the complexity-factor (see \ref{def} for a more detailed description). Its maximum value is 5, describing the most complex extended structure.\\
Some individual sources merit comment before we proceed. We searched ADS and NED for large-scale information on the sources. For the following sources no VLA map could be found in the literature: 0344+405, 0615+820, 0800+618, 1305+804, 1306+360, 1818+356, 2116+818.
The X-ray/radio source 0450+844 appears to be associated with a background/foreground galaxy and can be understood as the active nucleus of a galaxy (17--18 mag) possibly
dominating a cluster; there are several faint objects in the vicinity of the optical galaxy on the POSS plate (Johnston et al. 1984).
The same seems to be true for the source 1010+350: it is point-like in VLA observations, and its X-ray emission seems to come from a surrounding cluster.\\
We wish to mention that the information on the extended structure has been collected from the literature (cited in Table~\ref{comments}) and due to the different map qualities results in an inhomogeneous data base. Point-like structures seen in these maps do not necessarily imply that these sources have no large-scale structure; any such putative large-scale structure may fall below the surface-brightness threshold of the VLA maps. The large-scale position angles have been measured from the published maps.\\
In column 11 we give the position angle of the VLA jet ($\theta_{\rm VLA}$), followed by the position angle of the VLBI jet (column 12, $\theta_{\rm VLBI}$). Column 13 lists the difference between the two position angles ($\Delta$PA). A "/" in column 10 or 11 denotes that no information is available on the large-scale structure, and in column 12, on a point-like structure in the VLBI-scale structure.
In several cases two values are given in columns 11, 12, or 13 indicating jet and counter-jet position angles. For further calculations we used the position angle difference determined for the one of these with higher total flux-density.
For more complete information on the pc-scale radio properties of the
sources we refer to Paper II for details on the jet-component kinematics and to Paper I for details on the VLBI maps and model-fits.\\
\subsection{X-ray properties of the sample}
\begin{figure}[htb]
\begin{center}
\includegraphics[clip,width=8.3cm]{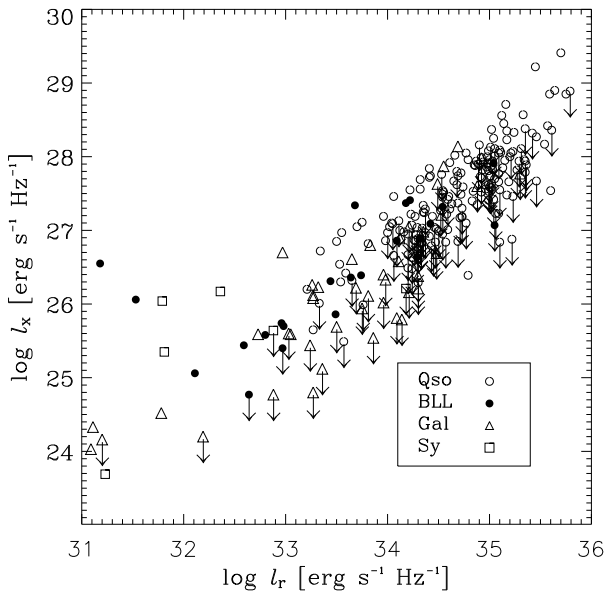}
\end{center}
\caption{The monochromatic X-ray luminosity (at 2 keV) as a function of the radio luminosity. The arrows denote
the upper limits of non-detected objects.}
  \label{figure:lumlum}
   \end{figure}

Before discussing the broad-band properties of
individual subclasses of objects we will give a general
 overview of the source content of the sample. 
In total, the CJF-survey consists of 293 sources; 167 have been detected by {\it ROSAT} and 126 have not been detected. The CJF contains 198 quasars (including Seyfert 1s), of which 93 were not detected
in the RASS survey, 53 radio galaxies (including Seyferts) of which
33 were not detected, 32 BL Lacs (8 non-detections), and 10
objects which have not yet been classified.
The highest rate of non-detection is among the radio galaxies, 
whereas most of the BL Lacs have been found as strong
X-ray emitters. In Fig.~\ref{redshift}(a) we show the redshift distributions of the CJF sources detected (solid black line) and not detected (shaded) by {\it ROSAT}. The biggest difference between the two distributions is among the nearby objects: lower $z$ objects (0$<z<$0.7) seem to have a higher likelihood to be detected by {\it ROSAT}.\\
Figs.~\ref{redshift}(b)--(d) show the distribution of Fig.~\ref{redshift}(a) separately for the three different classes of objects. In the case of the quasar distribution, higher-$z$ quasars are (preferentially) detected. BL Lac objects show the highest rate of detections among the low-$z$ objects. However, BL Lac objects are less numerous and only appear in the CJF at smaller redshifts than the quasars. The number of low-$z$ detected and non-detected radio galaxies is identical. \\
The count rates determined individually as described above 
are generally consistent (inside the mutual errors) with
the results of the standard processing of the RASS data.
Of particular interest is the group of 54 objects for
which pointed observations have been performed in the years
after the All Sky Survey.
\begin{figure}
\begin{center}
{\includegraphics[clip,width=7.3cm,angle=-90]{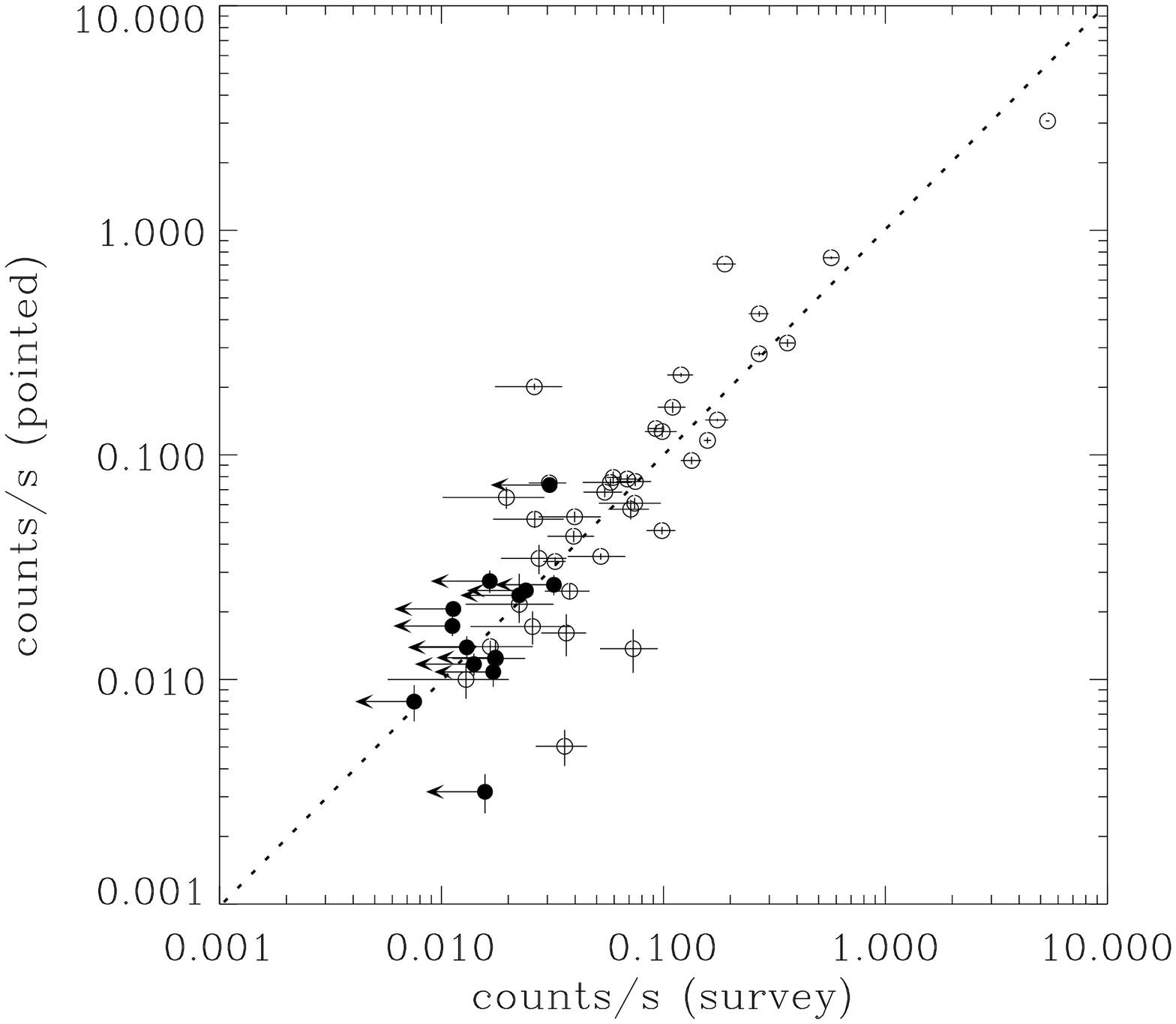}}
         \caption{Comparison of count rates obtained in pointed
	           observations with count rates from the All Sky Survey.
		                  The arrows denote upper limits. The equality line is shown dotted.}
				                      \label{figure:ctscts}
						      \end{center}
						       \end{figure}
Fig.~\ref{figure:ctscts} shows the count rate during the pointed observations as a function of the count rates obtained from the survey data. One object, the BL Lac object
Mrk 421 is outside the upper plot boundary.
Arrows at the symbols indicate 2$\sigma$ upper limits for sources non-detected in the survey observations.
The agreement between the two count rates is excellent; larger 
differences between the individual observations  
must be attributed mainly to variability of the objects. 
It should be noted that some quasars and BL Lac objects, observed repeatedly in
pointed observations, show variations of their count rates by a factor
of two to three, often accompanied by spectral changes as well.
For example, both the two extreme BL Lac objects Mrk 421 (1101+384) and 
 Mrk 501 (1652+398) are
known to show flux variations of about 50\% in different {\it ROSAT} observations.\\
In Fig.~\ref{figure:lumlum} we plot the K-corrected 
monochromatic X-ray luminosities
(at 2~keV) as a function of the 5~GHz radio luminosities for different
classes of objects in the sample. 
The arrows denote upper limits of non detected objects.
Three sources -- the radio galaxies 0651+410 and 1146+596 and the BL Lac object 
1357+769 -- have upper limits below the plot boundary and thus the
tentative identification of 1357+769 as BL Lac object might have to be revised.

\begin{figure}
\begin{center}
\includegraphics[clip,width=8.3cm]{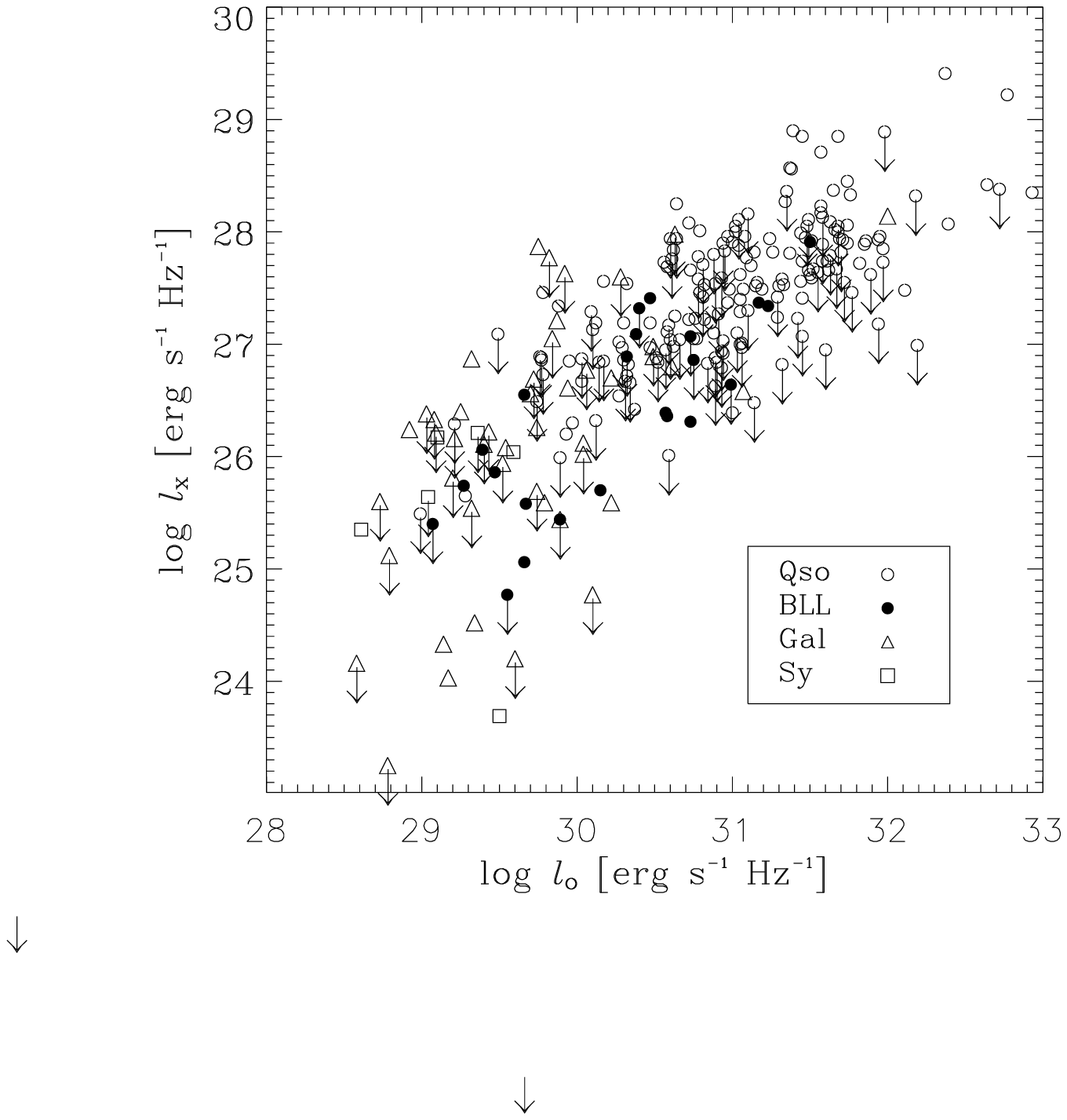}
 \end{center}
\caption {The monochromatic X-ray luminosity (at 2 keV) as a function of the optical luminosity. The arrows denote the upper limits of non-detected objects.}
\label{figure:lumlo}
\end{figure}
Interestingly, most of the objects (the radio galaxies included) exhibit a 
nearly linear relation between X-ray and radio luminosities suggesting a common mechanism for the production of X-ray and radio photons. In addition, we find a trend for the quasars to collect preferentially towards the upper part of the linear distribution while the radio galaxies seem to collect at the lower part. 
Far above this general trend in Fig.~\ref{figure:lumlum} 
we find the three extreme BL Lac objects
Mrk 421, Mrk 501 (at low radio luminosities), and 3C 66A. These three sources belong to the class of HBL where the X-ray emission is thought to be due to synchrotron radiation. 
The three Seyfert galaxies 2116+818, 0402+379, and 0309+411 show
excess X-ray emission at low radio luminosities.
0251+393 was classified as quasar by Marcha et al. (1996) and appears even as quasar over-luminous in X-rays.\\
\begin{figure}
\psfig{figure=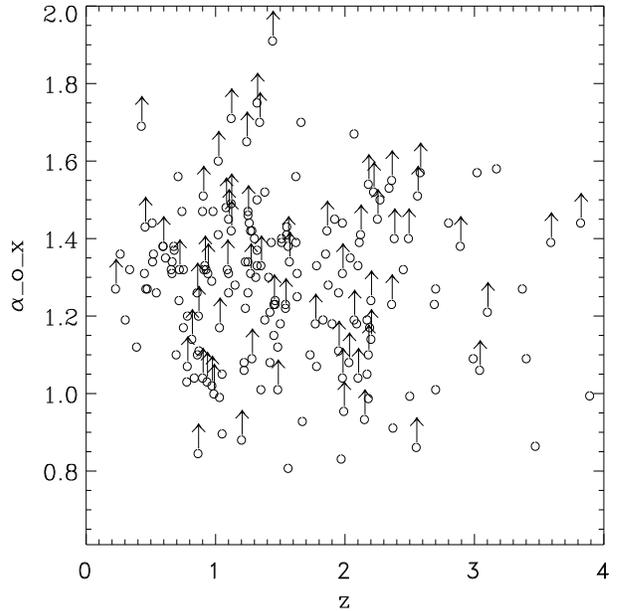,width=8.3cm,height=8.3cm,%
   bbllx=75pt,bblly=367pt,bburx=420pt,bbury=702pt,clip=}
   \caption {The X-ray loudness $\alpha_{\rm ox}$ of the quasars
    as a function of the redshift $z$. The arrows denote
     the lower limits of non-detected objects.}
     \label{figure:alpox}
     \end{figure}

Similar behaviour is seen in the plot of the optical vs. the 
 X-ray luminosities (Fig.~\ref{figure:lumlo}).
The source 1357+769 appears to be extremely under-luminous, both in
X-rays and in the optical and falls outside the lower plot boundaries.
Thus, either its classification as BL Lac object or its redshift might be
incorrect. Further, the upper limit for the X-ray luminosity of the optically rather bright galaxy 1456+375 falls below the plot boundary of Fig.~\ref{figure:lumlo}. The remarkably strong correlation between radio and X-ray luminosity and the weaker correlation between the radio and the optical luminosity have been discussed earlier by e.g., Browne \& Murphy (1987).\\
 \begin{figure*}[htb]
 \subfigure[]{\includegraphics[clip,width=5.3cm,angle=-90]{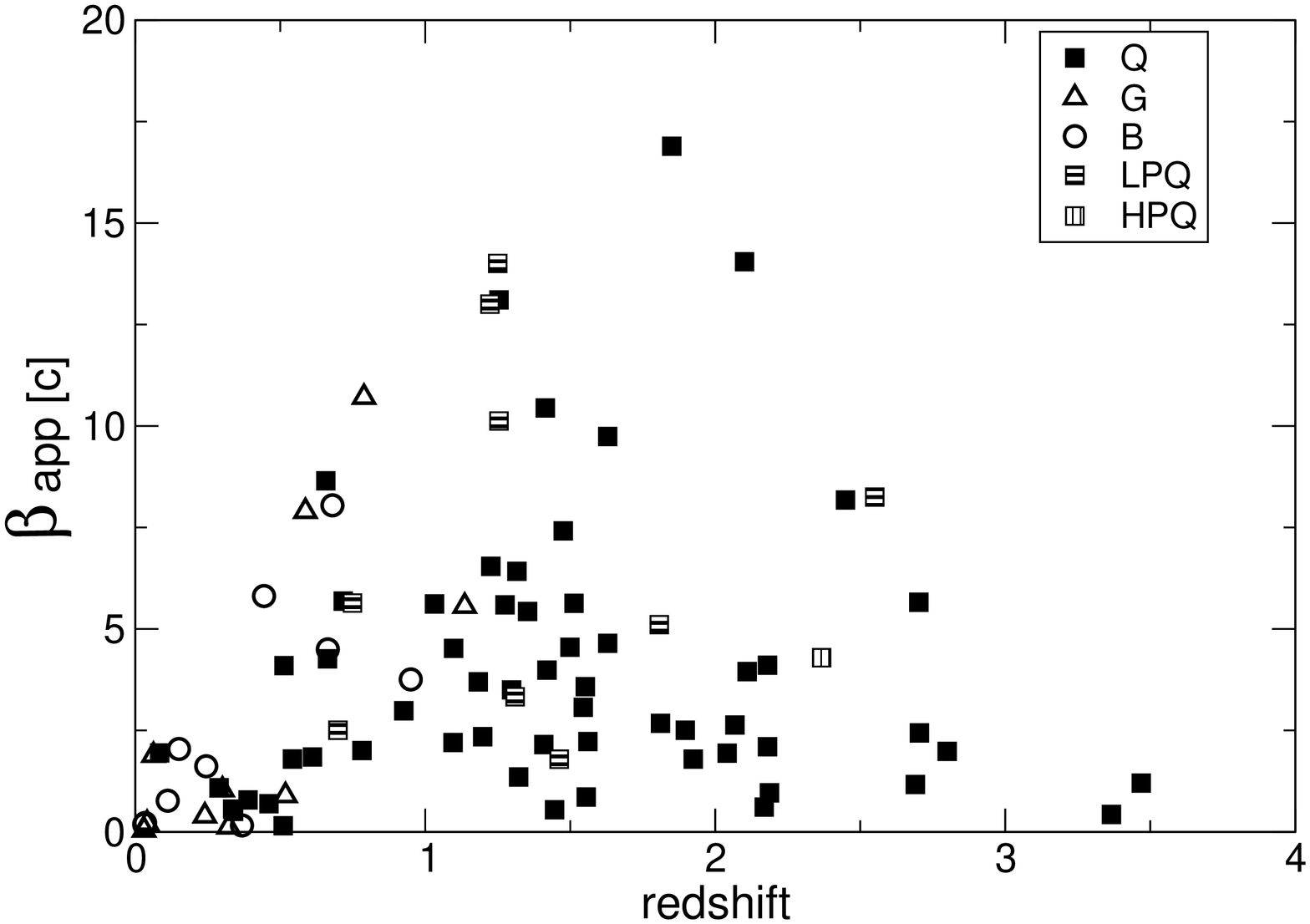}}
 \hspace*{0.5cm}\subfigure[]{\includegraphics[clip,width=5.3cm,angle=-90]{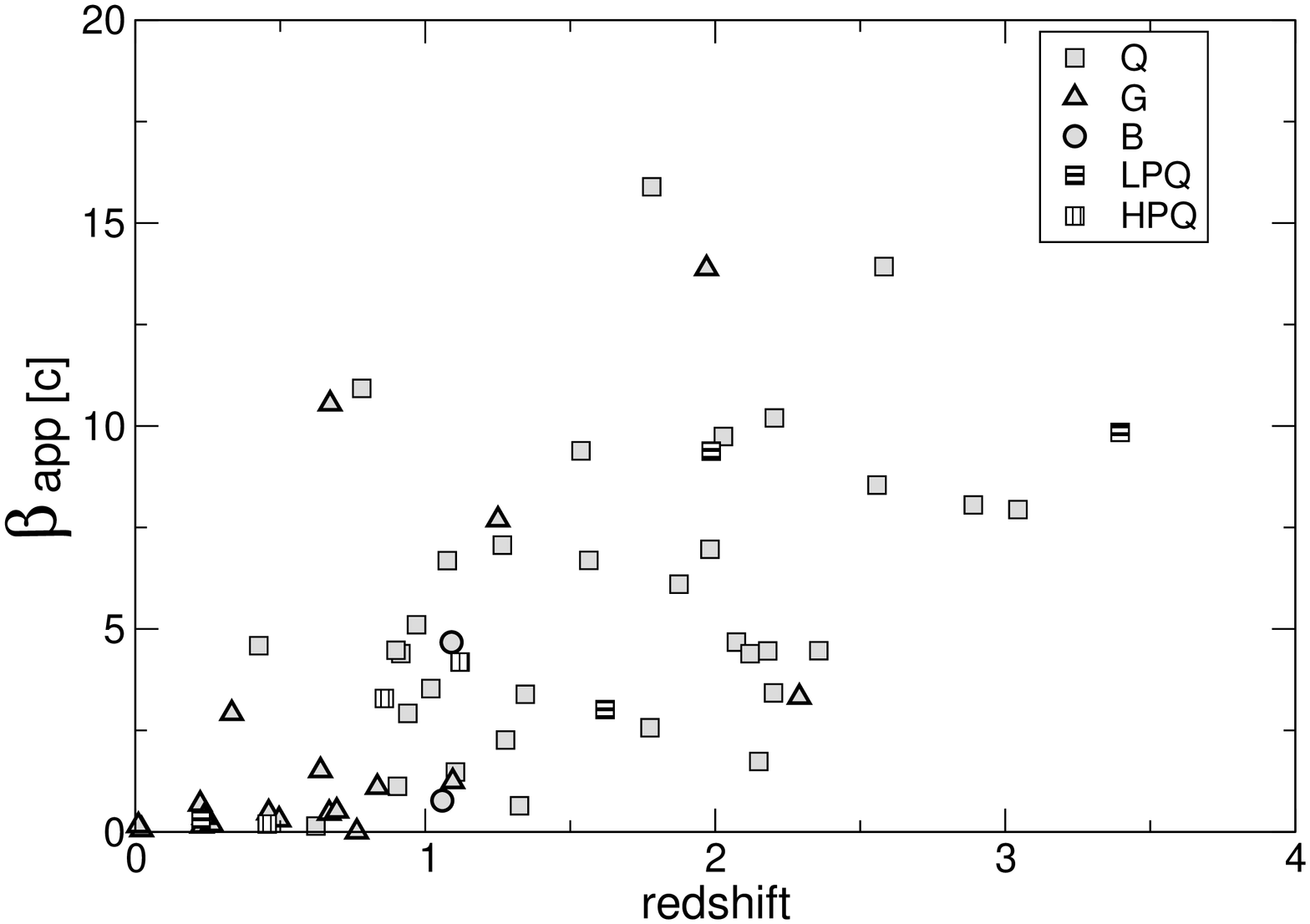}}\\
 \hspace*{-1cm}\subfigure[]{\includegraphics[clip,width=4.3cm,angle=-90]{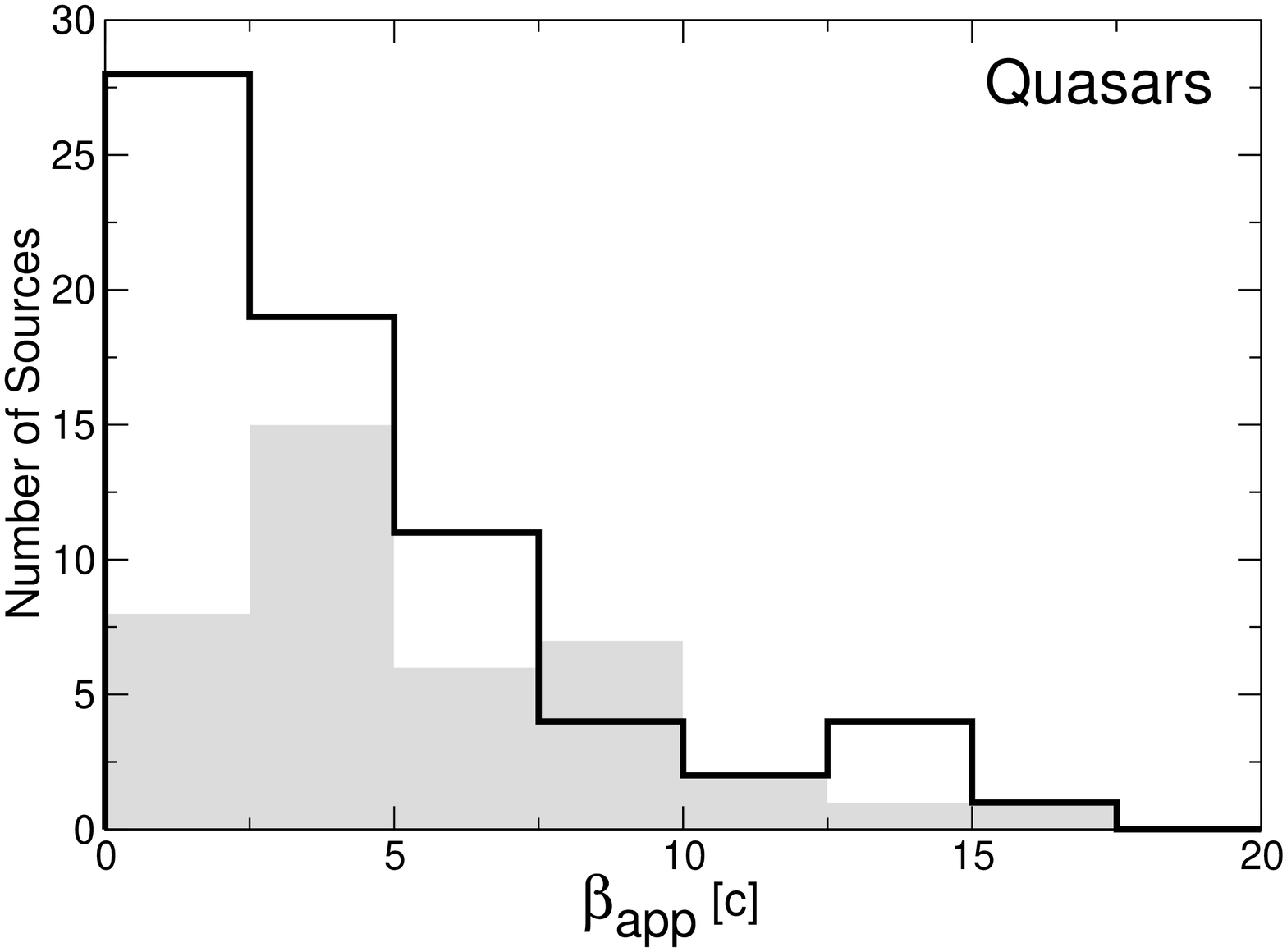}}
 \hspace*{0.7cm}\subfigure[]{\includegraphics[clip,width=4.3cm,angle=-90]{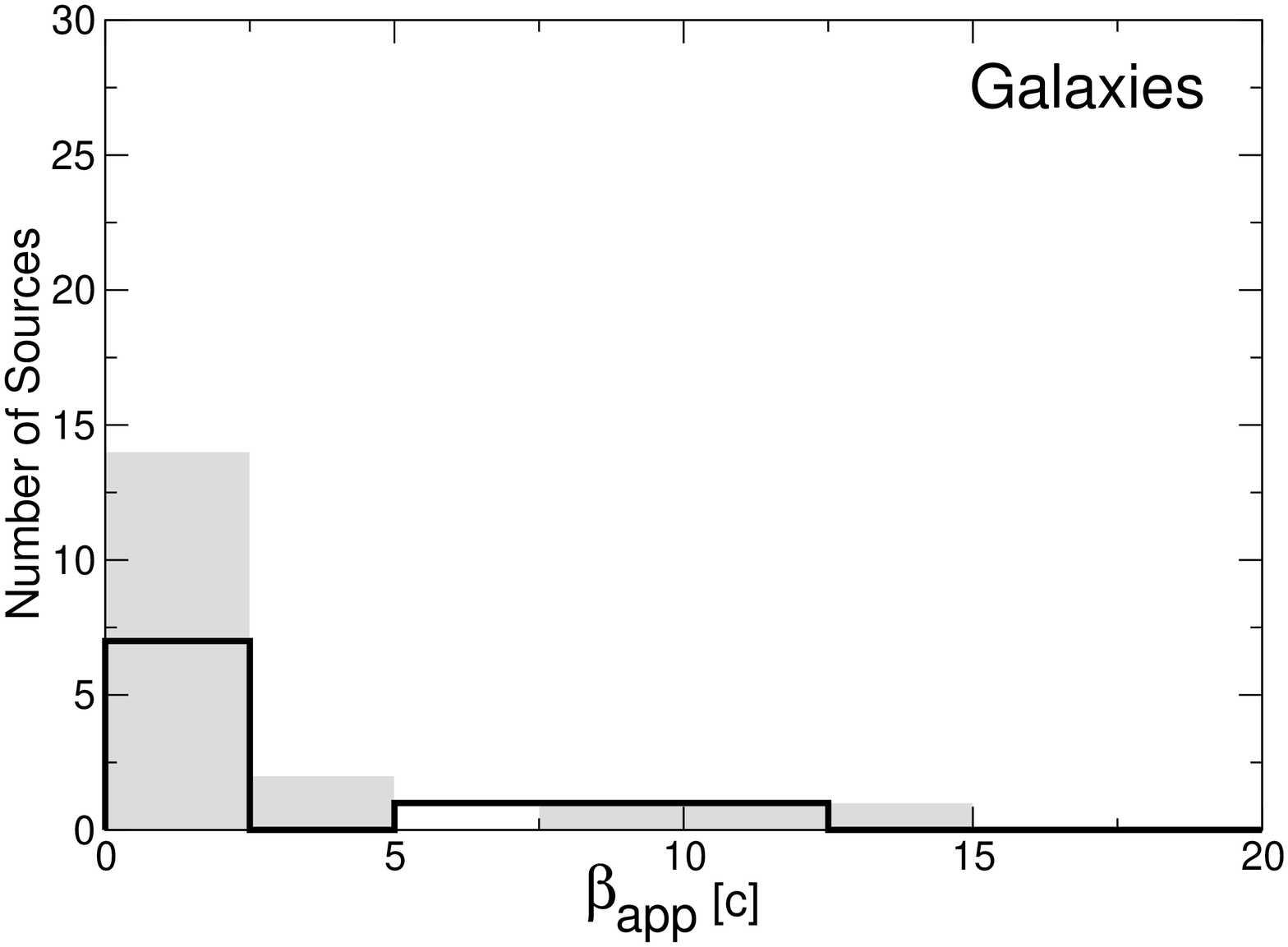}}
 \hspace*{1.0cm}\subfigure[]{\includegraphics[clip,width=4.3cm,angle=-90]{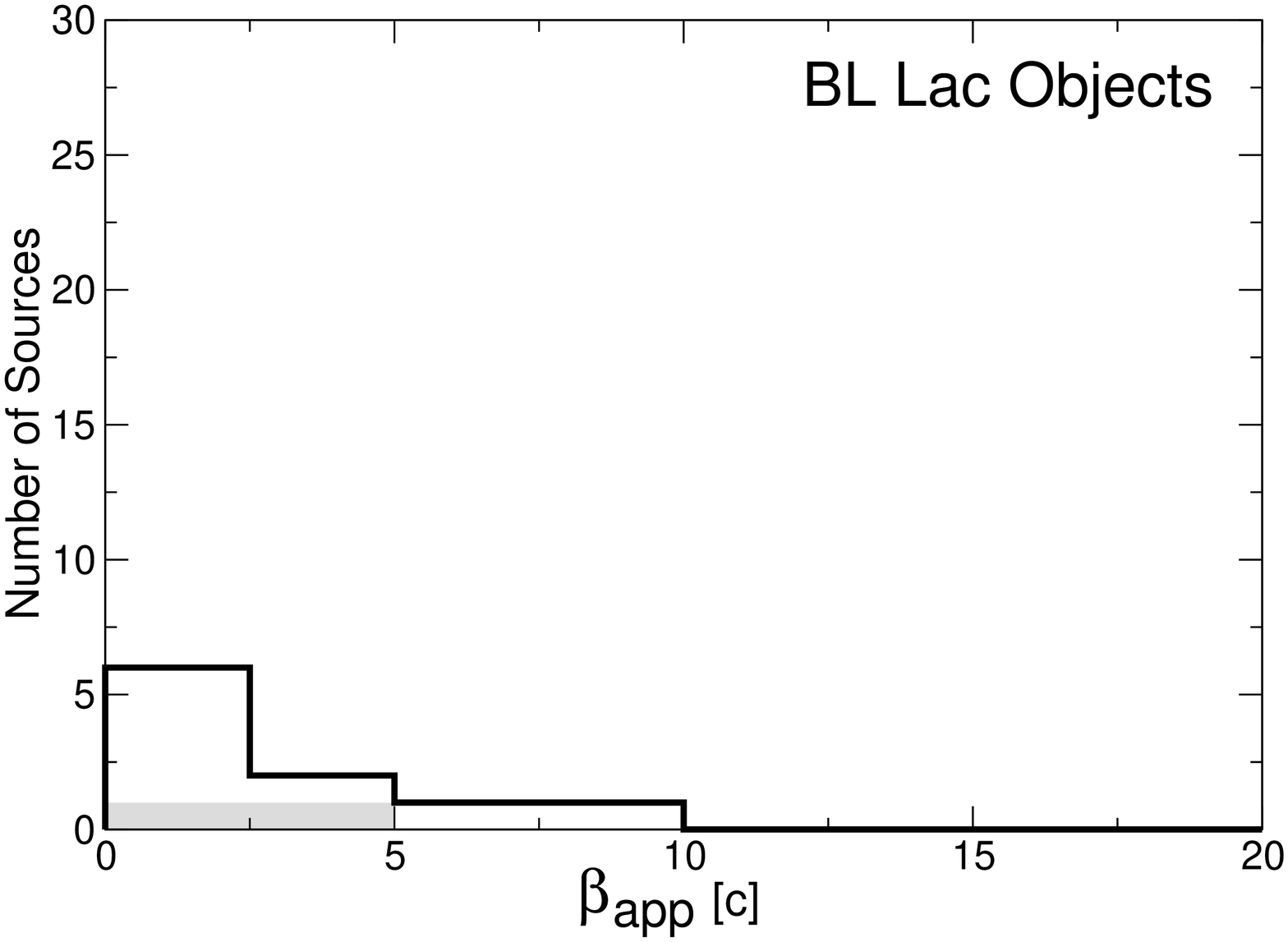}}\\
 \caption[]{The $\beta_{\rm app}$-$z$ relation is shown for those sources (source selection explained in the text) detected by {\it ROSAT} in panel (a) and those that are non detected in panel (b) (symbols identical but filled with grey).  The figures (c)--(e) show the $\beta_{\rm app}$-distribution of the sources in quasars, radio galaxies, and BL Lac objects. The non-detections are shown in grey.}
  \label{beta.classes}
  \end{figure*}

\subsubsection{Quasars}    
Flat-spectrum radio quasars are brighter in X-rays than are steep-spectrum
objects of comparable optical luminosities (Brinkmann et al. 1997).
The X-ray loudness
$\alpha_{\rm ox}= - 0.384 $~log~$(l_{2~{\rm keV}}/l_{2500{{\rm \AA}}})$ --
$l_{\rm x}$ and $l_{\rm o}$, respectively, in Fig.~\ref{figure:lumlo} --
has been used frequently in the past for the discussion of the
relative fraction of X-ray to optical emission in an evolving
quasar source population.
The average value $\langle\alpha_{\rm ox}\rangle$ = 1.28 found for the CJF quasars 
is nearly identical to the result of Brinkmann et al. (1997) for 
much larger sample of flat spectrum quasars. However, the 
dispersion in Fig.~\ref{figure:alpox} is rather large and the 
lower limits with  $\alpha_{\rm ox} > 1.7$ indicate that these 
quasars are either highly variable or X-ray quiet, like 
BAL quasars (Brinkmann et al. 1999). A further study of these
sources is highly desirable. First results from XMM-Newton observations (Brinkmann et al. 2003) indicate that internal absorption is the main cause for the X-ray quiescence of these objects.
\subsection{The determination of beaming parameters based on X-ray and radio data}
\subsubsection{The $\beta_{\rm app}$-relation for {\it ROSAT} detected and non-detected objects}
The apparent velocities observed in the core regions of AGN are thought to be caused by relativistic outflows. Relativistic motion of synchrotron-emitting plasma will result in the Doppler boosting of the synchrotron radiation.\\ 
We concentrate in this section on those sources where we have obtained the most reliable kinematic information (for details see Paper II). This is a subsample of 150 AGN with 89 sources detected by {\it ROSAT} (59 Q, 10 B, 10 G, 9 LPQ, 1 HPQ) and 61 non-detected sources (33 Q, 2 B, 19 G, 4 LPQ, 3 HPQ). In order to investigate the evolution of the apparent velocities with redshift for the detected and non-detected sources, we plot this relation in Fig.~\ref{beta.classes} in panels (a) and (b), respectively. At higher redshifts (z$\geq$1.5), we find some evidence for a lack of apparently slow sources for the non-detected sources and a tendency towards higher apparent velocities when compared to the detected sources. We plot histograms
of $\beta_{\rm app}$ for the quasars (together with LPQ and HPQ), radio galaxies, and BL Lac objects in panels (c)--(e).  The following median values for $\beta_{\rm app}$ are found for these three classes of objects in the detected sample: 3.33$c$, 0.97$c$, and 1.83$c$, respectively.  For the LPQs we find 5.64$c$. For the non-detections the values are: 4.47$c$ (Q), 0.51$c$ (G), and 2.72$c$ (B), respectively. For the LPQs and HPQs we find 6.20$c$ and 3.29$c$, respectively. The numbers of sources are small, nevertheless there is evidence in both samples for higher values for the quasars compared to the radio galaxies. We performed a K-S test comparing the distributions of $\beta_{\rm app}$ for both the quasars and the radio galaxies. 
Here and subsequently when we refer to K-S test results, the number
quoted is the probability of observing a value the K-S statistic $D$ from
two distributions drawn from the same parent that would be higher than the
one actually computed; thus a low number means that we can reject the null
hypothesis of two similar distributions with high significance.
For the comparison of the {\it ROSAT}-detected subsample, the K-S test yields 0.059,
and for the non-detected subsample 0.00013. Radio galaxies show the slowest apparent motions. This is in agreement with the results for the kinematic analysis of the complete survey CJF (see Paper II). The quasars show fastest apparent motions, faster than the BL Lac objects. We find evidence that the detected and non-detected quasars show a different $\beta_{\rm app}$-distribution (K-S test:  0.0095)
We list the median values in Table~\ref{dop.tab}. 

\subsubsection{Synchrotron Self-Compton Limit}
A fundamental parameter describing relativistic motion in AGN is the Doppler factor of the
flow,
\begin{equation}
\delta =[\Gamma(1-\beta\cos \phi)]^{-1},\\
\end{equation}
where $\beta$ is the speed (in units of the speed of light), $\Gamma=(1-\beta^{2})^{-1/2}$ is
the Lorentz factor of the flow, and $\phi$ is the angle between the direction of the flow and the line of sight.
Various Doppler factors can be calculated, based on different physical assumptions. 
Assuming that the observed X-rays are of IC origin, one can compute the IC
Doppler factor $\delta_{\rm IC}$ (e.g., Jones et al. 1974, Marscher 1987); 
this equals the real Doppler 
factor $\delta$ of the source flow only if all of the observed X-ray flux is produced 
through IC scattering.
 If part of the X-ray flux is produced by some other mechanism, 
then $\delta_{\rm IC}$ is a lower limit to $\delta$.
Ghisellini et al. (1993) and G\"uijosa \& Daly (1996) make use of almost 
the same formalism to derive this Doppler
factor for a sample of 105 sources for which the radio and X-ray data 
\begin{figure}[htb]
\begin{center}
\includegraphics[clip,width=5.8cm,angle=-90]{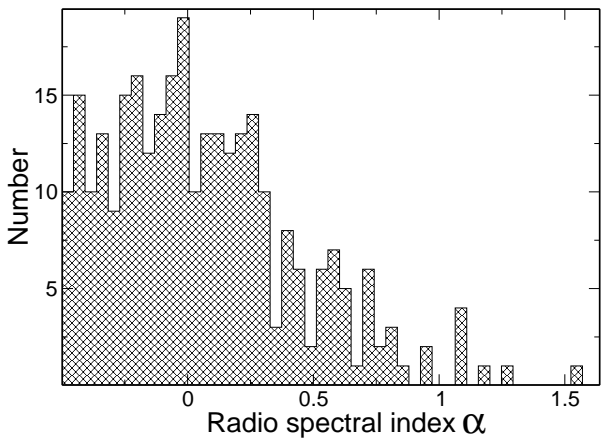}
\end{center}
\caption{The distribution of the observed spectral indices is shown (data have been taken from Taylor et al. 1996). The sign convention in Taylor et al. (1996) is different from the sign convention adopted in this paper.}
\label{index}
\end{figure}
were collected from the literature.
Here we apply the same formalism to the CJF-data and compare the results
 with those of Ghisellini et al. and G\"uijosa \& Daly.\\
Assuming the ideal case of a uniform spherical source of angular 
diameter $\theta_{\rm d}$, and a power-law energy distribution of the radiating particles 
moving in a tangled homogeneous magnetic field (in their rest frame),
one can predict the expected IC X-ray flux density, given the relevant radio and X-ray data. Using the observed fluxes this can in turn be used to determine the Doppler factor as
\begin{equation}
\delta_{\rm IC}=f(\alpha)S_{\rm m}[\frac{{\rm ln}(\nu_{\rm b}/\nu_{\rm m})\,\nu^{\alpha}_{\rm x}}{f_{\rm x}\,\theta^{6-4\alpha}_{\rm d}\nu^{5-3\alpha}_{\rm m}}]^{1/(4-2\alpha)}\,(1+z).
\end{equation}
\begin{figure*}[htb]
\vspace*{0.5cm}
\hspace*{-0.1cm}\subfigure[]{\rotate[r]{\psfig{figure=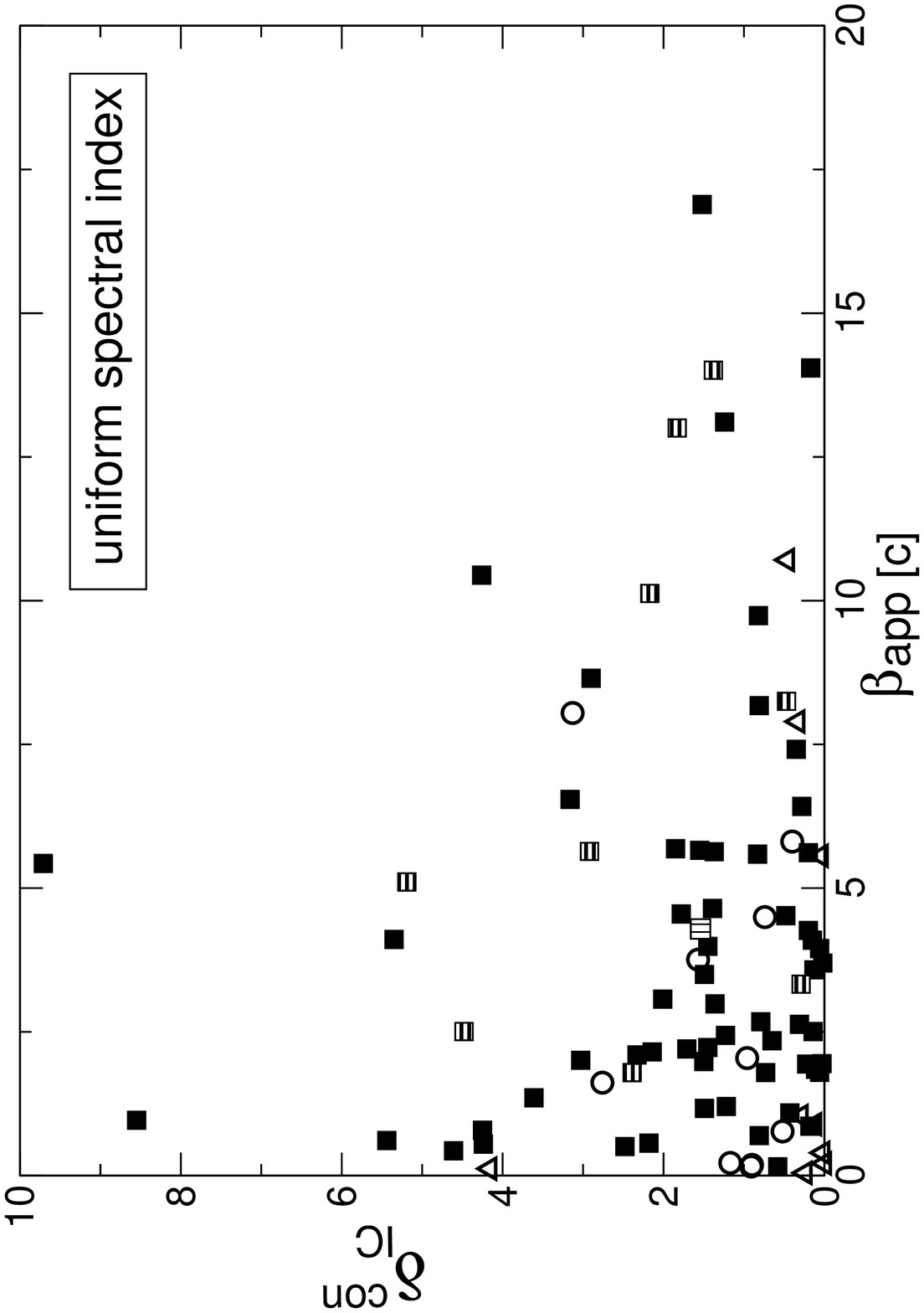,width=6.1cm}}}
\hspace*{0.3cm}\subfigure[]{\rotate[r]{\psfig{figure=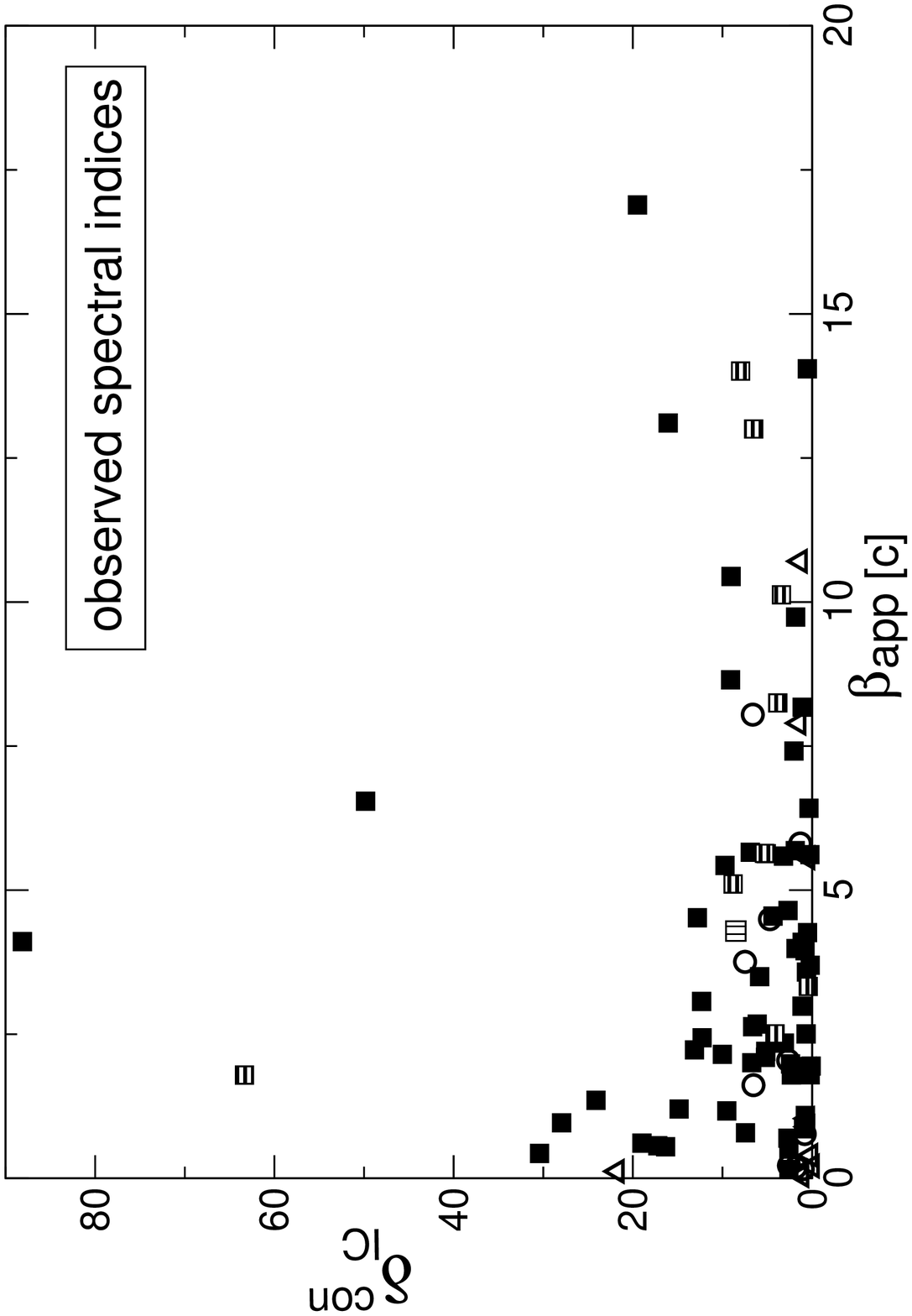,width=6.1cm}}}\\
\caption[]{The panels show the relation between $\beta_{\rm app}$ and $\delta_{{\rm IC}}^{\rm con}$ for a uniform spectral index (a) and for the individually determined spectral indices (b) in those CJF sources detected by {\it ROSAT}, including only the most reliably determined jet component motions. In both images, the same symbols as in Fig.~\ref{beta.classes} have been used for the individual source classes.}
\label{dop.beta}                                                                                                 \end{figure*}

Here $f_{\rm x}$ is the observed X-ray flux density (in Jy) at frequency $\nu_{\rm x}$ (keV),
$\nu_{\rm m}$ is the frequency at the radio peak (in GHz), $\theta_{\rm d}$ is the angular
diameter of the source (in milliarcseconds), $\nu_{\rm b}$ is the synchrotron high-frequency cutoff 
(assumed to be $10^{5}$ GHz), and $f(\alpha)\simeq -0.08\alpha+0.14$ according to Ghisellini et al. (1987). 
To convert $f_{\rm x}$ (see Table~1) from erg/cm$^{2}$/s to Jy we multiplied $f_{\rm x}$ by a factor of 0.2066 $\times$10$^{-6}$ taking emission at 2 keV into account (2 keV $\sim$ 4.84$\cdot$10$^{17}$ Hz). The flux density $S_{\rm m}$, is the value that would be obtained at $\nu_{\rm m}$ by extrapolating the optically thin spectrum (Marscher 1987). For $\alpha=-0.75$, this is about a factor of 2 larger than the observed peak flux density $S_{\rm op}$ (Marscher 1977, 1987).\\
Derivation of this formula assumes a single, spherical source. For most AGN, we observe instead a chain of jet components. In this case, according to Ghisellini et al. (1993):
\begin{equation}
\delta_{\rm continuous}=\delta_{\rm sphere}^{(4-2\alpha)/(3-2\alpha)}.
\end{equation}
All values for the Doppler factors have been calculated twice:  once for a uniform spectral index $\alpha=-0.75$ in order to enable a comparison with results presented in the literature, and once for the observationally determined individual spectral indices (Taylor et al. 1996, please see Fig.~\ref{index} for the spectral index distribution of the CJF sources).
All values are listed in the two tables \ref{non} and \ref{det}. Median values are listed in Table~\ref{dop.tab}. \\
In Fig.~\ref{dop.beta} we compare the apparent velocities for the CJF objects with the values for $\delta_{\rm IC}^{\rm con}$ derived from the SSC argument for the same components per source.  Panel (a) shows $\delta$ calculated using the same spectral index for all sources, and panel (b) using the individual spectral indices taken from Taylor et al. (1996). 
From this figure and Table~\ref{dop.tab} we expect to see whether beaming, i.e., the bulk velocity, is sufficient to explain the observed X-ray flux, since the pattern velocity does not contribute to this $\delta_{\rm IC}$.\\ According to Lind \& Blandford (1985) and Cohen \& Vermeulen (1992), the bulk velocity responsible for the boosting of the radiation could be smaller than the pattern velocity responsible for the superluminal motion (see also Ghisellini et al. 1993). In this case, the average Doppler factor of a sufficiently large sample of sources should be smaller compared to the average apparent velocity.
The presented numbers of sources are too small to allow any statistically significant conclusion. 
For either method of incorporating the spectral indices, we find some evidence for larger values of $\delta_{\rm IC}^{\rm sp}$ and $\delta_{\rm IC}^{\rm con}$ for the quasars compared to the radio galaxies (K-S test: 0.002).
This is consistent with Ghisellini et al. (1993), who derived Doppler factors for about 100 sources with known VLBI structures by comparing predicted and observed X-ray flux in the synchrotron self-Compton model. 
The derived Doppler
factors are largest for core-dominated quasars, intermediate for BL Lac
objects, and smallest for lobe-dominated quasars and radio galaxies. For a
subsample of 39 superluminal sources, Ghisellini et al. (1993) find that
apparent expansion speeds and Doppler factors correlate and have similar
average numerical values.\\
We find a better match between the median $\beta_{\rm app}$ and both $\delta_{\rm IC}$ calculated on the basis of the observed spectral indices compared to a uniformly $\alpha = -0.75$.
Although not statistically significant we find a similar median value of $\delta_{\rm IC}^{\rm con}$ and $\beta_{\rm app}$ for the quasars and of $\delta_{{\rm IC}^{\rm con}}$ and $\beta_{\rm app}$ for the BL Lac objects. This might support the conclusion that $\delta_{\rm IC}$ and $\beta_{\rm app}$ are of similar value and that there is no need to invoke other scenarios. For a statistically significant investigation of this question clearly a higher number of objects is required, as well as a significantly larger number of VLBI observations for the individual sources.
\begin{table*}[htb]
\setcounter{table}{6}
      \caption[]{Median values calculated for a set of parameters for those sources that have been detected by {\it ROSAT}.}
         \label{dop.tab}
	 \begin{center}
\begin{tabular}{|c|c|cccccc|}
            \noalign{\smallskip}
            \hline
              &Source class& Q&B&G& LPQ&HPQ& \\
            \noalign{\smallskip}
            \hline
            \noalign{\smallskip}
            &Number of objects &59&10&10&9&1&\\
	    \hline
&median $\beta_{\rm app}$  & 3.33$c$&1.83$c$& 0.97$c$& 5.64$c$& 4.29$c$&\\
\hline
$\alpha=-0.75$  & median $\delta_{\rm IC}^{\rm sp}$   & 1.3& 1.0& 0.2& 1.9& 1.4&\\
 &median $\delta_{\rm IC}^{\rm con}$  & 1.4& 0.9&0.2&2.2& 1.5&\\
\hline
observed $\alpha$&median $\delta_{\rm IC}^{\rm sp}$  & 3.1 & 2.1 & 0.6 & 3.7 & 4.8&\\
&median $\delta_{\rm IC}^{\rm con}$ & 4.4& 2.6& 0.5& 5.3& 8.5 &\\
            \noalign{\smallskip}
            \hline
         \end{tabular}
	 \end{center}
 \end{table*}
\subsubsection{Equipartition Doppler Factors}
The equipartition Doppler factors measure the ratio of the particle and magnetic energy densities.
By definition is $\delta_{\rm EQ}=\delta_{\rm IC}$ if the source is at equipartition. Otherwise, the ratio $\delta_{\rm EQ}/\delta_{\rm IC}$ measures the source's deviation from equipartition. $\delta_{\rm EQ}$ can be calculated from single-epoch radio observations by assuming that the particles and magnetic field are in equipartition (Readhead 1994).
G\"uijosa \& Daly (1996) find a strong correlation
between $\delta_{\rm EQ}$ and $\delta_{\rm IC}$ and suggest that they 
both represent reliable estimates of the true Doppler factor. 
We use the formula for the equipartition Doppler factor given by Readhead (1994) and G\"uijosa \& Daly (1996):\\
\vspace*{0.30cm}
\begin{eqnarray}
\delta_{\rm EQ}=\{[10^{3}F(\alpha)]^{34}{([1-(1+z)^{-1/2}]/2h)}^{-2}(1+z)^{(15-2\alpha)} \\
\nonumber
\times S_{\rm op}^{16}\theta^{-34}_{\rm d}(\nu_{\rm op}\times10^{3})^{-(2\alpha+35)}\}^{1/(13-2\alpha)}
                               \end{eqnarray}
\vspace*{0.2cm}

\begin{figure}[htb]
\vspace*{1.0cm}
\subfigure[]{\includegraphics[clip,width=5.8cm,angle=-90]{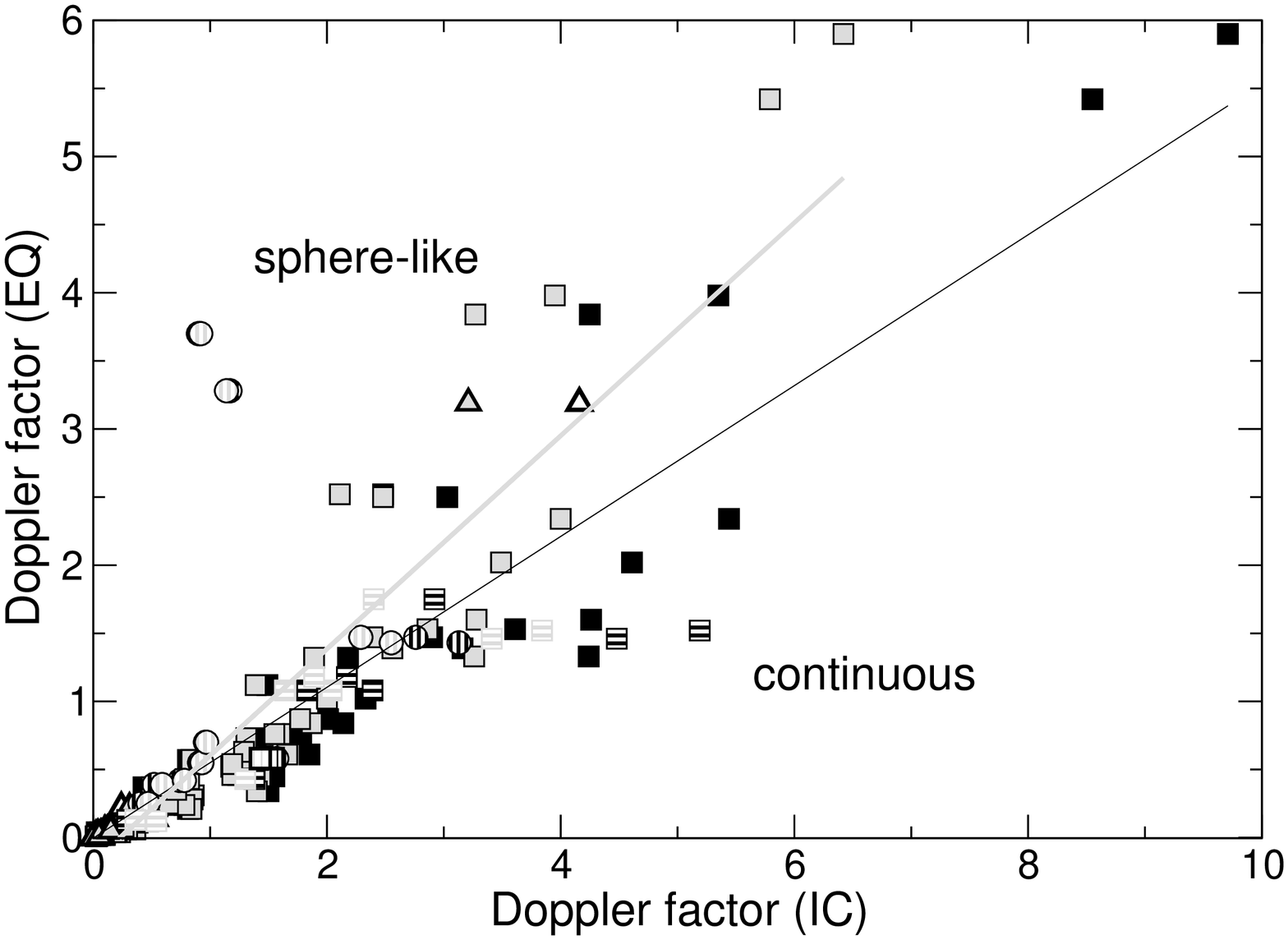}}
\subfigure[]{\includegraphics[clip,width=5.8cm,angle=-90]{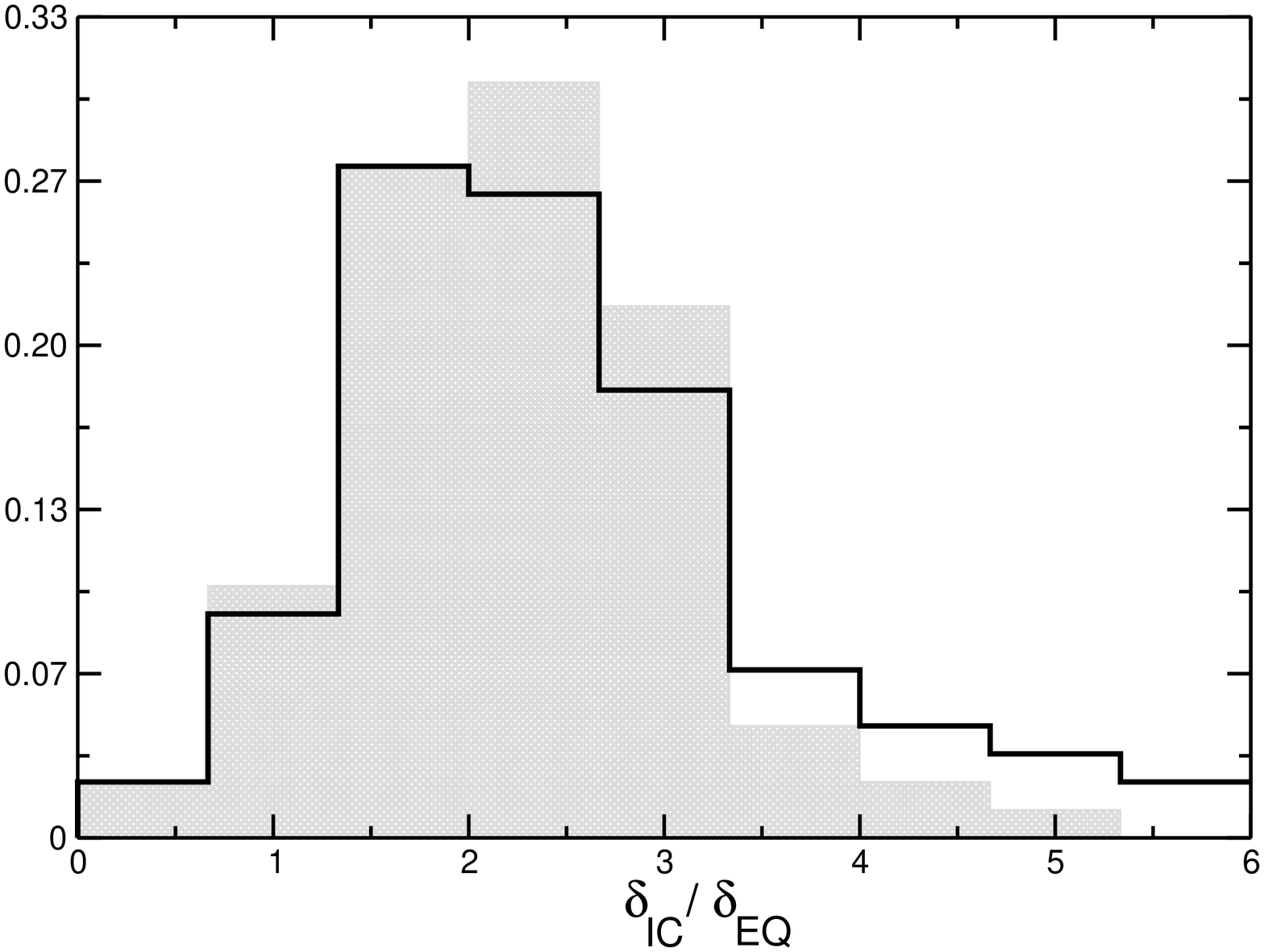}}
\caption[]{The relation between the equipartition $\delta_{\rm EQ}$ and IC Doppler $\delta_{\rm IC}$ factor for spherical and continuous jets is shown in (a). The continuous-jet case is displayed in black symbols (same symbol forms as in Fig.~\ref{beta.classes}). The solid lines represent linear regressions to the data (black: continuous jets, grey: spherical jets). In (b) a histogram shows the ratio between $\delta_{\rm IC}$ and $\delta_{\rm EQ}$ for continuous jets (black) and spherical jets (grey).}
\label{deic.deeq}
\end{figure}

\noindent
The equation as well as a graph for $F(\alpha)$ are given in Scott \& Readhead (1977). Here
we only need $F(-0.75)=3.4$. 
The calculated values for $\delta_{\rm EQ}$ are listed in Table~\ref{non} and \ref{det}.\\
In Fig.~\ref{deic.deeq} (a) we show the relation between $\delta_{\rm IC}$ and $\delta_{\rm EQ}$, and we calculate $\delta_{\rm IC}$ for both sphere-like and continuous jets.  
We find a good correlation between both Doppler factors. The probability that the correlation is spurious is 1$\times 10^{-7}$ for the continuous jet case and 5$\times 10^{-7}$ for the sphere-like jet case. $\delta_{\rm IC}$ and $\delta_{\rm EQ}$ are about equal as shown in the histogram displaying the ratio between these two values in Fig.~\ref{deic.deeq} (b). Both seem to present reliable estimates of the true Doppler factor, as found by G\"uijosa \& Daly (1996). We list the values in Table~\ref{non} and \ref{det}. We wish to stress that the calculation of both Doppler factors depends critically on $\theta_d$ which is raised to the biggest exponent in equations (2) and (4) when each is re-factored into products of powers of the individual observables, considering the case
of uniform $\alpha = -0.75$.
\subsubsection{Bulk Lorentz Factor and Viewing angle}
In the ballistic model of knot motion ($\Gamma_{\rm pattern}=\Gamma_{\rm bulk}$), the Lorentz factor $\Gamma$ and the viewing angle $\phi$ can be calculated with the help of $\delta _{\rm IC}$ and $\beta_{\rm app}$ (e.g., Ghisellini et al. 1993):
\begin{equation}
\Gamma=\frac{\beta_{\rm app}^{2}+\delta_{\rm IC}^{2}+1}{2\delta_{\rm IC}}\\
\end{equation}
\begin{equation}
\tan(\phi)=\frac{2\beta_{\rm app}}{\beta_{\rm app}^{2}+\delta_{\rm IC}^{2}-1}.
\end{equation}
Although we do not find any significant trend with regard to these two parameters, we include the calculated values for $\Gamma$ and $\phi$ for completeness in Table~\ref{non} and \ref{det}.\\

\subsubsection{Brightness temperature}
\begin{figure}[htb]
\begin{center}
\includegraphics[clip,width=5.8cm,angle=-90]{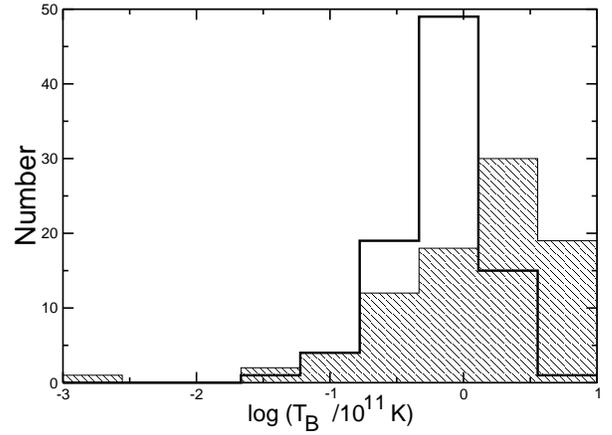}
\end{center}
\caption[]{The brightness temperature distribution for the observed brightness temperatures (solid line) and the intrinsic brightness temperatures (hatched area). In both cases the values for continuous jets have been determined.}
\label{tem}
\end{figure}

We observationally determine the brightness temperature via the following relation:\\
\begin{equation}
T_{\rm B}=1.77 \times 10^{12}\frac{F_{\rm m}}{\theta_{\rm d}^{2}\nu_{\rm m}^{2}}(1+z)
\end{equation}
for continuous jets (taken from Ghisellini et al. 1993). The intrinsic brightness temperature can be determined from the observed one using the relation $T_{\rm B}=T_{\rm Bi}\delta$ for the case of the moving sphere or $T_{\rm B}\propto\delta^{(4-2\alpha)/(5-2\alpha)}T_{\rm Bi}$ for the continuous case.
In Fig.~\ref{tem} the distributions of the brightness temperatures (observed and intrinsic) for continuous jets are shown.
\subsubsection{Core dominance parameter}
\begin{figure*}[htb]
\begin{center}
\hspace*{-1.0cm}\subfigure[]{\includegraphics[clip,width=4.4cm,angle=-90]{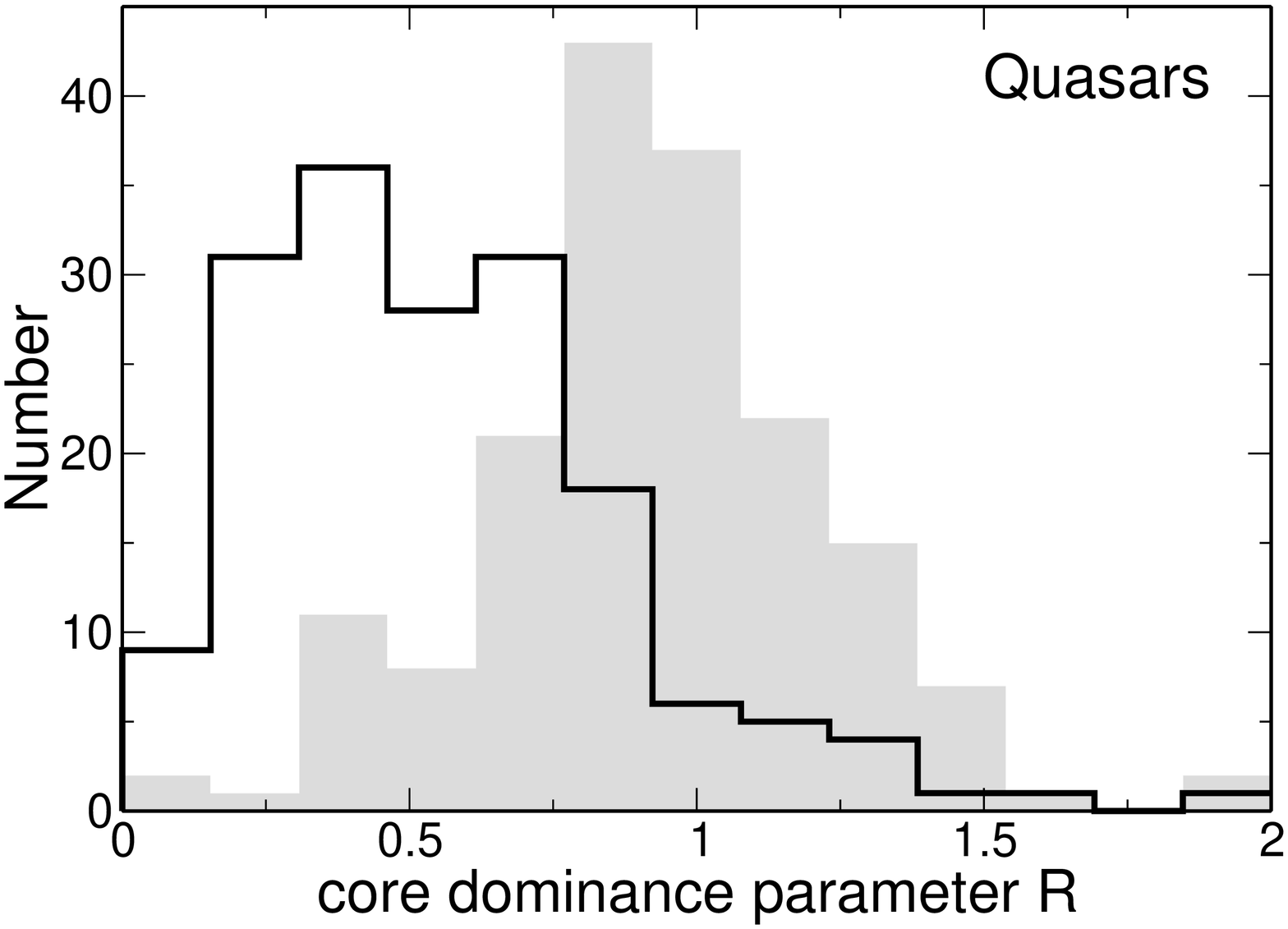}}
\hspace*{0.1cm}\subfigure[]{\includegraphics[clip,width=4.4cm,angle=-90]{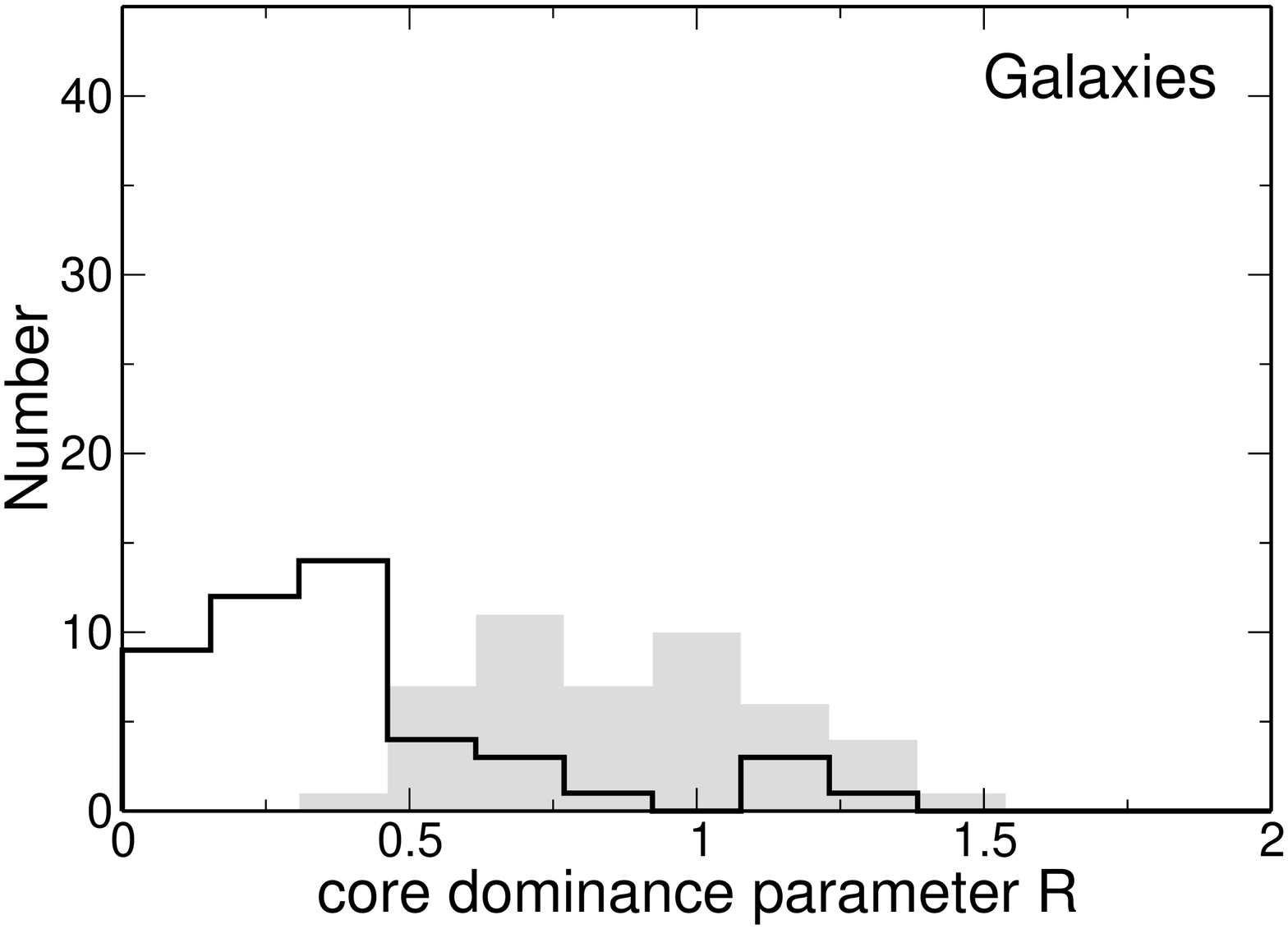}}
\hspace*{0.1cm}\subfigure[]{\includegraphics[clip,width=4.4cm,angle=-90]{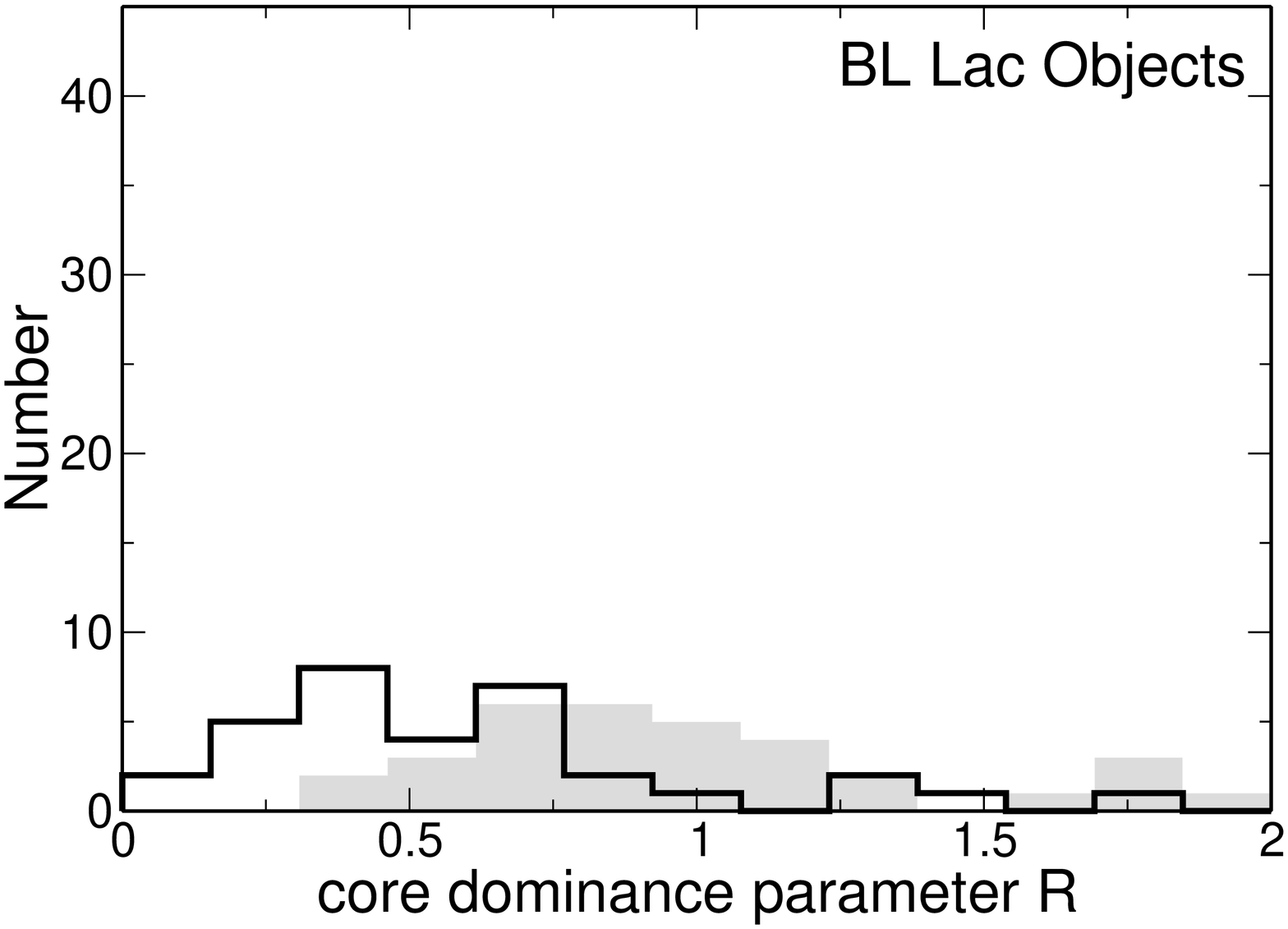}}\\
\hspace*{-1cm}\subfigure[]{\includegraphics[clip,width=4.4cm,angle=-90]{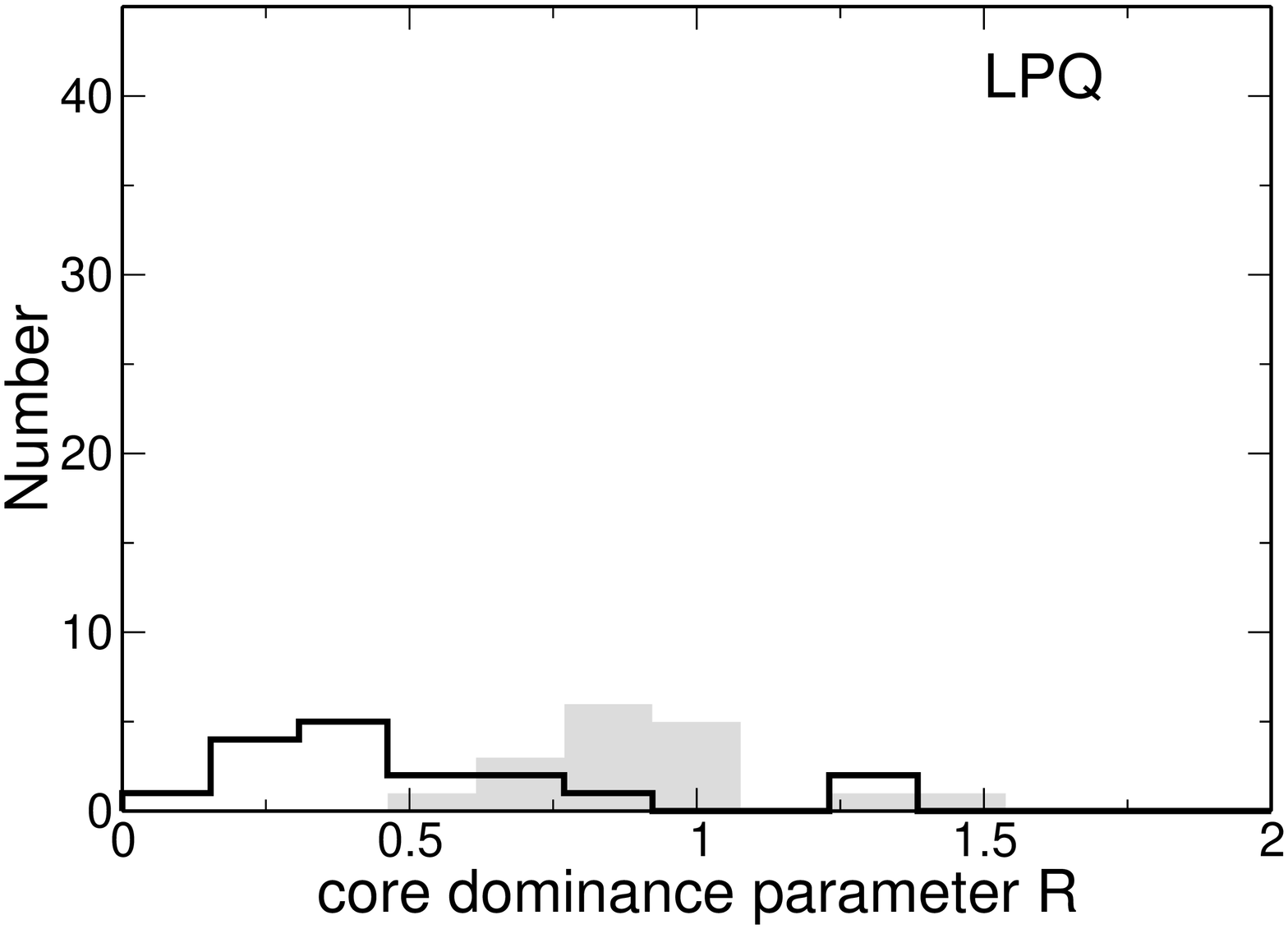}}
\hspace*{0.1cm}\subfigure[]{\includegraphics[clip,width=4.4cm,angle=-90]{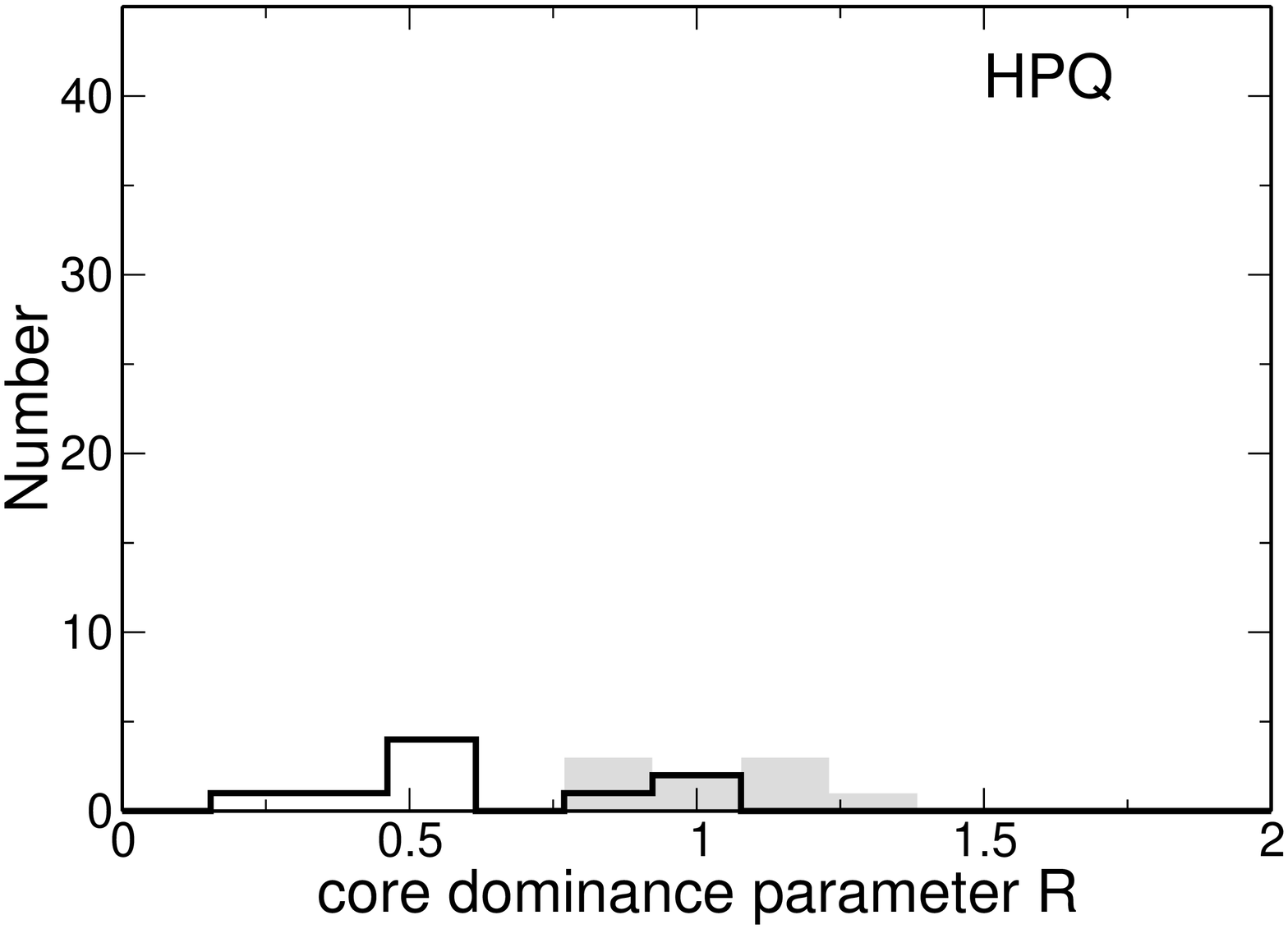}}
\end{center}
\caption[]{The distribution of the core dominance parameter $R$ is shown for different classes of objects. We show the distributions of $R$ defined as ratio between the VLBI core flux-density ($R_{\rm C}$, solid line) and the total VLBI flux-density ($R_{\rm V}$, grey), and the Green Bank 5 GHz flux-density.}
\label{cored}
\end{figure*}

\begin{figure}[htb]
\vspace*{7cm}
\hspace*{1.2cm}
\includegraphics[width=12.1cm]{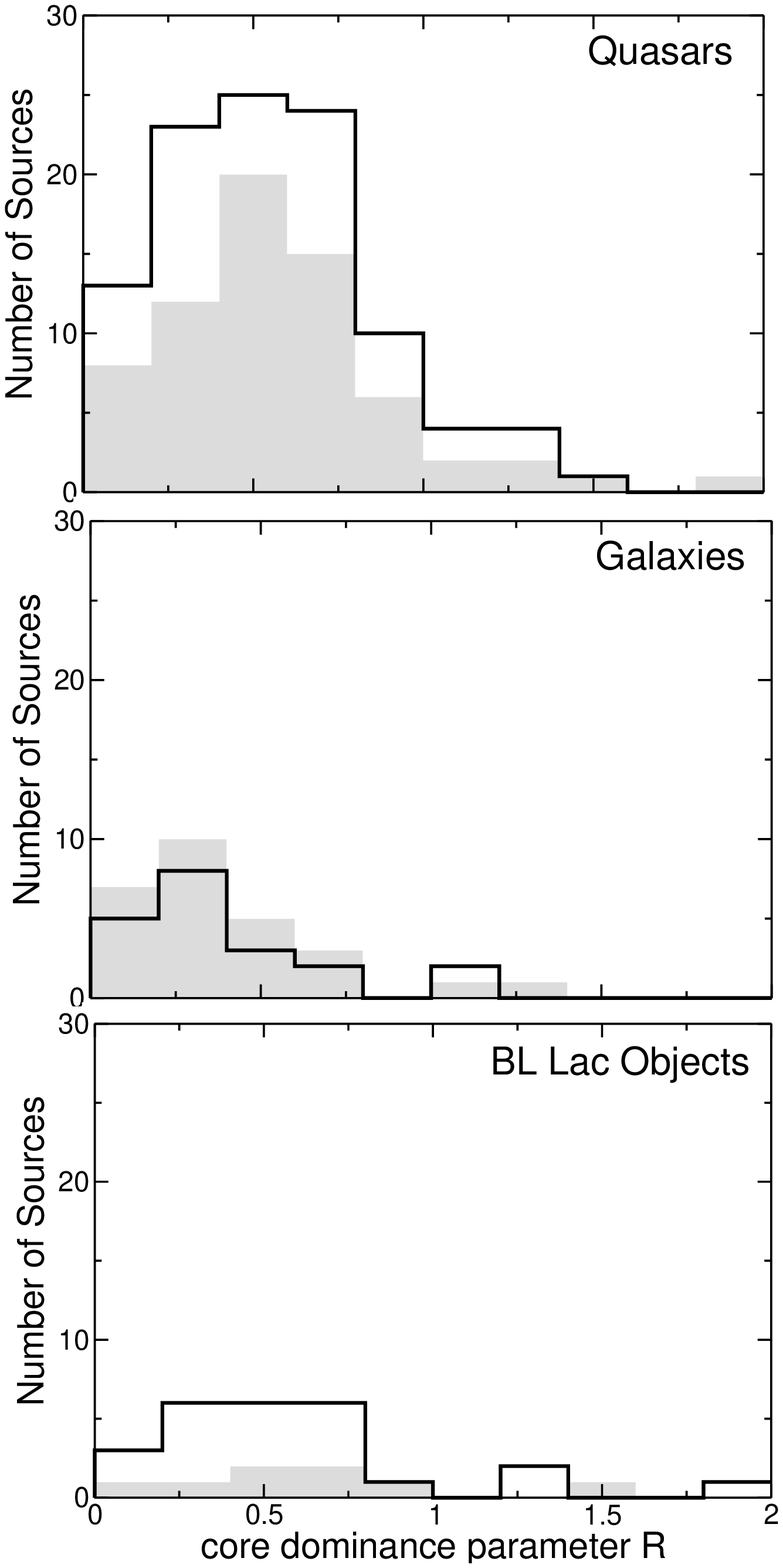}
\caption[]{The core dominance parameter $R$ (based on the CJF core flux-density) is shown for the {\it ROSAT} detected objects and for the non-detections (grey) for the main three classes of objects.}
\label{coredetnon}
\end{figure}

Within the currently accepted scenario of relativistic beaming, the emission of AGN is composed of two components. The pc-scale emission arises from Doppler-boosted jet emission while the extended emission from mainly isotropic radiation. The ratio of the two emissions, in the sense compact/extended, is defined as the core dominance parameter (hereafter: $R$; e.g., Orr \& Brown 1982). In the literature, $R$ is either calculated as ratio of flux densities or as ratio of luminosities (e.g., Punsly 1995). 
$R$ is defined here as the ratio of the flux densities.
In both cases, $R$ gives a measure for the role beaming plays in the appearance of this source.
In Fig.~\ref{cored} we show the $R$-distributions for the three optical classes
of objects. 
We calculate and compare two core dominance parameters, both using the Green Bank 5 GHz flux density as the denominator:  $R_V$ uses the total VLBI flux density from Taylor et al.\ (1996), and $R_C$ uses the VLBI core flux density taken from the model-fit parameters of Paper I.  The value for $R$ is expected to be smaller than 1 since the Green Bank flux contains the complete flux of the source while the 'compact' portion only contains the flux from the central region of the source. The reason for
calculating and comparing both values is that we expect to see an even
clearer trend in using the VLBI core flux, although usually the total VLBI flux
is used. Optically thin components contribute to the total VLBI flux while
these components do not contribute to $R$ calculated with the core flux only.
In Table~\ref{coredominance} we list the median values
for the different types of objects and mark the different types of $R$.
We find in general smaller median values for those $R$ that have been
calculated based on the VLBI core flux for all classes of objects.
In addition $R_C$ shows larger differences between
the values for the individual classes. We find significantly higher values 
of $R_C$ for
the quasars compared to the radio galaxies (K-S test: 0.0001) and 
for the BL Lac objects compared to the radio galaxies (K-S
test:0.009) when considering all sources, detected by {\it ROSAT} or not.
This finding also applies both for the detected sample (KS-Test
Q/G: 0.032; B/G: 0.050) as for the non-detected subsample (Q/G: 0.006,
B/G: 0.056). Fan \& Zhang (2003) find smaller $R$ for radio galaxies than for
quasars, which are even smaller than for BL Lac objects. This is in agreement
with our results presented here.
\begin{table*}
\setcounter{table}{7}
\begin{center}
\caption{Median values for the core dominance parameter $R$ based on two sorts of VLBI flux-density measures and the Green Bank 5 GHz flux-densities. The Green Bank and total VLBI flux-densities have been taken from Taylor et al. (1996). For those sources where no total VLBI flux-densities were given in Taylor et al., we obtained the values from Paper I. The VLBI core flux-densities have been taken from Paper I. 
\label{coredominance}}
\begin{tabular}{|c|c|ccccc|}
\hline\noalign{\smallskip}
&   Sample    &Q&G&B&LPQ&HPQ \\ \hline
$R_{\rm C}$ (VLBI core flux)&All &0.52 &0.33 &0.54 &0.39&0.52\\ 
$R_{\rm V}$ (total VLBI flux)&All &0.92  &0.84  &0.90&0.90&1.06\\
\hline
\hline
$R_{\rm C}$ (VLBI core flux)&detections&0.50&0.35&0.54&0.38&0.53\\
$R_{\rm C}$ (VLBI core flux)&non-detections&0.54&0.33&0.56&0.48&0.52\\
\hline

\hline
\noalign{\smallskip}
\end{tabular}
\end{center}
\end{table*}
The distributions of $R$ for all the detected and non-detected objects per source class are shown in Fig.~\ref{coredetnon}.  We find no significant differences between the detected and non-detected subsamples of classes of objects.   
\begin{figure}[htb]
\includegraphics[clip,width=5.8cm,angle=-90]{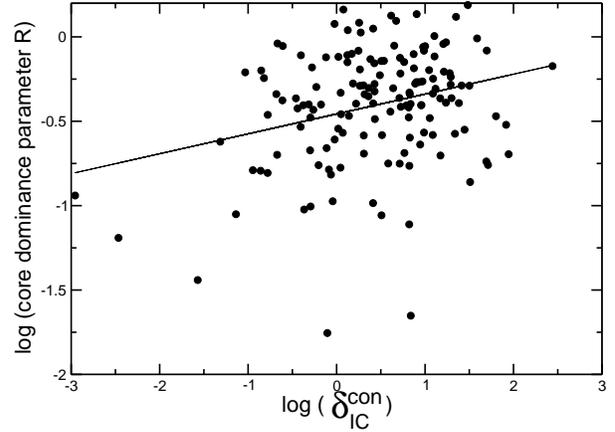}
\caption[]{The relation between the logarithms of $\delta_{\rm IC}^{\rm con}$ and the core dominance parameter $R_{\rm C}$ is shown.}
\label{coredreg}
\end{figure}
In Fig.~\ref{coredreg} we show the relation between the logarithms of $\delta_{\rm IC}^{\rm con}$ and $R_{\rm C}$. A correlation coefficient of 0.28 is not significant.

\subsubsection{Tabulated values}
\label{tab}
All values calculated in this and the following sections are listed in Tables ~\ref{non}--\ref{detrest} (these tables are only available in the online edition of  the Journal). The values shown in the plots of this paper are listed in Table~\ref{non} (for the sources not detected by {\it ROSAT}), and in Table~\ref{det} (for the sources detected by {\it ROSAT}). Not all of the CJF sources could be included in the correlation analysis since
$\beta_{\rm app}$ could not be determined for all the sources. A detailed
overview of the sources and jet components that have been used for the
kinematic analysis, and the reasons for source elimination from this kinematic
analysis, is given in Paper II. For completeness we list in Table~\ref{nonrest}
and Table~\ref{detrest} values for those sources without $\beta_{\rm app}$.
These values are not shown in the plots of this paper. Tables~\ref{non} and
\ref{det} list the source name (1) and $\beta_{\rm app}$ (2).  Columns
(3)--(9) list parameters obtained under the assumption of a uniform
spectral index $\alpha = -0.75$:  $\delta_{\rm IC}$
(3), the continuous $\delta_{\rm IC}$ (4), $\delta_{\rm EQ}$ (5), the logarithm
of the observed brightness temperature $T_{\rm B}$ (6), the logarithm of the
intrinsic brightness temperature $T_{\rm Bi}^{\rm con}$ (7), the bulk Lorentz
factor $\Gamma_{\rm sl}$ (8), and the angle to the line of sight $\phi$ (9).
Columns (10)--(16) list the same parameters computed using each source's
observed spectral index (Taylor et al. 1996). Continuing, the remainder
of the columns list the
VLBI flux (17, taken from Taylor et al. 1996), the flux of the core
determined from the model-fit parameters (18, taken from Paper I), the logarithm of the core dominance $R_{\rm C}$ calculated on the basis of the VLBI core flux-density (19), and the core dominance $R_{\rm V}$ calculated 
on the basis of the total VLBI flux-density (20). 
We used the 1-year WMAP data (Spergel et al., 2003) to obtain values for the cosmological parameters ($h=0.71$, $\Omega_m h^2=0.135$, and $\Omega_{\rm tot}=1.02$); differences in the apparent velocities due to differences between the 1-year and 3-year WMAP parameters are negligible with respect to the formal measurement errors of the velocities. This is discussed in more detail in Paper II. \\ 

Tables~\ref{nonrest} and \ref{detrest} list the source name in column (1).
Columns (2)--(6) list parameters obtained under the assumption of a uniform
spectral index $\alpha=-0.75$: $\delta_{\rm IC}$ (2), the 
observed brightness temperature $T_{\rm B}$ (5), and the logarithm of the intrinsic
brightness temperature $T_{\rm Bi}^{\rm con}$ (6).  Columns (7)--(11) list the
same parameters computed using each source's observed spectral index.
Continuing, the remainder of the columns list the VLBI flux (12, taken
from Taylor et al. 1996), the flux of the core determined from the model-fit
parameters (13, taken from Paper I), the logarithm of the core dominance
$R_{\rm C}$ (14), and the core dominance $R_{\rm V}$ (15).

\subsection{kpc-scale morphologies of CJF-AGN and the misalignment between pc- and kpc-scale structure}
\begin{figure}[htb]
\begin{center}
\vspace*{0.5cm}
\subfigure[]{\includegraphics[clip,width=5.8cm,angle=-90]{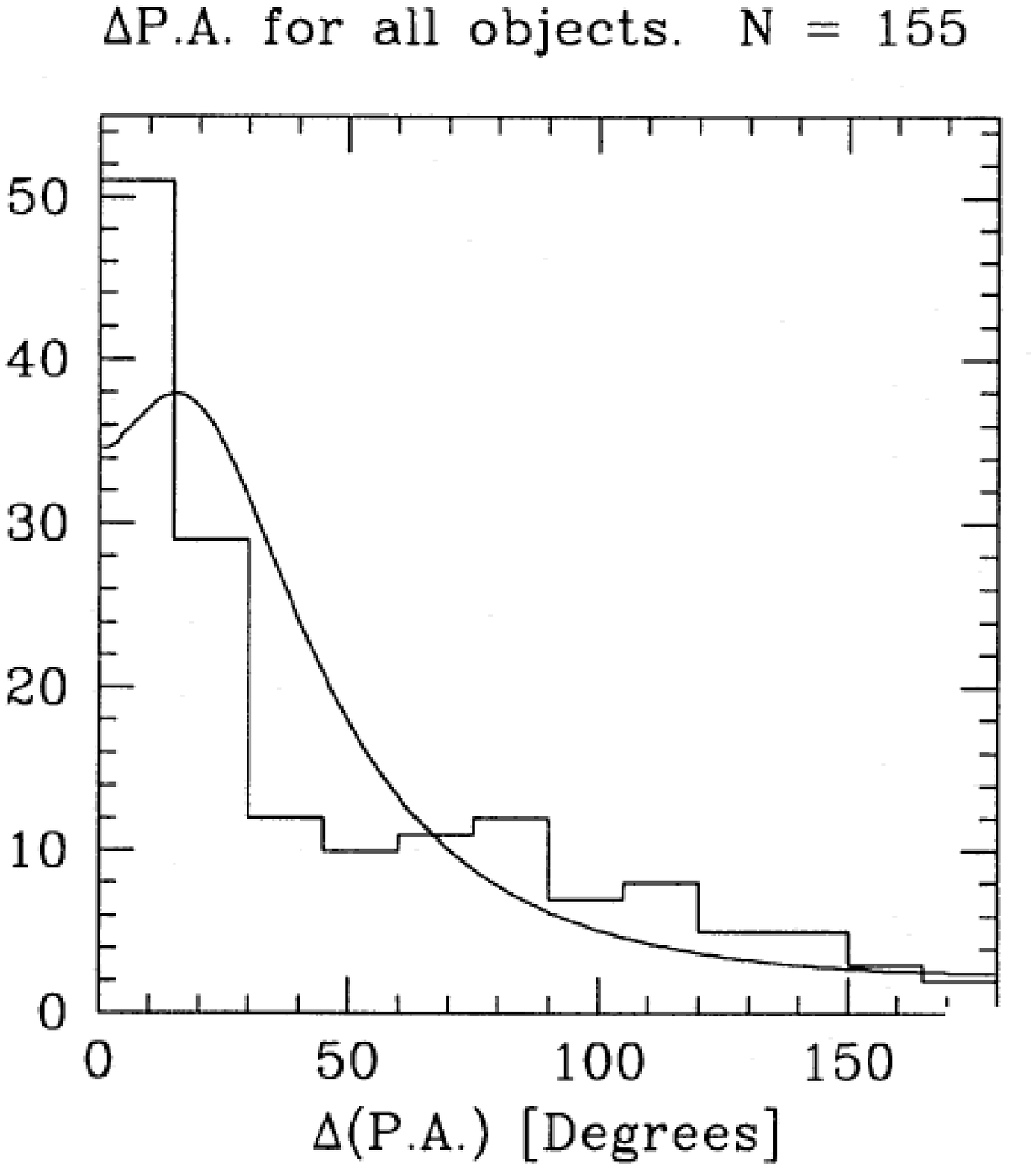}}\\
\vspace*{-0.6cm}
\hspace*{-1.5cm}
\subfigure[]{\includegraphics[clip,width=11.7cm]{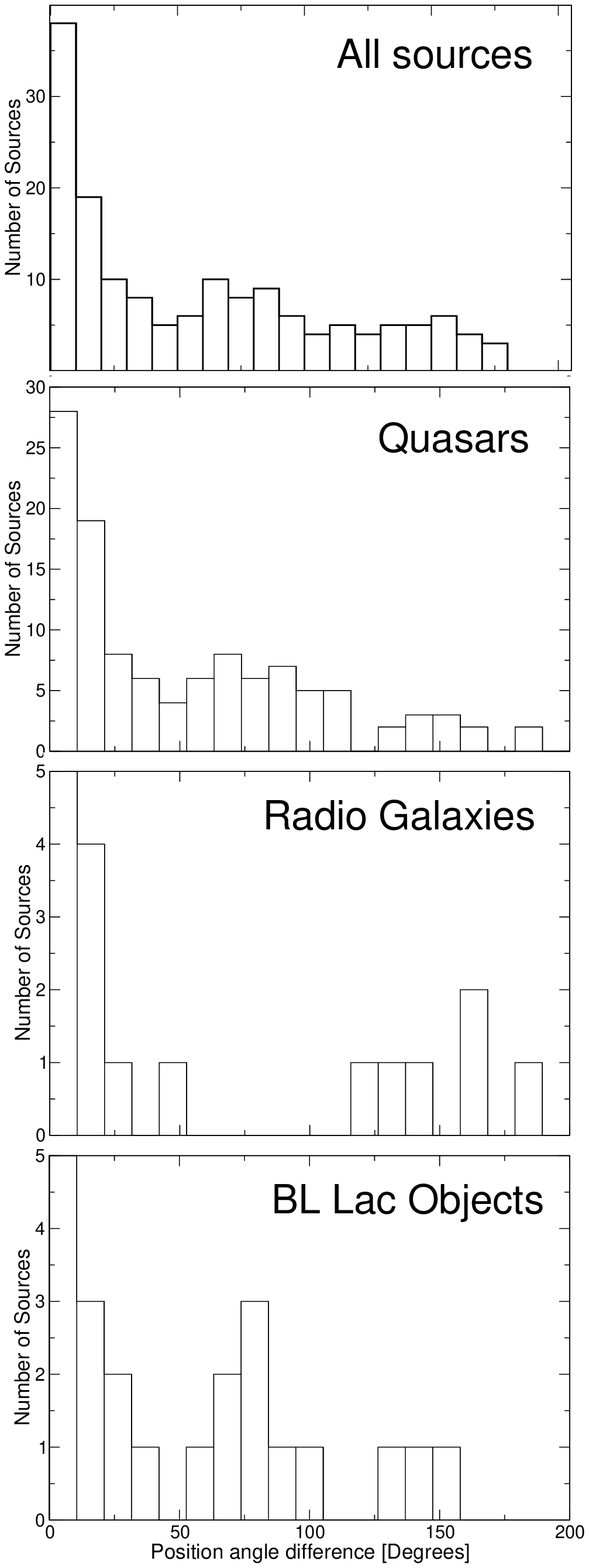}}\\
\end{center}
\vspace*{-0.6cm}
\caption{The panels (a)--(b) show histograms of the misalignment distribution. Figure (a) is taken from Appl et al. (1996) for a compilation of 155 sources. The superimposed smooth line shows the best fit by simple bend models performed by Appl et al. In (b) we show for the complete CJF sample (all sources with VLA- and VLBI-information) the misalignment distribution for the three main classes of sources (quasars: 114 sources, radio galaxies: 17 sources, and BL Lac objects: 22).}
\label{allmis}
\end{figure}            
\clearpage
\subsubsection{Misalignment}
Pearson \& Readhead (1988) found in the distribution of position angle differences between pc- and kpc-scales ($\Delta$PA) a highly unexpected bimodal pattern of relatively well aligned and roughly orthogonal jets.
More current investigations of larger samples by, e.g., Conway \& Murphy (1993) proved this excess to be statistically significant compared to the predictions of simple models.
The so-called ``misaligned population" of core-dominated AGN reveals a $\Delta$PA of $70^{\circ}$ to $90^{\circ}$ (termed the ``secondary peak" by Appl et al. 1996).  The position angle distributions of large samples of AGN have been studied in detail by e.g., Pearson \& Readhead 1988; Wehrle et al. 1992; Conway \& Murphy, 1993; Appl et al. 1996. In Fig.~\ref{allmis}(a) we show the histogram of misalignment angle for a sample of 155 radio sources together with the best fit by simple bend models by Appl et al. (1996).  A K-S test (0.016) suggests that
it is unlikely for the misalignment data to have been drawn from a parent population represented by the maximum likelihood model.\\
Small apparent misalignments can be explained by small random bends. Small intrinsic bends between pc- and kpc-scales will give the large $\Delta$PAs that are observed if sources are viewed almost along the direction of the VLBI jet. Assuming similar intrinsic bends, sources in which
the VLBI jet is oriented closer to the line of sight should show more extreme misalignment angles. The orthogonal misalignments, however cannot be explained by these processes (Conway \& Murphy 1993).\\
Several $\Delta$PA-distributions of different AGN-samples have been published; however, no complete compilation of $\Delta$PA values is currently available. We performed a literature search for kpc-scale morphological information on the CJF sources. In Table~1 (column 10) we describe the VLA structure of the sources. The major part of the information on the large scale structures has been derived by T. Pearson in VLA observations of the CJ-sources (http://www.astro.caltech.edu/\raisebox{-0.9ex}{\~{ }}tjp/cj/). However, VLA maps were not available for all sources; this is indicated for such cases in the table. To build up a homogeneous database for this kind of analysis, we redetermined the orientation of the large-scale structures from published maps and compared them with the pc-scale orientation derived directly from the CJF VLBI survey results (Paper I). This sample is the largest homogeneous sample for this kind of misalignment study in AGN being so far available.\\
114 of the 293 CJF-sources reveal a point-like VLA structure and do not contribute to this analysis. In Fig.~\ref{allmis}(b) we display the distribution of $\Delta$PA for those CJF sources that reveal kpc-scale extended emission. The figure clearly shows the expected peak around 0$^{\circ}$ (and a smaller peak around 150$^{\circ}$-180$^{\circ}$) for the aligned objects and an indication for the secondary peak around 75$^{\circ}$ for the misaligned objects. \\
\noindent
In Fig.~\ref{allmis}(c)--(e) we show the $\Delta$PA-distribution for the three classes of objects: quasars, BL Lac objects, and radio galaxies, respectively. The quasar distribution (c) contains more objects and is broader, covering the complete range of misalignment angles. Here we find some indication for the secondary peak. The radio galaxies (e) show primarily aligned kpc- and pc-scale jets: the distribution peaks around 0$^{\circ}$ and 180$^{\circ}$, while no misaligned objects have been observed. A fraction of the BL Lac objects (d) seems to be misaligned. \\
\begin{figure}[htb]
\vspace*{0.7cm}\rotate[r]{\psfig{figure=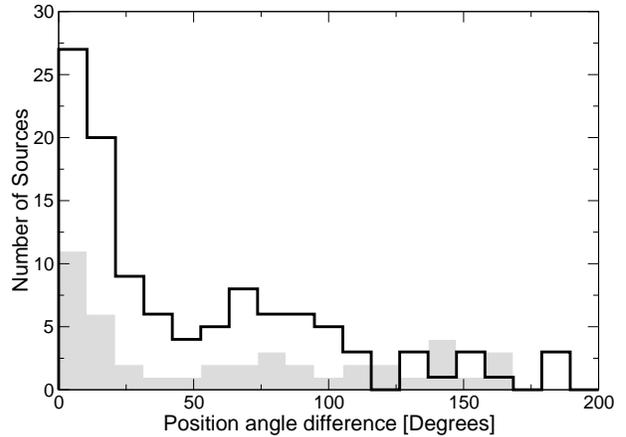,width=5.8cm}}
\caption[]{The distribution of the misalignment angle for the {\it ROSAT}-detected (solid line) and non-detected (grey) subsample.}
\label{mis}
\end{figure}
In Fig.~\ref{mis} we compare the distributions of the misalignment angle for the {\it ROSAT}-detected and non-detected sources. Based on a K-S test the distributions are not significantly different (0.260).\\
\subsubsection{\sf Investigating the relation between the large scale structure of AGN and X-ray emission}
\label{def}
\begin{figure*}[htb]
\subfigure[]{\psfig{figure=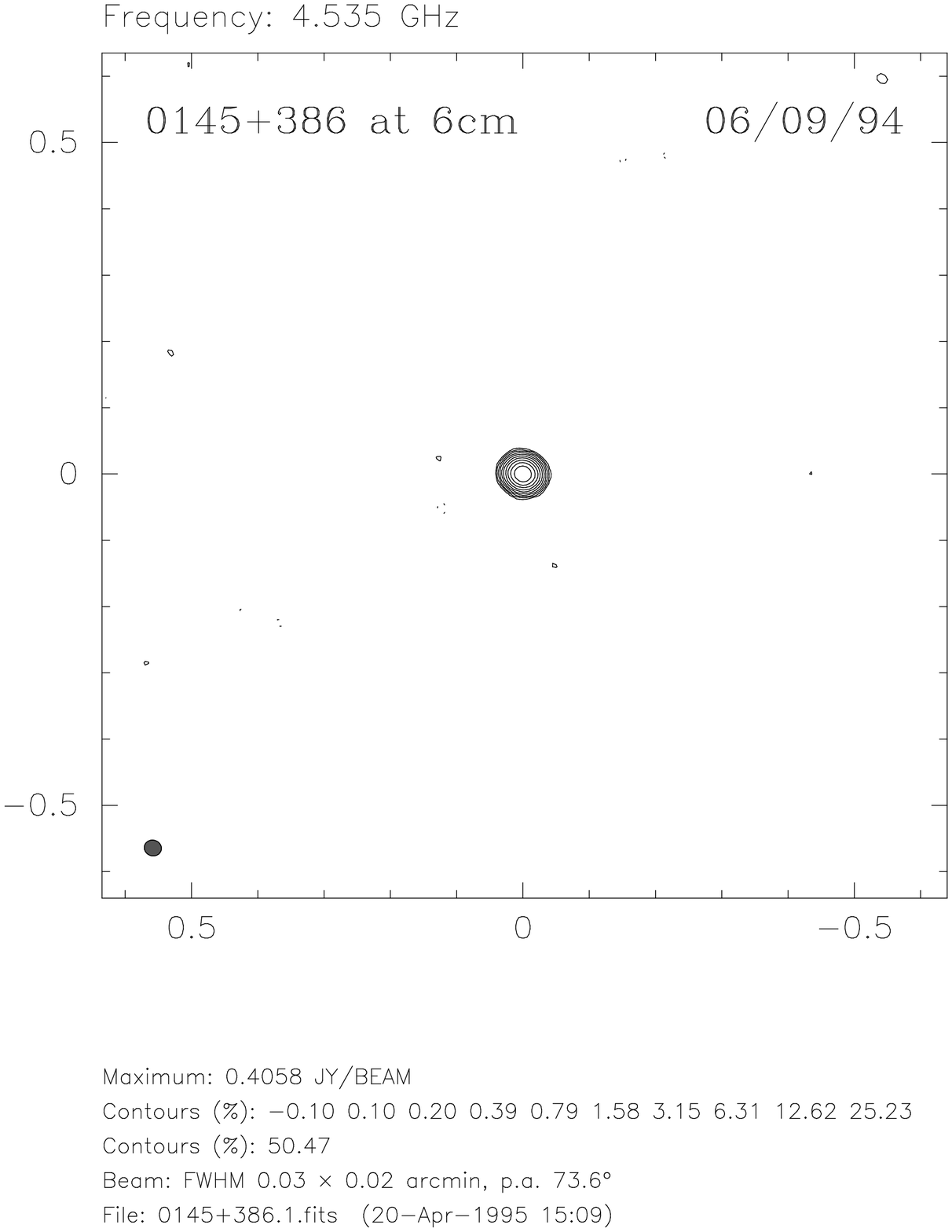,width=5.6cm}}
\subfigure[]{\psfig{figure=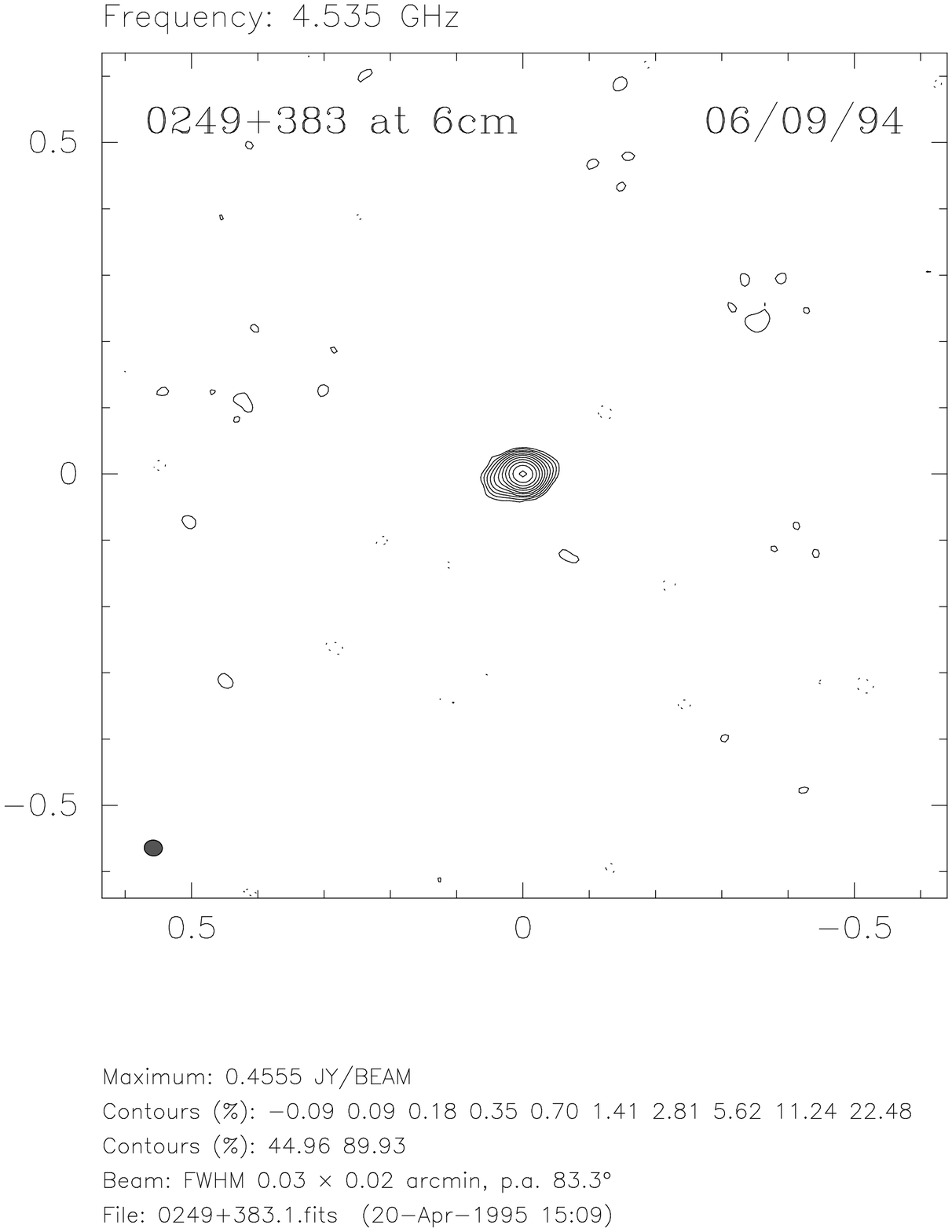,width=5.6cm}}
\subfigure[]{\psfig{figure=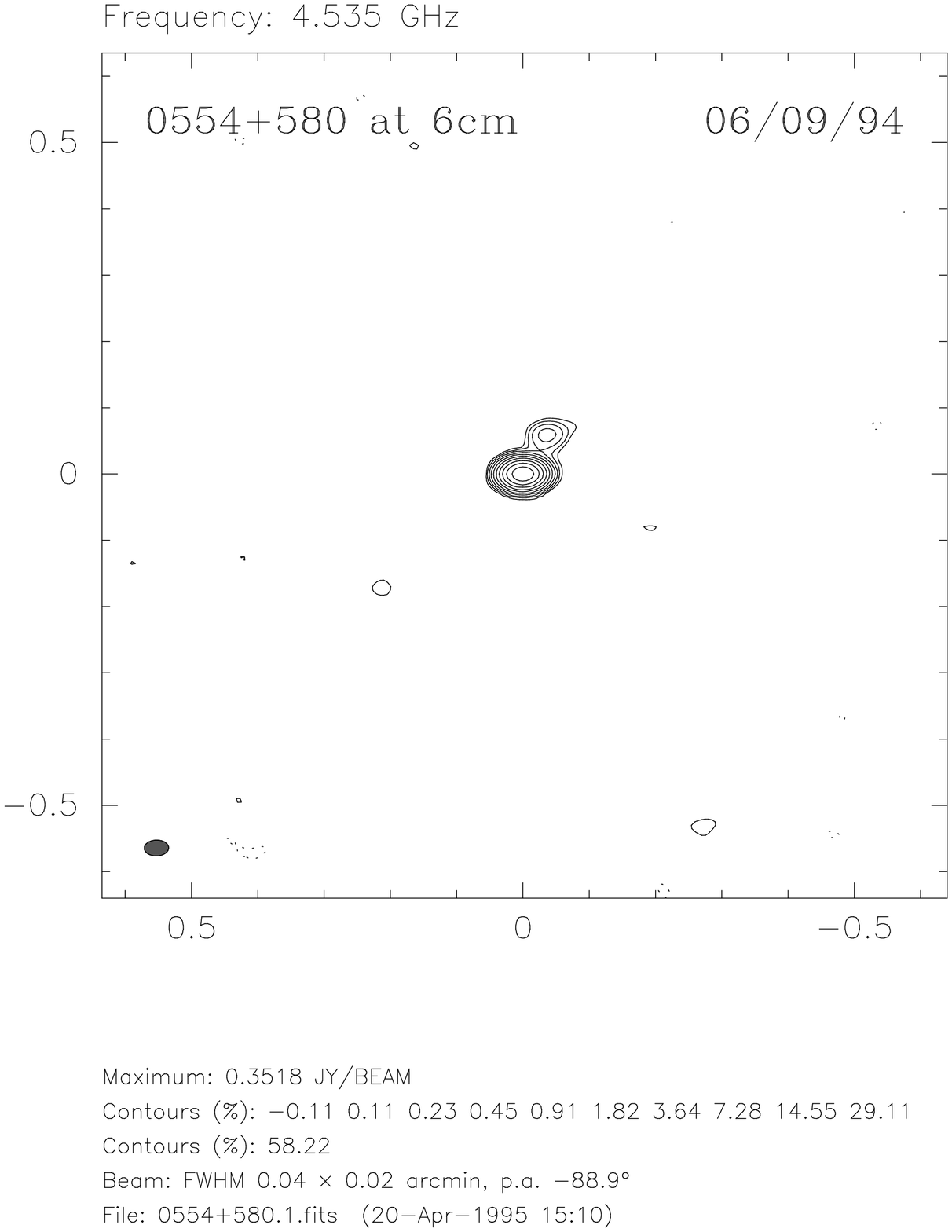,width=5.6cm}}
\subfigure[]{\psfig{figure=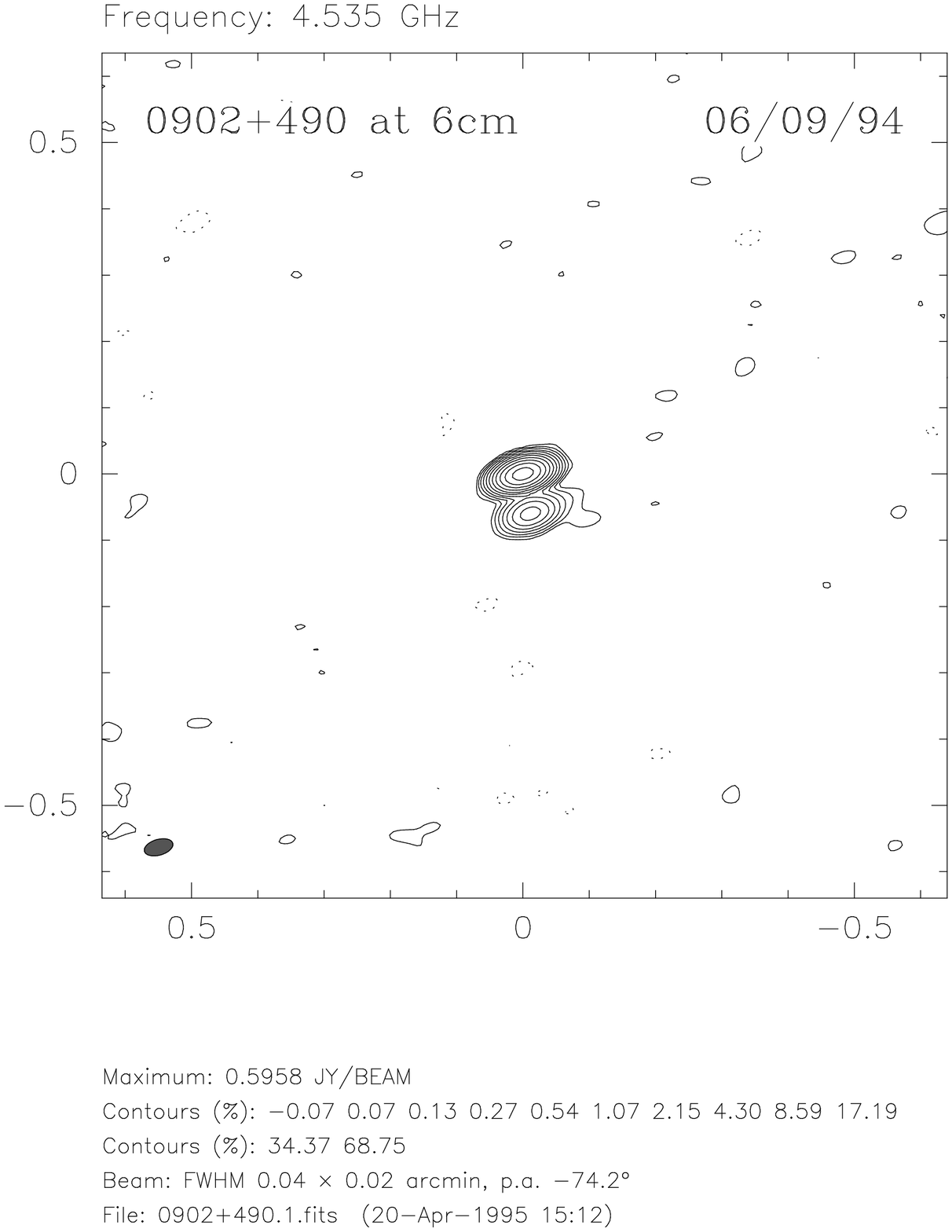,width=5.6cm}}
\hspace*{0.5cm}\subfigure[]{\psfig{figure=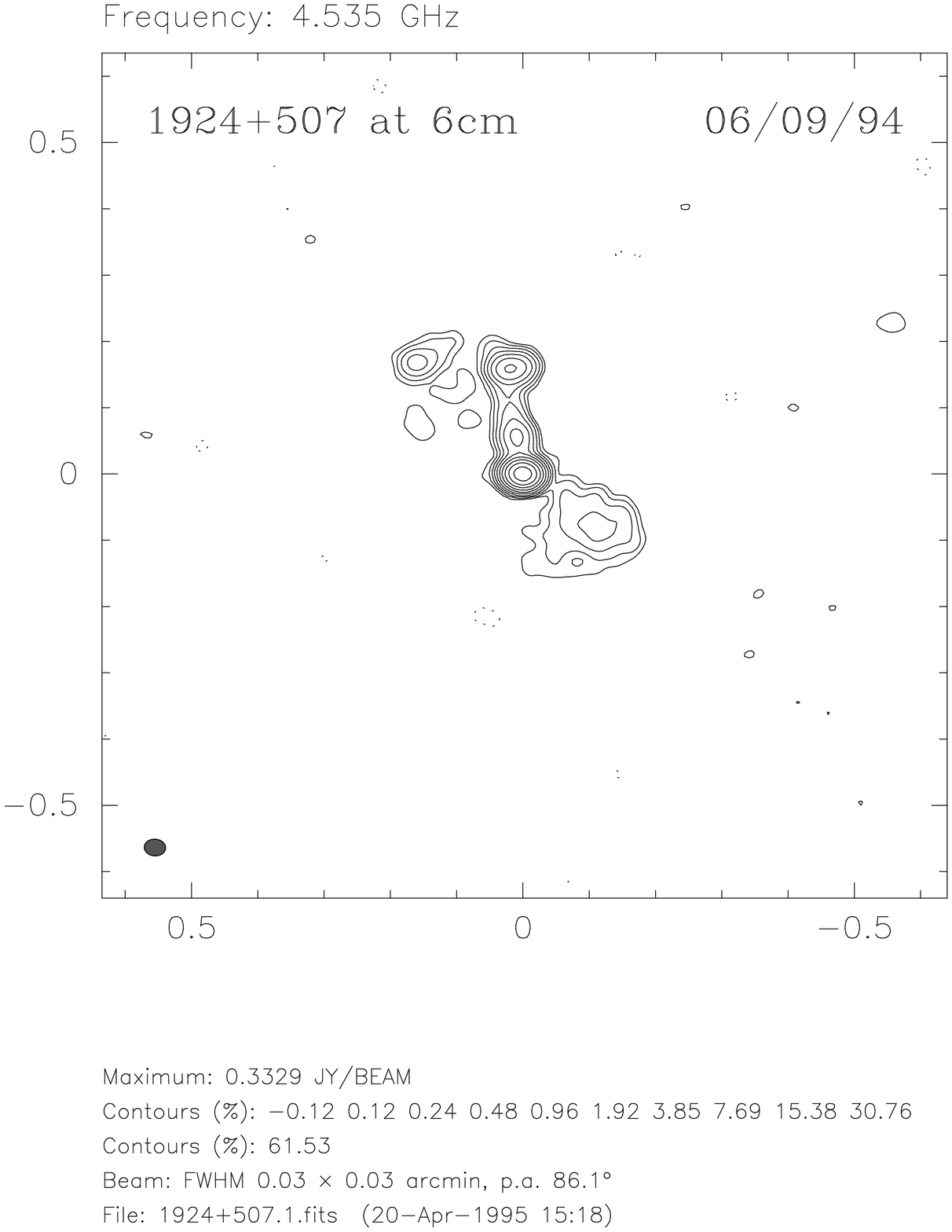,width=5.6cm}}
\hspace*{0.5cm}\subfigure[]{\psfig{figure=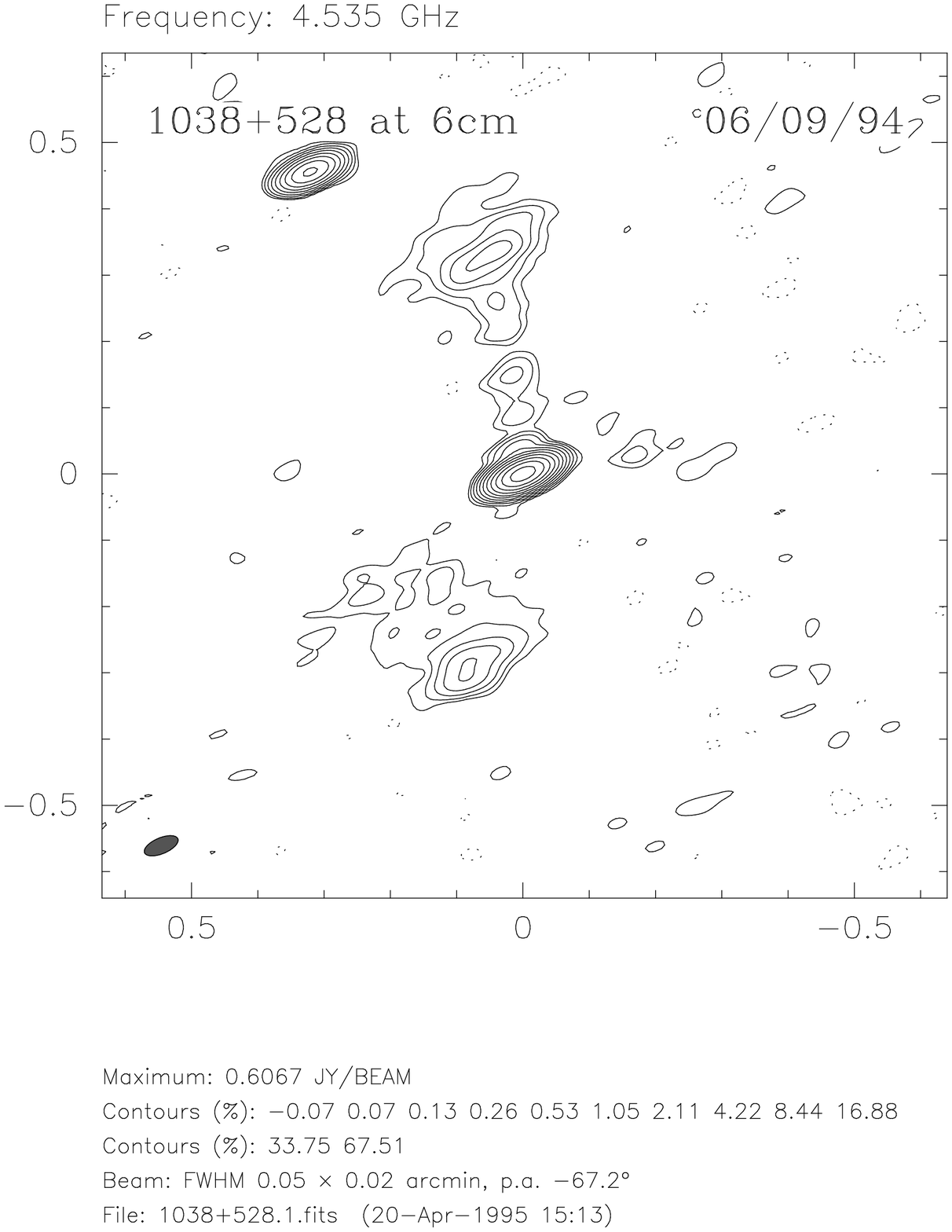,width=5.6cm}}
\caption{Six maps (images have been taken from http://www.astro.caltech.edu/$\sim$tjp/cj/) showing examples for the structures described by the complexity-factor (from a--f and {\bf 0} to {\bf 5}, respectively).}
\label{factor}
\end{figure*}

At least part of the X-ray emission from radio-loud AGN is thought to arise from the jet, because radio-loud objects have stronger X-ray emission and rather different X-ray spectra than do radio-quiet objects (e.g., Mushotzky 1993 and references therein). 
Surveys of extended radio jets with {\it CHANDRA} yield evidence that the X-ray emission is related to the kpc-scale morphology (e.g., Sambruna et al. 2004; Gelbord et al. 2004). 
In this section we therefore contrast the kpc-scale radio morphology of the {\it ROSAT} detected and non-detected objects to search for signs of this assumed correlation.\\                                 
114 of the 293 CJF-sources reveal a point-like VLA structure. Among those, 65 have not been detected by {\it ROSAT}, but 49 have been detected. Except for two objects (0014+813, 1246+586), the point-like sources have relatively low X-ray fluxes. We find significant evidence that {\it ROSAT}-detected sources tend to show extended radio emission on large scales. In order to be able to classify the extended morphology of the CJF kpc-scale jets, we adopted a classification scenario. 
The large-scale structures appear to be either unresolved, slightly resolved, jet-like and extended, double, or more complex. Sometimes complex jet- and counter-jets are visible in the large-scale maps; halo emission can appear along with jets or be the only large-scale component.\\ 
To quantify the large-scale structure, we adopted complexity-factors, where the number increases with the complexity of the morphology: the unresolved sources were classified as {\bf 0}, the slightly resolved sources as {\bf 1}, sources with a clearly resolved jet as {\bf 2}, double-source morphologies as {\bf 3}, jet- counter-jet structures as {\bf 4}, and the most complex morphologies as {\bf 5}. Examples for the differently complex structures and the assignment of the complexity-factors are shown in Fig.~\ref{factor}. \\                   
In Table~\ref{complexity} we list the distribution of these complexity-factors for the detected CJF sources and the non-detections. There are significant differences between the two distributions: the non-detected objects tend to show less complex kpc-scale structure, while the {\it ROSAT} detected CJF sources tend to have more complex kpc-scale structures. 
A K-S test (0.005) comparing the binned detected and non-detected complexity factors allows us to reject strongly the hypothesis that these two distributions are the same. Sources with the most complex
large-scale structures ({\bf 4}, {\bf 5}) are almost always ($\sim$ 97\%) detected by {\it ROSAT}. However, there may be a redshift-dependent effect at work here.  The average redshift of {\it ROSAT}-detected sources that have the most complex large-scale radio morphologies is 0.67, whereas for all sources detected by {\it ROSAT} it is 1.20, and for non-{\it ROSAT}-detected sources it is 1.34.\\
The large-scale jet structure might play an important role in contributing to the X-ray emission. Jets are very likely relativistic on kiloparsec scales as well. Large-scale relativistic proper motions have been directly observed in the nearby Radio Galaxy M87 (Biretta \& Junor 1995). The most plausible explanation of some of the newly discovered extended X-ray jets requires that the plasma have bulk relativistic motions on scales of hundreds of kiloparsecs (e.g., Tavecchio et al. 2000; Celotti et al. 2001; Sambruna et al. 2002). More observations are definitely required to search for and confirm large scale motion in these {\it ROSAT} detected quasars and BL Lac objects.\\
\begin{table}[htb]
\setcounter{table}{8}
      \caption[]{The numbers of objects with the complexity-factors describing the kpc-scale structure.}
         \label{complexity}                                                                             
	 \begin{tabular}{|ccccccc|}
            \noalign{\smallskip}
            \hline
            \noalign{\smallskip}
              complexity factor& {\bf 0}&{\bf 1}&{\bf 2}&{\bf 3}&{\bf 4}&{\bf 5} \\
            \noalign{\smallskip}
            \hline
            \noalign{\smallskip}
            detected by {\it ROSAT}& 49& 12& 35& 27& 20& 15\\
            not detected by {\it ROSAT}& 65& 10& 32& 11& 1& 0\\
            \noalign{\smallskip}
            \hline
         \end{tabular}
 \end{table}
\section{Discussion}
Although AGN form the ideal class of objects for multi-wavelength studies -- as their emission can cover almost 20 orders of magnitude in frequency from the radio to the $\gamma$-ray band -- our knowledge about their physics is limited by observational constraints and the inherent complex physical processes. Studies with multifrequency coverage for a single object are rare (e.g., NGC 3783, 3C273, 3C279). While {\it CHANDRA} observations determined the X-ray production mechanisms in a growing number of AGN (e.g., Harris \& Krawczynski 2002), this information is not available for a substantial number of sources in any large survey such as the CJF. However, assuming that the dominant mechanism is IC emission, we can place lower limits on beaming parameters derived from {\it ROSAT} observations for the complete CJF. To date this kind of analysis has relied on smaller samples and/or data taken from the literature. With the {\it ROSAT} observations of a complete and homogeneous VLBI survey, beaming indicators relying on radio and X-ray data can be estimated on an improved statistical basis. We draw the following conclusions:\\
\begin{itemize}
\item More than half of the CJF sources have been detected by {\it ROSAT}.
The good correlation between the radio- and X-ray luminosities of the CJF sources on the one hand and between the optical- and the X-ray luminosities on the other hand support the explanation that a common origin of the radiation can be derived, whereas differences exist between radio galaxies and quasars, beyond their different luminosities, which might be explained by an additional cluster emission in the case of the radio galaxies. \\
\item We find higher apparent velocities for the {\it ROSAT}-detected and non-detected quasars compared to the radio galaxies. Radio galaxies reveal the lowest values in both subsamples. We find some evidence for different apparent velocity distributions between those quasars detected by {\it ROSAT} and those not detected. A statistically significant analysis of the apparent motions of the BL Lac objects is hampered by small numbers of objects.\\
\item The calculation of the Doppler factors making use of the observationally determined spectral indices leads to different distributions compared to the calculations based on a uniform spectral index.\\
We find a better match between the median $\beta_{\rm app}$ and both $\delta_{\rm IC}$ calculated with the observed spectral indices compared to the uniformly assumed spectral index for all classes of objects.
The comparison of the IC Doppler factor with $\beta_{\rm app}$ (for those sources detected by {\it ROSAT}) seems to indicate that the bulk relativistic motion without any additional pattern speeds can explain the observations for quasars and BL Lac objects. However, larger source numbers (especially for the BL Lac objects) would be beneficial. The quasars have significantly larger values of $\delta_{\rm IC}^{\rm sp}$ and $\delta_{\rm IC}^{\rm con}$ compared to radio galaxies. \\
\item We find a good agreement between the Doppler factors derived from equipartition arguments and from IC calculations, especially in the case of a continuous jet model. $\delta_{\rm IC}$ is about 2-2.5 times $\delta_{\rm EQ}$ in the case of the continuous and the sphere-like jet case.\\
\item By investigating the distribution of the core dominance parameter $R$ calculated on the one hand by the ratio between the total VLBI flux-density at 5 GHz and the Green Bank 5 GHz flux-density, and on the other hand by the ratio between the VLBI core flux-density and the Green Bank flux-density, we find that the latter is better suited to discriminate between differently beamed objects. We find evidence for stronger beaming in quasars and BL Lac objects compared to the radio galaxies for all samples investigated. Within the individual classes, we find no significant differences between the {\it ROSAT}-detected and non-detected subsamples.\\
\item Pronounced extended radio structure is detected for most of the CJF sources. Roughly only one-third
of the sources are strongly core-dominated and do not reveal kpc-scale structure. Misalignments between the preferential direction on pc-scales and the direction of the overall structure have been found to be more common than expected in ``simple'' beaming models for some investigated AGN samples by e.g., Conway \& Murphy (1993) and Appl et al. (1996). Whereas radio galaxies do not show misaligned sources in our CJF sample, quasars and BL Lac objects show this effect. This can be explained by amplification of small intrinsic bends due to projection effects (e.g., Conway \& Murphy 1993, Conway \& Wrobel 1995).\\
\item 114 out of 293 sources from the complete CJF show no extended radio emission. For these objects the probability of {\it ROSAT} detectable X-ray emission is significantly lower than for the whole sample. There exist clear correlations between the extended radio structure and the likelihood of {\it ROSAT} detection. Sources detected by {\it ROSAT} show a higher degree in the complexity of their large-scale radio morphologies. However, we cannot exclude that a redshift-dependent effect influences this relation. The average redshift of {\it ROSAT}-detected sources with most complex large-scale radio structure is 0.67 compared to 1.20 for all sources detected by {\it ROSAT}. 
\end{itemize}
\begin{acknowledgements}
We wish to thank A. Witzel and T.P. Krichbaum for thought-provoking discussions and many useful comments. The {\it ROSAT} project is supported by the Bundesministerium f\"ur
Bildung, Wissenschaft, Forschung und Technologie (BMBF) and
the Max-Planck-Gesellschaft.  
Part of this work was supported by the European
Commission, TMR Programme, Research Network Contract
ERBFMRXCT97-0034 CERES and by the DLR, project 50QD0101.
S. Britzen acknowledges support by the Claussen-Simon Stiftung.
This research has made use of the NASA/IPAC Extragalactic Data Base
(NED) which is operated by the Jet Propulsion Laboratory, California
Institute of Technology, under contract with the National Aeronautics
and Space Administration.
\end{acknowledgements}

\onecolumn
\tabcolsep1.3mm
{\scriptsize
\textheight24cm
\textwidth17.5cm
\oddsidemargin-2cm
\tabcolsep0.7mm
\hspace*{1.3cm}\vspace*{1cm}
\tablecaption{CJF sources in the {\it ROSAT} All Sky Survey.} {\smallskip}
\setcounter{table}{1}
\tablehead{\noalign{\smallskip} \hline \noalign{\smallskip}
~CJF name~ & &~type~ & ~~  z~ & ~m$_v$~ & \multicolumn{1}{c}{S$_m$} &
~ EXML~ & \multicolumn{1}{c}{$f_x$ $\times 10^{-12}$} & ~OBS~& VLA structure&$\theta_{\rm VLA}$ &$\theta_{\rm VLBI}$&$\Delta$PA&\\
\multicolumn{1}{c}{\small ~~~ } & \multicolumn{1}{c}{\small ~~~ } &
\multicolumn{1}{r}{\small ~~~ } & \multicolumn{1}{c}{\small ~~~} &
\multicolumn{1}{c}{\small ~~~}&
\multicolumn{1}{c}{\tiny ~[mJy]~  } & \multicolumn{1}{c}{\small ~~~} &
\multicolumn{1}{c}{\small ~[{\footnotesize erg/cm$^2$/s}]~ } &
\multicolumn{1}{c}{\small ~~~}&
\multicolumn{1}{c}{\small }&
\multicolumn{1}{c}{\tiny ~~~~~~~~[deg]}&
\multicolumn{1}{c}{\tiny ~~~~~~~[deg]}&
\multicolumn{1}{c}{\tiny ~~~~~~~[deg]}\\
(1)&(2)&(3)&(4)&(5)&(6)&(7)&(8)&(9)&(10)&(11)&(12)&(13)\\
\noalign{\smallskip} \hline \hline  \noalign{\smallskip}}
\tabletail{\hline\multicolumn{12}{r}{continued on next page}\\}
\begin{supertabular}{clcllrrrrrrrrrrrrrrrrlll}
\label{all}
 0003+380&S4 0003+38, JVAS J0005+3820&G,Sy  &0.229 & 19.4 &  549 &  5.83 &            0.49 &     & R$^{1,4}$, {\bf 1}&110&120&10& \\
 0010+405&4C +40.01 ,JVAS J0013+4051&    G  &0.255 & 17.9 & 1040 & 34.60 &  1.61$\pm$ 0.38 &   s &R$^{27}$, {\bf 1}&335&330&5&\\
 0014+813&S5 0014+81, JVAS J0017+8135&    Q  &3.366 & 16.5 &  551 & 43.03 &  2.40$\pm$ 0.47 &  ps &U$^{1}$, {\bf 0}&/&189&/& \\
 0016+731&S5 0016+73, JVAS J0019+7327&    Q  &1.781 & 18.0 & 1712 &  4.45 &            0.58 &  p  &J$^{1}$, {\bf 2}&4&170&166&\\
 0018+729&S5 0018+72, JVAS J0021+7312&   G  & & 22.4 &  397 &  0.79 &            0.11 &     &U$^{1}$, {\bf 0}&/&277&/& \\
 0022+390&B3 0022+390, JVAS J0025+3919&   Q  &1.946 & 19.8 &  663 &  4.14 &            0.39 &     &J$^{1}$, {\bf 2}&21&164&143&\\
 0035+367&4C +36.01, [VCV2001] J003746.2+365911  &   Q &0.366 & 18.0 &  482 &  8.40 &  0.50$\pm$ 0.21 &   s  &J$^{27}$, {\bf 3}&13&357&16&\\
 0035+413&S4 0035+41, JVAS J0038+4137 &   Q  &1.353 & 19.9 & 1114 & 10.36 &  0.77$\pm$ 0.29 &   s &J$^{1}$\, {\bf 1}&115&106&9&\\
 0102+480&     JVAS J0105+4819& &      &      & 1088 &  2.89 &            0.33 &     &U$^{1}$, {\bf 0}&/&47&/&\  \\
 0108+388&OC +314, JVAS J0111+3906  &   G  &0.669& 22.0 & 1321 &  1.14 &            0.21 &     &J$^{5}$\, {\bf 2}&90&249&159&\\
 0109+351&B2 0109+35, JVAS J0112+3522 &   Q  &0.45  & 17.8 &  362 & 19.03 &  1.23$\pm$ 0.33 &   s &JCJ$^{1}$, {\bf 4}&40/278&205&73& \\
 0110+495&S4 0110+49, JVAS J0113+4948 &   Q  &0.389 & 18.4 &  710 & 31.87 &  2.31$\pm$ 0.58 &   s &?$^{27}$, {\bf ?}&55&329&86& \\
 0133+476&DA 55, JVAS J0136+4751      &   HPQ&0.859 & 18.  & 1816 &  2.04 &            0.75 &  p  &U$^{1}$, {\bf 0}&/&328&/&  \\
 0145+386&   JVAS J0148+3854        &   Q  &1.442 & 16.  &  370 &  1.05 &            0.12 &     &U$^{1}$, {\bf 0}&/&311&/& \\
0151+474&B3 0151+474, JVAS J0154+4743&  Q &1.026 & 20.5?&  505 &  2.22 &  0.17 &     &U$^{1}$, {\bf 0}&/&184&/&\\
 0153+744&S5 0153+74, JVAS J0157+7442 &   Q  &2.338 & 16.0 & 1549 & 12.82 &  0.95$\pm$ 0.34 &  ps &U$^{1,15,25}$, {\bf 0}&/&$\sim$160&/&\\
 0205+722&S5 0205+72, [VCV2001] J020952.2+722924 &   G  &0.895 & 20.7 &  560 &  2.45 &            0.47 &     &U$^{1}$, {\bf 0}&/&252&/& \\
 0212+735&S5 0212+73, JVAS J0217+7349 &   HPQ&2.367 & 19.  & 2278 & 30.11 &  2.45$\pm$ 0.58 &  ps &R$^{1}$, {\bf 1}&147&103&44&\\
 0218+357&S4 0218+35, JVAS J0221+3556 &   G$_{L}$&0.936 & 20.0 & 1498 &  3.30 &        0.26 &     &JCJ$^{14,27}$, {\bf 3}&70&/&/&\\
 0219+428&3C 066A, [VCV2001] J022239.6+430208    &   B &0.444 & 15.5 &  806 & 11.63 &  5.28$\pm$ 1.86 &   s &J$^{2}$, {\bf 2}&168&176&8&\\
 0227+403&B3 0227+403, JVAS J0230+4032&   Q  &1.019 & 17.0 &  436 &  0.40 &            0.35 &     &U$^{1}$, {\bf 0}&/&142&/& \\
 0248+430&B3 0248+430, JVAS J0251+4315&   LPQ&1.310 & 17.45& 1414 & 14.37 &  1.31$\pm$ 0.41 &   s &cJ$^{1}$, {\bf 3}&107&143&36&\\
 0249+383&B3 0249+383, JVAS J0253+3835&   Q  &1.122 & 18.5 &  450 &  1.19 &            0.25 &     &R$^{1}$, {\bf 1}&/&341&/& \\
 0251+393&B3 0251+393, JVAS J0254+3931&Q &0.289 & 17.0 &  408 & 65.39 &  3.28$\pm$ 0.57 &s&J$^{1}$, {\bf 2}&102&86&16& \\
 0256+424&B3 0256+424, JVAS J0259+4235&   Q  &0.867 & 19.5?&  366 &  3.12 &            0.42 &     &R$^{1}$, {\bf 1}&280&/&/&\\
 0307+380&B3 0307+380, JVAS J0310+3814&   Q  &0.816 & 20.0?&  760 &  2.54 &            0.39 &     &U$^{1}$, {\bf 0}&/&48&/& \\
 0309+411&S4 0309+41, JVAS J0313+4120&Q,Sy1 &0.134 & 18.0 &  516 &$>$ 100&  5.47$\pm$ 0.64 &  ps &J$^{1}$, {\bf 2}&310&307&3&\\
 0316+413&NGC 1275, JVAS J0319+4130   &   G/C&0.018& 12.64&42370 &$>$ 100& 283.9$\pm$ 5.97 &  ps &JCJ,H$^{12,13}$, {\bf 5}& 160(235)&180&20& \\
 0340+362& JVAS J0343+3622          &   Q  &1.485 & 20.0$^{r}$&  376 &  3.01 &       0.66 &   s &JCJ$^{1}$, {\bf 3}&50&31&19& \\
 0344+405&4C +40.12,   &   G  &0.039 & 16.5 &  478 &  3.83 &            0.75 &     &/&/&/&/&\\
0346+800&S5 0346+80, JVAS J0354+8009  &  Q$^{r}$ & & 21.3 &  396 &  0.91 &   0.09 &     &U$^{1}$, {\bf 0}&/&$\sim$130&/& \\
 0402+379&B3 0402+379, JVAS J0405+3803&   G,Sy &0.055 & 17.2 &  937 & 91.29 &  5.70$\pm$ 0.82 &   s &JCJ,H$^{1,27}$, {\bf 5}&223/$\sim$22&$\sim$250/34&27/12&\\
 0424+670&  JVAS J0429+6710         &Q &0.324 & 21.0?&  362 &  1.37 &            0.20 &     &R$^{27}$, {\bf 1}&280&/&/&\\
 0444+634&S4 0444+63, JVAS J0449+6332 &   Q  &0.781 & 19.7 &  606 &  3.73 &            0.79 &   s &R$^{1}$, {\bf 1}&176&174&2&\\
 0454+844&S5 0454+844, JVAS J0508+8432&   B& 0.112     & 16.5 & 1398 & 26.14 &  1.03$\pm$ 0.27 &  ps &U$^{1}$, {\bf 0}&/&161&/& \\
 0537+531&S4 0537+53, JVAS J0541+5312 &   Q  &1.275 & 18.0 &  665 &  0.26 &            0.41$\pm$0.09 &  p  &JCJ$^{1}$, {\bf 3}&34&318&76&\\
 0546+726&S5 0546+72, [VCV2001] J055253.0+724045 &   Q,C.g.l.  &1.555 & 17.0 &  401 &  8.56 &  0.84$\pm$ 0.37 &   s &J$^{1}$, {\bf 2}&150&300&150&\\
 0554+580&  JVAS J0559+5804         &   Q  &0.904 & 18.0 &  906 &  0.70 &            0.25 &     &J$^{1}$, {\bf 2}&328&285&43&\\
 0600+442&B3 0600+442, JVAS J0604+4413&   G$^{r}$  &1.136 & 21.5?$^r$&  705 & 24.67 &  1.66$\pm$ 0.46 &     &U$^{1,27}$, {\bf 0}&/&316&/& \\
 0602+673&S4 0602+67, JVAS J0607+6720 &   Q  & 1.95  & 20.6 &  657 &  9.24 &  0.99$\pm$ 0.39 &   s &U$^{1,27}$, {\bf 0}&/&172&/& \\
 0604+728&4C +72.10, JVAS J0610+7248  &   Q  &0.986 & 20.3 &  654 &  4.90 &            0.63 &   s &JCJ$^{1}$, {\bf 3}&72/286&105/270&33/16&\\
 0609+607& JVAS J0614+6046          &   Q  &2.702 & 19.0 & 1059 &   7.0 &             0.38$\pm$0.20 &   s &U$^{1}$, {\bf 0}&/&152&/& \\
 0620+389&B3 0620+389, JVAS J0624+3856&   Q,cluster?  &3.469 & 20.0 &  811 &  7.02 &  1.06$\pm$ 0.49 &   s &R$^{1,27}$, {\bf 1}&0&138&138& \\
 0621+446& JVAS J0625+4440          &   B&      & 18.0 &  369 &  7.59 &  1.03$\pm$ 0.46 &     &U$^{27}$, {\bf 0}&/&/&/& \\
 0615+820&S5 0615+82, JVAS J0626+8202 &   LPQ&0.710 & 17.5 &  999 &  2.45 &            0.23 &  p  & /&/&/&/&\\
 0627+532& JVAS J0631+5311          &   Q  &2.204 & 18.5 &  485 &  3.77 &            0.57 &     &U$^{1}$, {\bf 0}&/&$\sim$50&/& \\
 0633+596& JVAS J0638+5933         &      &      &      &  482 &  1.61 &            0.23 &     &U$^{1}$, {\bf 0}&/&241&/&\\
 0633+734&S5 0633+73, JVAS J0639+7324 &   Q  &1.85  & 17.8 &  748 &  9.96 &  0.57$\pm$ 0.26 &     &J$^{1}$, {\bf 2}&321&353&32&\\
 0636+680&S4 0636+68, JVAS J0642+6758 &   Q  &3.174 & 16.46&  499 &  0.31 &            0.14 &  p  &U$^{1}$, {\bf 0}&/&/&/&\\
 0641+393&B3 0641+391, JVAS J0644+3914 &   Q  &1.266 & 19.5 &  453 &  0.48 &            0.19 &     &J$^{1}$, {\bf 2}&67&4&63&\\
 0642+449&S4 0642+44, JVAS J0646+4451 &   LPQ&3.396 & 18.49& 1191 &  2.31 &            0.39 &  p  &U$^{1}$, {\bf 0}&/&90&/&\\
 0646+600&S4 0646+60, JVAS J0650+6001 &   HPQ&0.455 & 18.9 &  920 &  1.29 &            0.22 &     &U$^{1}$, {\bf 0}&/&216&/&\\
 0650+371&S4 0650+37, JVAS J0653+3705 &   Q  &1.982 & 18.0 &  977 &  4.02 &            0.62 &     &U$^{1}$, {\bf 0}&/&56&/&\\
 0650+453&B3 0650+45, JVAS J0654+4514 &   Q  &0.933 & 21.0 &  420 &  2.07 &            0.28 &   s &R$^{1}$, {\bf 1}&332&85&113&\\
 0651+410&CGCG 204-027, JVAS J0655+4100&  G  &0.022& 14.6 &  425 &  3.29 &            0.34 &     &U$^{1}$, {\bf 0}&/&156&/&\\
 0700+470&B3 0700+470, JVAS J0704+4700 &  G$^{r}$  &      & 20$^{r}$  &  443 &  1.83 & 0.28 &     &U$^{1}$, {\bf 0}&/&$\sim$90&/&\\
 0702+612&  JVAS J0707+6110          &  Q$^{r}$  &      & 17.0 &  370 &  4.59 &            0.24 &     &J$^{1,27}$, {\bf 2}&$\sim$70&70&0& \\
 0707+476&B3 0707+476, JVAS J0710+4732 &  LPQ&1.292 & 18.2 &  906 &  0.00 &            0.00 &     &JCJ$^{3}$, {\bf 4}&279/99&$\sim$25&74&\\
 0710+439&S4 0710+43, JVAS J0713+4349  &  G$^{r}$  &0.518 & 19.7 & 1629 &  1.03 &            0.59$\pm$0.06 & {\bf p}   &R$^{1}$, {\bf 1}&$\sim$170&180&10& \\
 0711+356&S4 0711+35, JVAS J0714+3534  &  LPQ&1.620 & 17.  &  901 &  5.04 &            0.33 &  ps &JCJ$^{3}$, {\bf 3}&145&157&12&\\
 0714+457&S4 0714+457, JVAS J0717+4538 &  Q  &0.940 & 19.0 &  480 &  2.24 &            0.32 &     &U$^{1}$, {\bf 0}&/&132&/& \\
 0716+714&S5 0716+714, JVAS J0721+7120 &  B&      & 15.5 &  788 &$>$ 100&  4.31$\pm$ 0.51 &  ps &J,H$^{11}$, {\bf 5}&300&15&75&\\
 0718+793&  JVAS J0726+7911          &     &      &      &  631 &  2.58 &            0.18 &   s &U$^{1}$, {\bf 0}&/&296&/&\\
 0724+571&   JVAS J0728+5701         &  Q  &0.426 & 17.0 &  393 &  1.80 &            0.27 &     &U$^{1}$, {\bf 0}&/&152&/&\\
 0727+409&S4 0727+40, JVAS J0730+4049  &  Q,cluster?  &2.500 & 19.3 &  468 & 15.09 &  1.10$\pm$ 0.35 &   s &U$^{1,27}$, {\bf 0}&/&305&/&\\
 0730+504&TXS 0730+504, JVAS J0733+5022&  Q  &0.72  & 19.0 &  890 &  1.25 &            0.34 &     &J$^{1}$, {\bf 2}&228&209&19& \\
0731+479&S4 0731+47, JVAS J0735+4750  &  Q   &0.782 & 18.  &  533 & 26.05 &  1.69$\pm$ 0.44 &   s &U$^{1}$, {\bf 0}&/&270&/&\\
 0733+597&UGC 03927, JVAS J0737+5941   &  G & 0.041 & 15.17&  357 & 21.78 &  1.03$\pm$ 0.29 &   s &U$^{1,27}$, {\bf 0}&/&$\sim$10&/&\\
 0738+491&  JVAS J0742+4900          &  Q& 2.32$^{r}$ & 21.0?$^{r}$&  352 &  5.08 &  0.35 &     &U$^{1}$, {\bf 0}&/&5&/&\\
 0740+768&  JVAS J0747+7639          &  G$^{r}$ &       & 19.5$^{r}$ &  592 &  1.08 &   0.06 &     &U$^{1}$, {\bf 0}&/&247&/& \\
 0743+744&S5 0743+74, JVAS J0749+7420  &  Q & 1.629 & 19.3 &  479 &  5.93 &  0.28$\pm$ 0.16 &  p  &U$^{1,27}$, {\bf 0}&/&25&/&\\
 0746+483&S4 0746+48, JVAS J0750+4814  &  Q  &1.951 & 18.5 &  860 &  6.94 &  0.51$\pm$ 0.24 &   s &U$^{1,21}$, {\bf 0}&/&274&/&\\
 0749+540&4C +54.15, JVAS J0753+5352   &  B & $>$ 0.2  & 18.5 &  877 &  9.26 &  0.85$\pm$ 0.30 &   s &U$^{1}$, {\bf 0}&/&$\sim$220&/&\\
 0749+426&B3 0749+426, JVAS J0753+4231 &  Q  &3.59  & 18.1 &  461 &  2.03 &            0.26 &     &U$^{1}$, {\bf 0}&/&220&/&\\
 0800+618& CLASS J0805+6144    &  Q$^{r}$  &3.044$^{r}$ & 19.8$^{r}$ &  981 &  4.02 &   0.42 &   s & / &/&152&/&\\
 0803+452&B3 0803+452, JVAS J0806+4504 &  Q  &2.102 & 19.6 &  414 &  4.08 &            0.67 &     &U$^{1}$, {\bf 0}&/&233&/&\\
 0804+499&S4 0804+49, JVAS J0808+4950  &  HPQ& 1.432  & 17.5 & 1222 & 14.76 &  0.68$\pm$ 0.24 &  ps &J$^{3}$, {\bf 2}&$\sim$200&145&55&\\
 0805+410&B3 0805+410, JVAS J0808+4052 &  Q  &1.420 & 19.0 &  743 & 27.83 &  1.15$\pm$ 0.31 &   s &J$^{1}$, {\bf 2}&240&34&154& \\
 0806+573&TXS 0806+573, JVAS J0811+5714&  Q  & 0.611 & 17.44&  405 & 26.53 &  1.20$\pm$ 0.30 &   s &J$^{1,27}$, {\bf 2}&$\sim$250&265&15&\\
 0812+367&B2 0812+36, JVAS J0815+3635  &  Q  &1.025 & 18.0 &  980 &  7.0 &            0.30$\pm$0.17 &   s &cJCJ$^{15,31}$, {\bf 5}&$\sim$350&349&1& \\
 0814+425&B3 0814+425, JVAS J0818+4222 &  B&0.245& 18.5 & 1891 &  9.63 &  0.69$\pm$ 0.33 &  ps &cJ$^{3}$, {\bf 3}&320&115&155&\\
 0820+560&S4 0820+56, JVAS J0824+5552  &  Q  &1.409 & 18.0 & 1199 &  2.60 &            0.82$\pm$0.09 & p   &J$^{16}$, {\bf 2}&309&82&133&\\
 0821+394&4C +39.23, CLASS J0824+3916   &  Q  &1.216 & 18.5 & 1012 & 42.49 &  2.11$\pm$ 0.44 &   s &J$^{1}$, {\bf 2}&139&318&179&\\
 0821+621&TXS 0821+621, JVAS J0825+6157&  Q  &0.542 & 17.6 &  615 & 41.02 &  1.90$\pm$ 0.38 &   s &JCJ$^{1,29}$, {\bf 5}&47/236&244&8&\\
 0824+355&6C 0824+35, JVAS J0827+3525  &  Q & 2.249 & 20.28&  746 &  0.05 &            0.03 &     &J$^{1}$, {\bf 2}&117&118&1&\\
 0831+557&4C +55.16, JVAS J0834+5534   &  G &  0.240 & 19.0 & 5780 & 46.96 &  2.45$\pm$ 0.53 &   s &JCJ$^{1, 27}$, {\bf 4}&172/325&$\sim$305&20&\\
 0833+416&B3 0833+416, JVAS J0836+4125 &  Q & 1.298 & 17.2 &  385 & 26.50 &  0.88$\pm$ 0.24 &   s &JCJ$^{1}$, {\bf 3}&338/189&180&9&\\
 0833+585&S4 0833+58, JVAS J0837+5825  &  Q & 2.101 & 18.0 &  669 &  8.53 &  0.54$\pm$ 0.24 &   s &J$^{3}$, {\bf 3}&152&78&74&\\
 0836+710&4C +61.07, JVAS J0841+7053   &  Q & 2.180 & 16.5 & 2423 &$>$ 100& 11.28$\pm$ 0.82 &  ps &JCJ$^{3}$, {\bf 3}&204&211&7&\\
 0843+575&   JVAS J0847+5723         &  G?$^{r}$&       & 23.0?$^{r}$&  384 &  0.25 &            0.09 &     &U$^{1}$, {\bf 0}&/&42/$\sim$320&/& \\
 0847+379&4C +37.25, JVAS J0850+3747   &  G & 0.407 & 19.5 &  382 &  8.45 &  0.36$\pm$ 0.17 &  ps &JCJ,H$^{17}$, {\bf 5}&165&7&158& \\
 0850+581&4C +58.17, JVAS J0854+5757   &  Q & 1.322 & 18.  & 1187 &  8.15 &  0.51$\pm$ 0.22 &  ps &JCJ$^{18,19}$, {\bf 5}&139&156&17&\\
 0859+470&4C +47.29, JVAS J0903+4651   &  LPQ&1.462 & 18.7 & 1285 & 18.54 &  0.60$\pm$ 0.18 &   s & D$^{3,16}$, {\bf 3}&330&352&22&\\
 0859+681&S4 0859+681, JVAS J0903+6757 &  Q  &1.499 & 19.5 &  751 &  8.38 &  0.38$\pm$ 0.17 &   s &U$^{27}$, {\bf 0}&/&14&/&\\
 0900+520&   JVAS J0903+5151         &  Q  &1.537 & 19.6 &  395 &  4.82 &            0.27 &     &U$^{1}$, {\bf 0}&/&254&/&\\
 0902+490&S4 0902+49, JVAS J0905+4850  &  Q  &2.690 & 18.5 &  547 & 10.24 &  0.54$\pm$ 0.21 &     &D$^{1}$, {\bf 3}&172&321&149&\\
 0917+449&B3 0917+449, JVAS J0920+4441 &  Q  &2.18  & 19.0 & 1033 & 68.05 &  1.61$\pm$ 0.26 &   s &JCJ$^{1}$, {\bf 4}&270/$\sim$170&190&20&\\
 0917+624&OK +630, JVAS J0921+6215     &  Q  &1.446 & 19.5 & 1322 &  1.12 &            0.37$\pm$0.02 &  p  &J$^{3}$, {\bf 2}&238&343&105&\\
 0923+392&4C +39.25, JVAS J0927+3902   &  LPQ& 0.699& 17.86& 7480 &$>$ 100&  3.56$\pm$ 0.36 &  ps &JCJ$^{3}$, {\bf 3} &259&277&18&\\
 0925+504&RGB J0929+502, JVAS J0929+5013& B,cluster?&      & 16.0 &  558 & 37.82 &  1.13$\pm$ 0.23 &   s &U$^{1}$, {\bf 0}&/&128&/&\\
 0927+352&B2 0927+35, JVAS J0930+3503   & B?$^{r}$&    & 19.2 &  383 & 28.88 &  0.79$\pm$ 0.20 &  ps &J$^{30}$, {\bf 2}&280&285&5&\\
 0929+533&S4 0929+53, JVAS J0932+5306   & Q  &0.595 & 19.0 &  384 &  1.43 &            0.24 &     &R$^{1}$, {\bf 1}&?&133&/&\\
 0930+493&TXS 0930+493, JVAS J0934+4908 & Q  &2.582 & 18.4 &  574 &  0.53 &            0.08 &     &J$^{1}$, {\bf 2}&147&225&78&\\
 0942+468&B3 0942+468, JVAS J0945+4636  & G  &0.639$^{r}$ & 20.61 &  354 &  3.40 &            0.16 &     &U$^{1,27}$, {\bf 0}&/&41&/&\\
 0945+408&4C +40.24, JVAS J0948+4039    & LPQ&1.252 & 17.5 & 1592 & 14.49 &  0.46$\pm$ 0.16 &   s &D$^{3}$, {\bf 3}&32&116&84&\\
 0945+664&4C +66.09, CLASS J0949+6614    & G  &0.85?$^{r}$& 21.6 & 1407 &  0.00 &            0.00 &     & D$^{27}$, {\bf ?}&35&/&/&\\
 0949+354& JVAS J0952+3512  & Q  &1.875 & 18   &  403 &  5.13 &            0.26 &     &J$^{1}$, {\bf 2}&160&165&5&\\
 0950+748&S5 0950+748, JVAS J0954+7435  & G  &0.695$^{r}$ & 21.7 &  738 &  0.83 &            0.05 &     &U$^{1}$, {\bf 0}&/&253&/& \\
 0954+556&4C +55.17, JVAS J0957+5522    & HPQ&0.909 & 17.7 & 2270 & 63.79 &  1.01$\pm$ 0.17 &  ps &JCJ$^{3,16,23}$, {\bf 4}&$\sim$290/$\sim$45&195&95&\\
 0955+476&B3 0955+476, JVAS J0958+4725  & Q  &1.873 & 18.0 &  834 & 37.75 &  0.77$\pm$ 0.16 &  ps &?$^{23}$, {\bf ?}&$\sim$120&127&7& \\
 0954+658&S4 0954+65, JVAS J0958+6533&B&0.368 & 16.7 & 1417 & 15.41 &  0.92$\pm$ 0.29 &  ps &J$^{1,23}$, {\bf 2}&205&$\sim$290&85&\\
 1003+830&S5 1003+83, JVAS J1010+8250  &  G  &0.322 & 20.5 &  716 &  29 &            0.92$\pm$0.24 &     &J$^{1}$, {\bf 2}&115&85&30& \\
 1010+350&B2 1010+35, JVAS J1013+3445  &  Q,cluster$^{\#,\$}$ &1.414 & 19.0 &  597 &  8.31 &  0.30$\pm$ 0.14 &     &U$^{1,27}$, {\bf 0}&/&96&/&\\
 1014+615&TXS 1013+615, JVAS J1017+6116&  Q  &2.80  & 18.12&  631 &  7.68 &  0.21$\pm$ 0.08 &     &U$^{1,27}$, {\bf 0}&/&254&/&\\
 1015+359&S4 1015+35, JVAS J1018+3542  &  Q  &1.226 & 18.09&  587 & 14.80 &  0.57$\pm$ 0.18 &   s &J$^{1}$, {\bf 2}&128&186&58&\\
 1020+400&4C +40.25, JVAS J1023+3948   &  Q  &1.254 & 17.5 &  785 & 16.88 &  0.96$\pm$ 0.28 &  ps &cD$^{1}$, {\bf 3}&22&318&64&\\
 1030+415&B3 1030+415, JVAS J1033+4116 &  HPQ&1.120 & 18.2 &  485 &  5.36 &            0.22 &     &J$^{1}$, {\bf 3}&103/0&354&6&\\
 1030+398&B3 1030+398, JVAS J1033+3935 &  G$^{r}$  &1.095 & 21.5 &  645 &  2.87 & 0.12 &     &U$^{1}$, {\bf 0}&/&$\sim$40&/& \\
 1030+611&S4 1030+61, JVAS J1033+6051  &  Q  &0.336 & 19.7 &  579 & 13.77 &  0.21$\pm$ 0.07 &     &R$^{1}$, {\bf 1}&334&171&163&\\
 1031+567&S4 1031+56, JVAS J1035+5628  &  G, CSO  & 0.46  & 20.3 & 1200 &  1.71 &            0.08 &     &U$^{1,27}$, {\bf 0}&/&220&/&\\
 1038+528&TXS 1038+528, JVAS J1041+5233&HPQ  &0.677 & 17.4 &  709 & 64.41 &  1.06$\pm$ 0.17 &   s &cJCJ$^{1,2}$, {\bf 5}&10,35,165&24&11&\\
 1041+536&7C 1041+5338, JVAS J1044+5322&  Q  &1.897 & 19.0 &  481 & 29.22 &  0.53$\pm$ 0.13 &   s &U$^{1,27}$, {\bf 0}&/&$\sim$180&/&\\
 1039+811&S5 1039+81, JVAS J1044+8054  &  LPQ& 1.254  & 16.5 & 1144 & 53.81 &  1.35$\pm$ 0.24 &  ps &J$^{15,23}$, {\bf 2}&220&$\sim$280&60&\\
 1044+719&S5 1044+71, JVAS J1048+7143  &  Q  &1.15  & 19.  & 2410 &  7.3 &           0.33$\pm$0.02 &   s &cJ$^{1}$, {\bf 3}&219/337&107&112&\\
 1053+704&S5 1053+70, JVAS J1056+7011  &  Q  &2.492 & 18.5 &  675 &  2.79 &            0.20 &     &J$^{1}$, {\bf 2}&69&208&139&\\
 1053+815&S5 1053+81, JVAS J1058+8114  &  G  &0.706 & 18.5 &  770 & 16.57 &  0.59$\pm$ 0.19 &  ps &J$^{1}$, {\bf 2}&104&226&122&\\
 1058+726&4C +72.16, JVAS J1101+7225   &  Q  &1.46  & 17.9 &  953 & 14.31 &  1.17$\pm$ 0.34 &  ps &JCJ$^{1,27}$, {\bf 5}&222/335&5&30&\\
 1058+629&4C +62.15, JVAS J1101+6241   &  Q  & 0.663 & 17.7 &  700 & 70.04 &  0.99$\pm$ 0.15 &  ps &JCJ$^{1}$, {\bf 5}&$\sim$20&25&5&\\
 1101+384&MRK 421, JVAS J1104+3812     &  B&0.0308& 13.3 &  722 &$>$ 100&  438$\pm$ 4.21 &  ps &JCJ$^{9,32}$, {\bf 4}&$\sim$50/$\sim$310&323&13&\\
 1105+437&B3 1105+437, JVAS J1108+4330 &  Q  &1.226 & 19.5 &  375 &  5.92 &  0.33$\pm$ 0.16 &     &R$^{1,27}$, {\bf 1}&/&227&/&\\
 1106+380&B2 1106+38, CLASS J1109+3744  &  G  &2.29  & 23.0?$^{r}$&  867 &  2.86 &            0.22 &     &U$^{4}$, {\bf 0}&/&16&/& \\
 1107+607&  JVAS J1110+6028          &     &      &      &  404 &  0.21 &            0.04 &     &U$^{1,27}$, {\bf 0}&/&27&/&\\
 1124+455&B3 1124+455, JVAS J1126+4516 &  Q$^{r}$  &1.811 & 17.0 &  355 & 22.26 &  0.95$\pm$ 0.25 &   s &U$^{1,27}$, {\bf 0}&/&351&/&\\
 1124+571&S4 1124+57, JVAS J1127+5650  &  Q  &2.890 & 19.0 &  597 &  5.19 &            0.13 &     &U$^{10}$, {\bf 0}&/&82&/&\\
 1125+596&TXS 1125+596, JVAS J1128+5925&  Q  & 1.779 & 20.0 &  393 & 23.78 &  0.44$\pm$ 0.11 &   s &U$^{1}$, {\bf 0}&/&261&/&\\
 1128+385&B2 1128+38, JVAS J1130+3815  &  Q  &1.733 & 19.4 &  746 & 12.11 &  0.62$\pm$ 0.24 &   s &J$^{3}$, {\bf 2}&175&241&66&\\
 1143+590& JVAS J1146+5848&  Q  &1.982 & 19.6 &  674 &  5.08 &            0.70 &     &U$^{1}$, {\bf 0}&/&62&/&\\
 1144+542&S4 1144+54, JVAS J1146+5356  &  Q  &2.201 & 20.5 &  484 &  3.83 &            0.16 &     &U$^{1}$, {\bf 0}&/&186&/&\\
 1144+402&B3 1144+402, JVAS J1146+3958 &  Q & 1.088 & 18.5 &  739 &  4.73 &            0.47 &     &J$^{2}$, {\bf 2}&35&$\sim$10&25&\\
 1144+352&B2 1144+35B, JVAS J1147+3501 &  G & 0.063 & 15.7 &  663 & 10.34 &  0.62$\pm$ 0.23 &   s &JCJ$^{20}$, {\bf 4}&120&$\sim$300&180&\\
 1146+596&NGC 3894, JVAS J1148+5924    &  G & 0.0108& 11.0 &  627 &  2.40 &            0.23 &     &U$^{1}$, {\bf 0}&/&$\sim$140&/&\\
 1150+812&8C 1150+812, JVAS J1153+8058 &  LPQ&1.250 & 18.5 & 1181 &  4.88 &            0.46$\pm$0.03 &  p  &J$^{3}$, {\bf 2}&260&167&93& \\
 1151+408&B3 1151+408, JVAS J1153+4036 &  Q  &0.916 & 19.5 &  380 &  1.22 &            0.18 &     &J$^{1}$, {\bf 2}&30&90&60&\\
 1155+486&TXS 1155+486, JVAS J1158+4825&  Q  &2.028 & 19.9 &  445 &  3.33 &            0.41 &     &J$^{1}$, {\bf 2}&192&256&64&\\
 1205+544&  JVAS J1208+5413    &     &      &      &  397 &  1.25 &            0.14 &     &U$^{1}$, {\bf 0}&/&120/225&/&\\
 1206+415&B3 1206+416, JVAS J1209+4119 & B  &      & 16.3 &  515 & 36.39 &  1.21$\pm$ 0.25 &   s &U$^{1,27}$, {\bf 0}&/&196&/&\\
 1213+350&4C +35.28, JVAS J1215+3448   &  Q  &0.857 & 20.1 & 1152 &  8.6 &            0.22$\pm$0.09 &     &$^{15,23}$, {\bf ?}&25&55/235&30&\\
 1216+487&S4 1216+48, JVAS J1219+4829  &  Q  &1.076 & 18.5 &  680 &  3.39 &            0.18 &     &JCJ$^{1}$, {\bf 3}&80/105/285&104&1&\\
 1218+444&B3 1218+444B, JVAS J1221+4411&  Q  &1.345 & 17.3 &  478 &  2.69 &            0.13 &     &U$^{1}$, {\bf 0}&/&318&/&\\
 1221+809&8C 1221+809, JVAS J1223+8040 &  B&  & 18.7 &  518 &  6.64 &  0.34$\pm$ 0.17 &   s &JCJ$^{1}$, {\bf 3}&353/215&354&1&\\
 1223+395&B2 1223+39, JVAS J1225+3914  &  Q$^{n}$ &0.623 & 20.6 &  438 &  1.73 &    0.09 &     &JCJ$^{1}$, {\bf 3}&50/230&37/207 &13/23&\\
 1226+373&TXS 1226+373, JVAS J1228+3706&  Q  &1.515 & 18.0 &  953 & 14.25 &  0.39$\pm$ 0.13 &   s &U$^{1,27}$, {\bf 0}&/&308&/&\\
 1239+376&B2 1239+37, JVAS J1242+3720  &  Q  &3.818 & 19.5 &  446 &  0.60 &            0.05 &     &J$^{21}$, {\bf 2}&167&16&151&\\
 1240+381&B2 1240+38, JVAS J1242+3751  &  Q  &1.316 & 19.1 &  768 &  6.32 &  0.24$\pm$ 0.11 &     &U$^{1,4,27}$, {\bf 0}&/&113&/&\\
 1246+586&PG 1246+586, JVAS J1248+5820 &  B&      & 14.0 &  414 &$>$ 100&  6.50$\pm$ 0.47 &   s &U$^{1,27}$, {\bf 0}&/&2&/&\\
 1250+532&TXS 1250+532, JVAS J1253+5301&     &      & 16.4 &  396 & 11.31 &  0.52$\pm$ 0.17 &   s &J$^{1,27}$, {\bf 2}&14,200&253&53&\\
 1254+571&MRK 231, JVAS J1256+5652     &Q,pec Sy1&0.04217& 14.41&419& 6.29  &  0.21$\pm$ 0.12 &  p  &SA(rs)c?, U$^{1,28}$, {\bf 0}&/&/&/&\\
 1258+507&TXS 1258+507, JVAS J1300+5029&  Q  &1.561 & 22.2$^{r}$ &  391 &  6.88 &  0.29$\pm$ 0.13 &     &J$^{1}$, {\bf 2}&170&166&4& \\
 1300+580& JVAS J1302+5748&  G$^{r}$  &1.088$^{r}$ & 21.1$^{r}$ &  758 & 12.37 &  0.41$\pm$ 0.14 &   s &U$^{1,27}$, {\bf 0}&/&14&/& \\
 1305+804&8C 1305+804, J130605.6+800820 &  Q  &1.183 &      &  375 & 19.93 &  0.89$\pm$ 0.25 &   s &/&/&60&/&\\
 1306+360&[VCV2001] J130823.7+354637&Q?$^{r}$&1.055$^{r}$& 20.4$^{r}$ &  437 & 11.51 &  0.42$\pm$ 0.14 &   s & /&/&348&/& \\
 1307+562&JVAS J1309+5557&  Q  &1.629 & 17.6 &  416 & 30.66 &  0.95$\pm$ 0.18 &   s &R$^{1}$, {\bf 1}&/&192&/&\\
 1308+471& JVAS J1310+4653   &     & 1.113& 19.1 &  393 &  8.38 &  0.24$\pm$ 0.11 &     & U$^{1,27}$, {\bf 0}&/&/&/&\\
 1309+555&TXS 1308+554, JVAS J1311+5513 & Q  &0.926 & 19.1 &  677 & 11.58 &  0.29$\pm$ 0.11 &   s &J$^{1}$, {\bf 2}&?&346&/&\\
 1312+533&  JVAS J1314+5306           &    &      &      &  433 &  2.49 &            0.15 &     &U$^{1}$, {\bf 0}&/&242&/&\\
 1322+835&S5 1322+83, VCS1 J1321+8316   &    &1.024,t$^{r}$ &      &  506 &  4.11 &            0.20 &     &R$^{1}$, {\bf 1}&70&$\sim$310&120&\\
 1323+800&S5 1323+79, [VCV2001] J132351.6+794252   & G&1.97 & 21.5 &  458 &  2.87 &            0.22 &     &U$^{1}$, {\bf 0}&/&89&/&\\
 1321+410& JVAS J1324+4048 & G$^{r}$  &0.496 & 19.5 &  413 &  1.36 &            0.09 &     &U$^{1}$, {\bf 0}&/&$\sim280$&/& \\
 1325+436&B3 1325+436, JVAS J1327+4326  & Q  &2.073 & 20.0 &  533 &  5.59 &            0.19 &     & U$^{1,27}$, {\bf 0}&/&225&/&\\
 1333+459&S4 1333+45, JVAS J1335+4542   & Q  &2.449 & 18.5 &  598 &  6.22 &  0.32$\pm$ 0.13 &   s &U$^{1,10,27}$, {\bf 0}&/&295&/&\\
 1333+589&    JVAS J1335+5844         & G  & & 21.9 &  820 & 10.23 &  0.34$\pm$ 0.12 &     &U$^{1,27}$, {\bf 0}&/&$\sim$20&/& \\
 1335+552& JVAS J1337+5501 & Q  &1.096 & 19.0 &  811 & 11.15 &  0.41$\pm$ 0.14 &   s &R$^{1,27}$, {\bf 1}&45&$\sim$110&65&\\
 1337+637& JVAS J1339+6328 & Q  &2.558 & 18.5 &  431 &  3.10 &            0.10 &     &U$^{1}$, {\bf 0}&/&213&/&\\
 1342+663&S4 1342+663, JVAS J1344+6606  & Q  &1.351 & 20.0 &  510 &  1.75 &            0.10 &     &J$^{1}$, {\bf 2}&136&/&/&\\
 1347+539&4C +53.28, JVAS J1349+5341    & Q  &0.980 & 17.3 &  635 & 15.97 &  0.58$\pm$ 0.17 &   s &JCJ$^{1}$, {\bf 4}&317&138&179&\\
 1355+441&B3 1355+441, CLASS J1357+4353  & G$^{r}$  &0.646$^{r}$ & 21.0 $^{r}$ &  464 &  1.94 &    0.12 &     &/&/&296&/& \\
 1357+769&S5 1357+76, JVAS J1357+7643   & B$^{r}$ & & 19.4 &  844 &  0.45 &            0.05 &     &U$^{1}$, {\bf 0}&/&249&/& \\
 1356+478&S4 1356+47, CLASS J1358+4737   & G  &0.230 & 19.5 &  428 &  0.00 &            0.00 &     &/&/&247&/&\\
 1413+373& JVAS J1415+3706            & Q  &2.36  & 18.33&  383 &  1.49 &             0.30$\pm$0.04& p   &U$^{1}$, {\bf 0}&/&126&/&\\
 1415+463&4C +46.29, JVAS J1417+4607    & Q  &1.552 & 17.9 &  904 & 34.46 &  0.42$\pm$ 0.09 &   s &JCJ$^{1}$, {\bf 4}&259&260&1&\\
 1418+546&PG 1418+546, JVAS J1419+5423  & B&0.151 & 15.65& 1707 &$>$ 100&  1.44$\pm$ 0.19 &  ps &J$^{2}$, {\bf 3}&263&127&136&\\
 1417+385&B3 1417+385, JVAS J1419+3821  & Q  &1.832 & 19.3 &  871 & 19.46 &  0.40$\pm$ 0.11 &   s &U$^{1}$, {\bf 0}&/&/&/&\\
 1421+482&  JVAS J1423+4802           & Q  &2.220 & 18.9 &  536 &  0.71 &            0.07 &     &U$^{1}$, {\bf 0}&/&278&/&\\
 1424+366&JVAS J1426+3625 & B&1.091 & 18.3 &  429 &  2.06 &            0.11 &   s &J$^{1}$, {\bf 2}&189&227&38&\\
 1427+543&S4 1427+543, JVAS J1429+5406  & Q  &2.991 & 19.8 &  718 & 13.27 &  0.36$\pm$ 0.11 &   s &U$^{1}$, {\bf 0}&/&138&/&\\
 1432+422&B3 1432+422, JVAS J1434+4203  & Q  &1.240 & 17.8 &  353 &  2.38 &            0.12 &     &U$^{27}$, {\bf 0}&/&$\sim$100&/&\\
 1435+638&S4 1435+63, JVAS J1436+6336   & Q  &2.068 & 15.0 &  795 & 14.20 &  1.09$\pm$ 0.34 &  ps &J$^{16}$, {\bf 3}&230&215&15&\\
 1438+385& CLASS J1440+3820& Q  &1.775 & 21.6 &  944 &  0.29 &            0.05 &     & cD$^{1,27}$, {\bf 3}&$\sim$350&$\sim$350&0&  \\
 1442+637&JVAS J1443+6332             & Q  &1.380 & 17.3 &  456 &  5.78 &            0.48$\pm$0.06 &  p  &U$^{1}$, {\bf 0}&/&183&/&\\
 1448+762&S5 1448+76, JVAS J1448+7601   & Q  &0.899 & 22.3 &  683 &  1.16 &            0.08 &     &U$^{1}$, {\bf 0}&/&83&/&\\
 1456+375& JVAS J1458+3720& G  &0.333 & 18.2 &  591 &  0.77 &            0.03 &     &U$^{1}$, {\bf 0}&/&118&/&\\
 1459+480&TXS 1459+480, JVAS J1500+4751 & B$^{r}$ &1.059$^{r}$ & 19.9 $^{r}$ &  489 &  3.29 &            0.21 &     & J$^{1}$, {\bf 2}&?&79&/& \\
 1504+377&S4 1504+377, JVAS J1506+3730  & G,Sy2&0.6715& 21.2 & 1003 &  3.76 &            0.14 &     &J$^{23}$, {\bf 2}&83&224&141&\\
 1505+428&B3 1505+428, JVAS J1506+4239  & G$^{r}$ &0.587 & 19.4 $^{r}$ &  404 & 11.11 &  0.50$\pm$ 0.16 &   s &J$^{1}$, {\bf 3}&250-270$^{1}$&259&0& \\
 1526+670& JVAS J1526+6650& Q  &3.02  & 17.2 &  417 &  9.82 &  0.22$\pm$ 0.08 &   s &U$^{1,27}$, {\bf 0}&/&45&/&\\
 1531+722&S5 1531+72, JVAS J1531+7206   & Q  &0.899 & 17.7 &  452 & 25.44 &  0.41$\pm$ 0.11 &   s &J$^{1}$, {\bf 2}&56&288&128&\\
 1534+501& JVAS J1535+4957 & Q  &1.121 & 18.0 &  359 &  0.97 &            0.07 &     &U$^{1}$, {\bf 0}&/&325&/&\\
 1543+517&JVAS J1545+5135 & Q  &1.924 & 17.3 &  544 & 25.50 &  0.50$\pm$ 0.12 &   s &J$^{1}$, {\bf 2}&180&178&2&\\
 1543+480&JVAS J1545+4751 & Q &1.277 & 21.7 &  441 &  2.27 &            0.09 &     &U$^{1}$, {\bf 0}&/&131&/&\\
 1545+497&4C +49.26, JVAS J1547+4937    & G  &0.70  & 19.6 &  549 &  1.32 &            0.08 &     &J$^{1}$, {\bf 3}&340&/&/&\\
 1547+507&S4 1547+50, JVAS J1549+5038   & Q  &2.169 & 18.5 &  724 & 49.91 &  0.74$\pm$ 0.13 &   s &J$^{1}$, {\bf 2}&120&215&95& \\
 1550+582&7C 1550+5815, JVAS J1551+5806 & Q  &1.324 & 16.7 &  367 & 70.16 &  0.79$\pm$ 0.12 &   s &U$^{1,27}$, {\bf 0}&/&156&/&\\
 1619+491&JVAS J1620+4901 & Q  &1.513 & 17.8 &  469 &  6.04 &  0.50$\pm$ 0.21 &     &U$^{1,27}$, {\bf 0}&/&$\sim$15&/&\\
 1622+665& JVAS J1623+6624& G  &0.201 & 17.2 &  520 & 36.95 &  0.60$\pm$ 0.11 &   s &U$^{27}$, {\bf 0}&/&57&/&\\
 1623+578&7C 1623+5748, JVAS J1624+5741 & G  &0.789 & 17.3 &  590 &  8.61 &  0.22$\pm$ 0.08 &   s &U$^{27}$, {\bf 0}&/&254&/&\\
 1624+416&4C +41.32, JVAS J1625+4134    & LPQ&2.550 & 22.  & 1362 &  4.84 &            0.11$\pm$0.04 &   {\bf p} &J$^{1}$, {\bf 2}&351&239&112&\\
 1629+495&JVAS J1631+4927& Q  &0.52  & 18.3 &  394 & 16.02 &  0.56$\pm$ 0.15 &   s &JCJ$^{1}$, {\bf 3}&$\sim$245-270 &$\sim$245&0&\\
 1633+382&B3 1633+382, JVAS J1635+3808  & LPQ& 1.807 & 18   & 3189 & 12.87 &  0.31$\pm$ 0.10 &  ps &JCJ$^{3}$, {\bf 4}&165/9&283&86&\\
 1636+473&B3 1636+473, JVAS J1637+4717  & Q  &0.740 & 17.5 & 1330 & 24.85 &  0.55$\pm$ 0.13 &  ps &J$^{1,16}$, {\bf 3}&10-30&334&36-56&\\
 1637+574&S4 1637+57, JVAS J1638+5720   & LPQ&0.749& 17.  & 1807 &$>$ 100&  2.12$\pm$ 0.23 &  ps &J,H$^{2,3}$, {\bf 4}&270-280&200&70-80&\\
 1638+540&JVAS J1639+5357 & Q  &1.977 & 19.7 &  369 &  4.40 &            0.04$\pm$0.01 &  p  &U$^{1}$, {\bf 0}&/&206&/&\\
 1638+398&B3 1638+398, JVAS J1640+3946  & Q  &1.66  & 16.5 & 1285 &  2.42 &            0.15$\pm$0.03 &  p  &JCJ$^{1}$, {\bf 3}&145&/&/&\\
 1642+690&S4 1642+69, JVAS J1642+6856   & HPQ&0.751 & 19.2 & 1516 & 59.99 &  0.68$\pm$ 0.10 &   s &JCJ,H$^{3}$, {\bf 5}&168&191&23&\\
 1641+399&3C 345, GB6 J1642+3948       & Q  & 0.595 & 15.96& 8363 &$>$ 100&  3.96$\pm$ 0.29 &  ps &J,H$^{3}$, {\bf 5}&328&230&98&\\
 1645+635&87GB 1645+63, JVAS J1645+6330 & Q  &2.379 & 19.4 &  444 &  1.94 &            0.09 &     &J$^{1}$, {\bf 2}&132&$\sim$30&102& \\
 1645+410&87GB 1645+41, JVAS J1646+4059 & Q  &0.835 & 20.7 &  388 &  9.92 &  0.36$\pm$ 0.13 &     &U$^{1}$, {\bf 0}&/&$\sim$130&/&\\
 1652+398&MRK 501, JVAS J1653+3945      & B&0.03366& 14.15& 1371&$>$ 100& 80.15$\pm$ 1.49 &  ps &JCJ,H$^{1,6}$, {\bf 4}&49&132&83&\\
 1656+571&4C +57.28, JVAS J1657+5705    & HPQ&1.281 & 17.4 &  844 & 27.41 &  0.65$\pm$ 0.15 &  ps &JCJ$^{1,16,27}$, {\bf 3}&52&54&2&\\
 1656+482&4C +48.41, JVAS J1657+4808    & G & & 20.0 $^{r}$ &  847 & 53.82 &  1.03$\pm$ 0.17 & s &JCJ$^{1,27}$, {\bf 3}&24-37&255&129-142&\\
 1656+477&S4 1656+47, JVAS J1658+4737   & Q  &1.622 & 18.0 & 1420 & 15.31 &  0.42$\pm$ 0.13 &   s &U$^{3,27}$, {\bf 0}&/&344&/&\\
 1700+685&87GB 1700+685, JVAS J1700+6830& G  &0.301 & 17.1 &  435 & 24.48 &  0.23$\pm$ 0.05 &   s &U$^{1,27}$, {\bf 0}&80&130&50&\\
 1716+686&HS 1716+6839, JVAS J1716+6836 & Q  & 0.339 & 18.5 &  988 &$>$ 100&  2.95$\pm$ 0.15 &   s &U$^{1}$, {\bf 0}&/&328&/&\\
 1719+357&S4 1719+35, JVAS J1721+3542   & Q  &0.263 & 17.5 &  874 & 70.37 &  1.32$\pm$ 0.22 &   s &JCJ$^{1}$, {\bf 5}&175&178&3&\\
 1722+401&B3 1722+401, JVAS J1724+4004  & Q?$^{r}$ &1.049 & 21.0 &  532 & 18.58 &  0.60$\pm$ 0.16 &   s & JCJ$^{1}$, {\bf 4}&228/327&305&22& \\
 1726+455&B3 1726+455, JVAS J1727+4530  & Q  &0.717 & 19.0 & 1066 & 19.66 &  0.53$\pm$ 0.13 &   s &J$^{1}$, {\bf 2}&326-345&278&48-67&\\
 1732+389&IRAS 17326+3, JVAS J1734+3857 & Q$^{r}$  &0.97  & 19.0 &  561 &  3.94 &      0.15 &  ps &J$^{1}$, {\bf 2}&180&110&70& \\
 1734+508&S4 1734+508, JVAS J1735+5049  & G? &0.835$^{r}$ & 22.4$^{r}$ &  798 &  2.83 &            0.08 &     &U$^{1}$, {\bf 0}&/&18&/&\\
 1738+499&S4 1738+499, JVAS J1739+4955  & Q  &1.545 & 19.0 &  478 & 15.48 &  0.43$\pm$ 0.11 &   s &R$^{1}$, {\bf 1}&16&23&7&\\
 1738+476&S4 1738+47, JVAS J1739+4737   & B&0.950 & 19.5 &  789 & 13.06 &  0.44$\pm$ 0.12 &   s &U$^{1,27}$, {\bf 0}&/&250&/& \\
 1739+522&4C +51.37, JVAS J1740+5211    & HPQ& 1.381 & 18.5 & 1133 & 60.93 &  0.93$\pm$ 0.15 &   s &J$^{16}$, {\bf 2}&261&7&106& \\
 1744+557&NGC 6454, CLASS J1744+5542     & G  &0.0306& 14.5 &  599 &$>$ 100&  0.91$\pm$ 0.12 &   s &JCJ$^{24}$, {\bf 4}&70/257&248&9&\\
 1745+624&4C +62.29, JVAS J1746+6226    & Q  &3.889 & 19.5 &  580 &$>$ 100&  0.72$\pm$ 0.08 &  ps &J$^{1}$, {\bf 2}&228&211&17&\\
 1746+470& JVAS J1747+4658  &    &      & 21.3 &  634 & 14.79 &  0.39$\pm$ 0.11 &   s &U$^{1,27}$, {\bf 0}&/&270&/&\\
 1749+701&HS 1749+7006, JVAS J1748+7005 & B&0.7699& 17.01&  728 &$>$ 100&  1.41$\pm$ 0.09 &  ps &J$^{1}$, {\bf 2}&25&308&77&\\
1747+433&  JVAS J1749+4321           &  B & & 17.0 &  367 & 12.61 &  0.38$\pm$ 0.12 &   s &J$^{1}$, {\bf 2}&180&170&10&  \\
 1751+441&B3 1751+441, JVAS J1753+4409  & Q  &0.871 & 19.5 &  998 & 28.77 &  0.69$\pm$ 0.15 &   s &JCJ$^{1}$, {\bf 5}&77&85&8&\\
 1755+578&JVAS J1756+5748& Q  &2.110 & 18.0 &  455 & 24.10 &  0.37$\pm$ 0.08 &   s &U$^{1,27}$, {\bf 0}&/&260&/&\\
 1758+388&S4 1758+38, JVAS J1800+3848   & Q  &2.092 & 18.0 &  722 & 57.94 &  1.33$\pm$ 0.22 &   s &R$^{1}$, {\bf 1}&255&265&10&\\
 1803+784&S5 1803+78, JVAS J1800+7828   &B&0.6797& 17.  & 2633 &$>$ 100&  1.63$\pm$ 0.18 &  ps &J$^{3,6}$, {\bf 3}&195&265&70&\\
 1800+440&B3 1800+440, JVAS J1801+4404  & Q  &0.663 & 17.5 & 1148 & 77.54 &  1.48$\pm$ 0.20 &   s &JCJ$^{1,16,27}$, {\bf 4}&240&205&35&\\
 1807+698&3C 371, JVAS J1806+6949      &B&0.051 & 14.4 & 2189 &$>$ 100&  3.40$\pm$ 0.13 &  ps &J,H$^{6}$, {\bf 4}&241&264&23&\\
 1809+568& JVAS J1810+5649           & Q?$^{r}$  &2.041$^{r}$ & 19? $^{r}$ &  576 & 11.30 &  0.19$\pm$ 0.07 &   s &U$^{1}$, {\bf 0}&/&324&/&\\
 1812+412&B3 1812+412, JVAS J1814+4113 &  Q  &1.564 & 18.9 &  534 &  4.57 &            0.20 &     &J$^{27}$, {\bf 2}&325-250&80&115-170&curv.\\
 1818+356&S4 1818+35, [VCV2001] J182042.1+354040  &  Q  &0.971 & 21.1 &  573 &  5.87 &            0.26 &   s &/&/&143&/&\\
 1826+796&S5 1826+79, JVAS J1823+7938  &  G  &0.224 & 16.7 &  577 &  2.16 &            0.07 &     &U$^{1}$, {\bf 0}&/&242&/& \\
 1823+568&4C +56.27, JVAS J1824+5651   &  B&0.664 & 18.4 & 1135 &$>$ 100&  2.25$\pm$ 0.17 &  ps &J,H$^{3}$, {\bf 5}&98&197&99&\\
 1828+399& JVAS J1829+3957  &     &      & 15.3 &  353 &  2.44 &            0.17 &     &U$^{1}$, {\bf 0}&/&302&/&\\
 1834+612& JVAS J1835+6119& Q  &2.274 & 17.6 &  590 & 17.89 &  0.27$\pm$ 0.08 &   s &J$^{1}$, {\bf 3}&154&190&36&\\
 1839+389&B3 1839+389, JVAS J1840+3900  & Q  &3.095 & 19.5 &  476 &  3.21 &            0.23 &     &U$^{1}$, {\bf 0}&/&359&/&\\
 1842+681&S4 1842+68, JVAS J1842+6809   & Q  &0.472 & 17.9 &  936 &$>$ 100&  1.36$\pm$ 0.13 &   s &J,H$^{1}$, {\bf 4}&$\sim$290&134&156&\\
 1843+356&  JVAS J1845+3541           & G? &0.764 & 21.9 &  794 &  1.70 &            0.09 &     &U$^{1}$, {\bf 0}&/&45&/&\\
 1849+670& JVAS J1849+6705& Q  &0.657 & 18.0 &  992 &$>$ 100&  0.86$\pm$ 0.10 &   s &JCJ$^{1}$, {\bf 3}&240&306&66&\\
 1851+488&S4 1851+48, JVAS J1852+4855   & Q  &1.25  & 19.0 &  351 &  1.20 &            0.12 &     &U$^{1}$, {\bf 0}&/&/&/&\\
 1850+402&S4 1850+40, JVAS J1852+4019   & Q  &2.12  & 18.5 &  535 &  2.93 &            0.20 &     &J$^{1}$, {\bf 2}&327&235&92&\\
 1856+737& JVAS J1854+7351& Q  &0.461 & 17.5 &  546 &$>$ 100&  2.03$\pm$ 0.19 &   s &J$^{1}$, {\bf 3}&24&30&6& \\
 1908+484& JVAS J1909+4834& Q  &0.513 & 19.0 &  423 &  6.40 &  0.33$\pm$ 0.16 &     &cJ$^{1}$, {\bf 5}&90 +&51&39&\\
 1910+375& JVAS J1912+3740& Q  &1.104 & 18.5 &  402 &  3.15 &            0.21 &     &R$^{1}$, {\bf 1}&?&176&/&\\
 1924+507&4C +50.47, JVAS J1926+5052    & Q  &1.098 & 17.9 &  354 & 17.70 &  0.87$\pm$ 0.25 &   s &cJCJ$^{1,29}$, {\bf 4}&9&2&7& \\
 1926+611&87GB 1926+611, JVAS J1927+6117& B& $>$0.2  & 17.5 &  618 & 21.64 &  0.58$\pm$ 0.13 &   s &J$^{1}$, {\bf 2}&195&127&68& \\
 1928+738&4C +73.18, JVAS J1927+7358    & LPQ&0.302 & 16.5 & 3561 &$>$ 100&  9.10$\pm$ 0.48 &  ps &cJCJ$^{3}$, {\bf 5}&189&167&22&\\
 1936+714&S5 1936+71, JVAS J1936+7131   & Q  &1.864 & 19.5 &  391 &  1.05 &            0.08 &     &J$^{1}$, {\bf 2}&351&189&162&\\
 1943+546&S4 1943+54, JVAS J1944+5448   & G  &0.263 & 17.6 &  938 &  5.22 &            0.23 &     &U$^{1}$, {\bf 0}&/&81&/&\\
 1946+708&87GB 1946+704, JVAS J1945+7055& G  &0.101 & 16.1 &  645 &  1.95 &            0.11 &     &U$^{1}$, {\bf 0}&/&205&/&\\
 1950+573& JVAS J1951+5727& G$^{r}$  &0.652 & 18.0 &  476 &  7.36 &  0.58$\pm$ 0.24 &   s &U$^{1,27}$, {\bf 0}&/&74&/& \\
 1954+513&S4 1954+51, JVAS J1955+5131   & LPQ& 1.223  & 18.5 & 1610 &  48.30 &  1.86$\pm$ 0.34 &   s&JCJ$^{1}$, {\bf 4}&345/180&302&43&\\
 2007+777&S5 2007+77, JVAS J2005+7752   & B&0.342 & 16.5 & 1279 & 48.53 &  1.01$\pm$ 0.19 &  ps &JCJ$^{3}$, {\bf 4}&90/250&264&14&\\
 2005+642&JVAS J2006+6424& Q? &1.574 & 19.0 &  739 &  10.8 &            0.31$\pm$0.12 &   s &U$^{1}$, {\bf 0}&/&/&/& \\
 2007+659&TXS 2007+659, JVAS J2007+6607 & Q  &1.325 & 16.4 &  756 &  3.90 &            0.23 &     &J$^{1}$, {\bf 2}&322&210&112&\\
 2010+723&4C +72.28, JVAS J2009+7229    & B&$>$0.2   & 19.0 &  910 &  3.16 &            0.39 &     &J$^{1}$, {\bf 2}&103&320-230&127-143&\\
 2017+745&4C +74.25, JVAS J2017+7440    & Q  &2.187 & 18.3 &  500 & 26.66 &  1.00$\pm$ 0.24 &   s &J$^{1}$, {\bf 2}&180&90&90&\\
 2021+614&87GB 2021+612, JVAS J2022+6136& LPQ&0.227 & 19.5 & 2743 &  4.91 &            0.36 &     &U$^{1}$, {\bf 0}&/&$\sim$35&/&\\
 2023+760&S5 2023+76, JVAS J2022+7611   & B&$>$0.2   & 17.8 &  426 &  2.55 &            0.09 &     &J$^{1}$, {\bf 2}&?&209&/&\\
 2054+611& JVAS J2055+6122            & Q?$^{r}$ &0.864$^{r}$ & 21.5?$^{r}$&  414 &  5.42 &     0.55 &     &U$^{1}$, {\bf 0}&/&161&/&\\
 2116+818&S5 2116+81, JVAS J2114+8204   & Q,Sy1&0.084 & 15.7 &  376 &$>$ 100& 11.44$\pm$ 0.61 &   s &/&/&333&/& \\
 2136+824&S5 2136+82, JVAS J2133+8239   & Q  &2.357 & 18.9 &  509 &  4.91 &            0.39 &     &J$^{1}$, {\bf 2}&133&143&10&\\
 2138+389&  JVAS J2140+3911           &    &1.306,t$^{r}$ & 19.0 &  502 &  3.50 &            0.54 &   s &U$^{1}$, {\bf 0}&/&91&/& \\
 2200+420&BL Lacertae, JVAS J2202+4216  &B&0.0686& 14.5& 3593 &$>$ 100&  4.33$\pm$ 0.62 &  ps &JCJ$^{7}$, {\bf 4}&$\sim$150/$\sim$300&$\sim$180&30&\\
 2214+350&B2 2214+35, JVAS J2216+3518   & Q  &0.510 & 18.0 &  477 &  6.53 &  0.45$\pm$ 0.24 &     &J$^{1}$, {\bf 3}&212&173&39&\\
 2229+695&S5 2229+69, JVAS J2230+6946   & B&1.413,t,$^{r}$ & 19.6 & 1365 &  0.38 &            0.15 &     &JCJ$^{1}$, {\bf 3}&70/167&$\sim$70&0& \\
 2235+731& JVAS J2236+7322& Q &1.345 & 21.5?&  424 &  3.21 &            0.63 &     &U$^{1}$, {\bf 0}&/&39&/&\\
 2238+410& JVAS J2241+4120     & B?$^{r}$&0.726$^{r}$ & 17.9 &  677 &  5.01 &            0.51 &     &J$^{1}$, {\bf 2}&277&135&142&\\
 2253+417&B3 2253+417, JVAS J2255+4202  & Q  &1.476 & 18.8 & 1120 & 10.67 &  1.06$\pm$ 0.36 &   s &U$^{1,27}$, {\bf 0}&/&$\sim$40&/&\\
 2255+416&4C +41.45, JVAS J2257+4154    & Q  &2.15  & 20.9 & 1111 &  3.95 &            0.39 &     &R$^{1}$, {\bf 1}&180&179&1&\\
 2259+371& JVAS J2301+3726& Q  &2.179 & 20.4 &  406 &  1.91 &            0.22 &     &JCJ$^{1}$, {\bf 3}&19&7&12&\\
 2309+454& JVAS J2311+4543& Q &1.447 & 20.0? $^{r}$&  597 &  1.23 &            0.18 &     &U$^{1}$, {\bf 0}&/&122&/&\\
 2310+385&B3 2310+385, JVAS J2312+3847  & Q  & 2.181 & 17.5 &  484 &  3.03 &            0.23 &     &U$^{1}$, {\bf 0}&/&238&/&\\
 2319+444& JVAS J2322+4445     & G  $^{r}$&1.251$^{r}$ & 19.9 &  366 &  0.31 &            0.17 &     &J$^{1}$, {\bf 2}&109&344&125&\\
 2346+385&B3 2346+385, JVAS J2349+3849  & Q  &1.032 & 19.1 &  640 & 21.63 &  1.97$\pm$ 0.57 &   s &J$^{1}$, {\bf 2}&263&329&66&\\
 2351+456&4C +45.51, JVAS J2354+4553  & LPQ& 1.986 & 20.6 & 1145 &  5.49 &            0.47 &     &J$^{1,27}$, {\bf 2}&200&288&88&\\
 2352+495& JVAS J2355+4950 & G  &0.237 & 20.1 & 1552 &  0.44 &            0.14 &     &U$^{1}$, {\bf 0}&/&163&/&\\
 2353+816&S5 2353+81, JVAS J2356+8152   & B&1.344 & 20.3 &  476 &  2.66 &            0.31 &     &JCJ$^{1}$, {\bf 2}&63&341&82& \\
 2356+390&B3 2356+390, JVAS J2358+3922  & Q, C.g.l.& 1.198 & 20.6 &  371 &   10.0 &            0.71$\pm$0.03 &   s &J$^{1,27}$, {\bf 2}&336&230&106&\\
 2356+385&S4 2356+38, JVAS J2359+3850   & Q & 2.704 & 19.0 &  449 &  7.97 &  1.27$\pm$ 0.48 &   s &R$^{1}$, {\bf 1}&90&180&90& \\
\end{supertabular}
}
\begin{table*}[htb]
\caption{References and abbreviations for Table 1.}
\setcounter{table}{2}
\label{comments}
\begin{flushleft}
\begin{tabular}{llll}
&&&\\
\noalign{\smallskip}
/ no VLA information found&&&\\
$^{1}$ T. Pearson&$^{8}$ Peacock \& Wall 1982&$^{15}$ Browne \& Perley 1986&$^{22}$ Machalski \& Condon 1983\\
$^{2}$ Price et al. 1993&$^{9}$ Antonucci \& Ulvestad 1985&$^{16}$ Reid et al. 1995&$^{23}$ Perley 1982\\
$^{3}$ Murphy et al. 1993&$^{10}$ Neff \& Hutchings 1990&$^{17}$ Machalski et al. 1982&$^{24}$ Bridle \& Fomalont 1978\\
$^{4}$ Vigotti et al. 1989&$^{11}$ Wagner et al. 1996&$^{18}$ Garrington et al. 1991&$^{25}$ K\"uhr et al. 1986\\
$^{5}$ Baum et al. 1990&$^{12}$ Pedlar et al. 1983&$^{19}$ Barthel et al. 1986&$^{26}$ Hummel et al. 1997\\
$^{6}$ Cassaro et al. 1999&$^{13}$ Pedlar et al. 1990&$^{20}$ Giovannini et al. 1999&$^{27}$ Patnaik et al. 1992\\
$^{7}$ Antonucci 1986&$^{14}$ Patnaik et al. 1993&$^{21}$ Machalski 1998&$^{28}$ Thean et al. 1999\\
      &                           &                      &                         \\
      $^{29}$ Owen \& Puschell 1984\\
      $^{30}$ Machalski et al. 1996\\
      $^{31}$ Perley et al. 1982\\
      $^{32}$ Ulvestad et al. 1983\\
      & & & \\
      $^{m}$ : Marcha et al. 1996&$^{r}$ : priv. comm. R. Vermeulen&$^{n}$ : NED information&\\
                           &                            &                     & \\
			   U: &unresolved component&R:& slightly resolved component\\
			   (c)J:&(complex) jet-like extended component&(c)D:&(complex) double source\\
			   (c) JCJ:&(complex) jet- and counter-jet&H:&halo emission\\
			   G$_{L}$:& gravitational lens system& C.g.l.& Candidate gravitational lens\\
			   &  &&\\
			   $^{\#}$ Kim et al. 1991&$^{\$}$ Slee \& Siegman 1983&\\
			   \noalign{\smallskip}
			   \end{tabular}
			   \end{flushleft}
			   \end{table*}

\clearpage
\onecolumn
\landscape
\textheight24cm
\textwidth17.5cm
\oddsidemargin-2cm
\tabcolsep0.3mm
\tablecaption{Calculated values for those sources that have not been detected by {\it ROSAT}. This table is only
available in the online edition of the Journal.} {\smallskip}
\setcounter{table}{3}
\tablehead{\noalign{\smallskip} \hline \noalign{\smallskip}
&      &&&& $\alpha=-0.75$ &&&& &&&&observed $\alpha$&&& &&&\\ \hline
Source & $\beta_{\rm app}$ & $\delta_{\rm IC}$ & $\delta_{\rm IC}^{\rm con}$ & $\delta_{\rm EQ}$ & log (T$_{\rm B}$/10$^{11}$)&log (T$_{\rm Bi}^{\rm con}$/10$^{11}$) & $\Gamma_{\rm sl}$ & $\phi$ & $\delta_{\rm IC}$ & $\delta_{\rm IC}^{\rm con}$ & $\delta_{\rm EQ}$ &log (T$_{\rm B}$/10$^{11}$) &log (T$_{\rm Bi}^{\rm con}$ / 10$^{11}$) & $\Gamma_{\rm sl}$ & $\phi$ & VLBI flux & VLBI core flux & log(R$_{\rm C}$) & R$_{\rm V}$ \\
\multicolumn{1}{c}{\tiny ~~~ } & 
 \multicolumn{1}{c}{\tiny ~[$c$]} &
\multicolumn{1}{c}{\tiny ~~~}&
\multicolumn{1}{c}{\tiny ~~~~~} & \multicolumn{1}{c}{\tiny ~~~} &
\multicolumn{1}{c}{\tiny ~~[K] } &
\multicolumn{1}{c}{\tiny ~~[K]}&
\multicolumn{1}{c}{\tiny ~}&
\multicolumn{1}{c}{\tiny ~~[deg]}&
\multicolumn{1}{c}{\tiny ~~~}&
\multicolumn{1}{c}{\tiny ~~~~~} & \multicolumn{1}{c}{\tiny ~~~} &
\multicolumn{1}{c}{\tiny ~~[K] } &
\multicolumn{1}{c}{\tiny ~~[K]}&
\multicolumn{1}{c}{\tiny }&
\multicolumn{1}{c}{\tiny ~~~~[deg]}&
\multicolumn{1}{c}{\tiny ~~~~[mJy]}&
\multicolumn{1}{c}{\tiny ~~~~[mJy]}&&\\
(1)&(2)&(3)&(4)&(5)&(6)&(7)&(8)&(9)&(10)&(11)&(12)&(13)&(14)&(15)&(16)&(17)&(18)&(19)&(20)\\
\noalign{\smallskip} \hline \hline  \noalign{\smallskip}}
\tabletail{\hline\multicolumn{12}{r}{continued on next page}\\}
\tablelasttail{\hline}
{\small
\begin{supertabular}{c|c|ccccccc|ccccccc|cccc}
\label{non}
0016+731& 15.893&  0.12&  0.07&  0.02& -0.93& -0.14&      &  7.20&  0.15&  0.09&  0.00& -0.93& -0.23&860.16&  7.20&  1056&  1673&-0.21& 0.98\\
0108+388&  0.460&  0.27&  0.21&  0.05& -0.62& -0.15&  2.35&-52.22&  0.23&  0.17&  0.01& -0.62& -0.07&  2.79&-51.29&   456&  1714&-0.46& 1.30\\
0133+476&  3.287&  4.78&  6.77&  2.41&  0.85&  0.28&  3.63& 11.37&  9.07& 17.26&  0.80&  0.85&  0.07&  5.18&  4.09&  1686&  1795&-0.03& 0.99\\
0227+403&  3.531&  0.34&  0.27&  0.10& -0.40&  0.00& 19.73& 31.36&  1.04&  1.05&  0.01& -0.40& -0.41&  7.01& 29.38&   332&   550&-0.12& 1.26\\
0444+634& 10.927&  2.60&  3.22&  1.56&  0.66&  0.30& 24.44&  9.90&  3.72&  5.27&  0.58&  0.66&  0.18& 18.03&  9.38&   368&   654&-0.22& 1.08\\
0554+580&  1.124&  0.19&  0.13&  0.04& -0.74& -0.13&  6.00& 82.40&  0.20&  0.14&  0.01& -0.74& -0.15&  5.78& 82.32&   146&   293&-0.79& 0.32\\
0627+532& 10.199&  1.85&  2.12&  1.38&  0.59&  0.37& 29.27& 10.85& 10.84& 32.34&  0.15&  0.59& -0.19& 10.26&  5.28&    67&   485&-0.86& 1.00\\
0641+393&  7.066&  3.42&  4.49&  1.53&  0.66&  0.21&  9.16& 13.13&  7.09& 12.73&  0.39&  0.66& -0.03&  7.14&  8.11&   458&   577& 0.01& 1.27\\
0642+449&  9.845&  9.98& 16.64&  4.47&  1.12&  0.27&  9.90&  5.75& 12.20& 22.39&  2.12&  1.12&  0.21& 10.11&  4.60&  1568&  1652& 0.12& 1.39\\
0646+600&  0.192&  1.49&  1.62&  0.55&  0.25&  0.10&  1.09& 17.11&  1.76&  2.01&  0.20&  0.25&  0.04&  1.17& 10.23&   475&  1055&-0.29& 1.15\\
0651+410&  0.054&  0.18&  0.13&  0.07& -0.67& -0.05&  2.80& -6.40&  0.33&  0.24&  0.01& -0.67& -0.28&  1.68& -6.94&   179&   341&-0.38& 0.80\\
0711+356&  3.015&  2.74&  3.43&  1.93&  0.72&  0.35&  3.21& 21.13& 15.37& 51.48&  0.26&  0.72& -0.19&  8.01&  1.41&   157&  1245&-0.76& 1.38\\
0714+457&  2.920&  2.83&  3.57&  2.05&  0.73&  0.35&  3.10& 20.57&  5.70&  9.72&  0.58&  0.73&  0.12&  3.69&  8.31&   397&   524&-0.08& 1.09\\
0724+571&  4.588&  1.04&  1.05&  0.48&  0.17&  0.16& 11.09& 23.46&  3.07&  4.49&  0.08&  0.17& -0.21&  5.13& 17.31&   349&   556&-0.05& 1.42\\
0800+618&  7.941&  9.86& 16.40&  5.59&  1.19&  0.34&  8.18&  5.69& 17.25& 38.75&  2.01&  1.19&  0.17& 10.48&  2.53&   961&  1187&-0.01& 1.21\\
0900+520&  9.389&  0.10&  0.06&  0.01& -1.13& -0.27&465.40& 12.16&  0.23&  0.15&  0.00& -1.13& -0.62&190.77& 12.15&   225&   302&-0.24& 0.77\\
0930+493& 13.929&  0.16&  0.11&  0.03& -0.87& -0.21&597.27&  8.21&  0.93&  0.90&  0.00& -0.87& -0.85&105.34&  8.18&    61&   559&-0.97& 0.97\\
0942+468&  1.513&  0.10&  0.06&  0.02& -1.06& -0.21& 16.72& 66.77&  0.22&  0.14&  0.00& -1.06& -0.53&  7.59& 66.15&   224&   353&-0.20& 1.00\\
0949+354&  6.103&  2.87&  3.63&  1.32&  0.61&  0.22&  8.10& 15.35&  6.31& 11.11&  0.28&  0.61& -0.04&  6.19&  9.12&   256&   374&-0.20& 0.93\\
0950+748&  0.515&  0.23&  0.16&  0.03& -0.80& -0.25&  2.91&-56.43&  2.63&  4.06&  0.00& -0.80& -1.12&  1.56&  9.43&   266&   662&-0.44& 0.90\\
1030+398&  1.241&  0.33&  0.26&  0.05& -0.59& -0.18&  4.05& 75.39&  0.50&  0.42&  0.01& -0.59& -0.34&  2.78& 72.30&   254&   692&-0.41& 1.07\\
1030+415&  4.186&  1.33&  1.42&  0.56&  0.26&  0.16&  7.62& 24.59&  9.51& 25.88&  0.05&  0.26& -0.48&  5.73&  4.48&   250&   390&-0.29& 0.80\\
1031+567&  0.472&  0.27&  0.20&  0.05& -0.68& -0.20&  2.41&-53.26&  2.30&  3.26&  0.00& -0.68& -0.96&  1.42& 11.83&   314&  1090&-0.58& 0.91\\
1106+380&  3.320&  0.27&  0.20&  0.08& -0.49& -0.01& 22.27& 33.33&  2.24&  3.22&  0.00& -0.49& -0.76&  3.80& 23.83&    76&   681&-1.06& 0.79\\
1124+571&  8.056&  0.93&  0.91&  0.22& -0.04& -0.02& 36.02& 13.97&  5.01&  9.36&  0.02& -0.04& -0.59&  9.08& 10.26&   325&   424&-0.26& 0.71\\
1143+590&  6.965&  1.38&  1.48&  0.57&  0.31&  0.19& 18.68& 15.75&  1.42&  1.53&  0.20&  0.31&  0.18& 18.18& 15.71&   357&   581&-0.28& 0.86\\
1144+542&  3.426&  1.99&  2.32&  0.80&  0.41&  0.16&  4.20& 25.00&  4.38&  6.87&  0.14&  0.41& -0.11&  3.64& 12.89&   194&   387&-0.40& 0.80\\
1146+596&  0.155&  0.00&  0.00&  0.00& -2.80& -0.60&200.13&-17.62&  0.00&  0.00&  0.00& -2.80& -0.88&107.42&-17.62&    72&   501&-0.94& 0.80\\
1151+408&  4.392&  0.07&  0.04&  0.01& -1.26& -0.29&142.53& 25.65&  1.21&  1.33&  0.00& -1.26& -1.32&  9.00& 23.98&   269&   546&-0.15& 1.44\\
1155+486&  9.743&  2.95&  3.75&  1.49&  0.66&  0.26& 17.75& 10.75&  9.05& 19.53&  0.25&  0.66& -0.10&  9.82&  6.32&   258&   351&-0.24& 0.79\\
1216+487&  6.684&  1.87&  2.14&  0.63&  0.33&  0.10& 13.17& 15.83&  8.72& 19.90&  0.08&  0.33& -0.41&  6.98&  6.37&   295&   618&-0.36& 0.91\\
1223+395&  0.144&  0.36&  0.29&  0.09& -0.44& -0.06&  1.60&-18.71&  1.70&  2.07&  0.01& -0.44& -0.62&  1.15&  8.57&   200&   506&-0.34& 1.16\\
1321+410&  0.297&  0.11&  0.07&  0.02& -1.08& -0.28&  4.95&-33.44&  0.30&  0.21&  0.00& -1.08& -0.66&  1.95&-35.90&   189&   413&-0.34& 1.00\\
1323+800& 13.881&  2.56&  3.16&  1.12&  0.55&  0.20& 39.10&  7.97&  4.88&  7.74&  0.25&  0.55& -0.01& 22.27&  7.34&   242&   543&-0.28& 1.19\\
1325+436&  4.676&  1.28&  1.36&  0.51&  0.24&  0.15&  9.54& 22.56&  6.04& 12.22&  0.05&  0.24& -0.37&  4.91&  9.26&   140&   595&-0.58& 1.12\\
1337+637&  8.546&  0.29&  0.22&  0.04& -0.69& -0.24&126.92& 13.33&  1.45&  1.66&  0.00& -0.69& -0.82& 26.23& 12.98&   173&   389&-0.40& 0.90\\
1356+478&  0.146&&&  0.01& -1.18&&&  0.00&&&  0.00& -1.18&&&  0.00&   237&   446&-0.26& 1.04\\
1424+366&  4.670&  1.02&  1.02&  0.25& -0.02& -0.02& 11.72& 23.15&  1.15&  1.19&  0.07& -0.02& -0.07& 10.46& 22.87&   623&   759& 0.16& 1.77\\
1438+385&  2.566&  1.03&  1.03&  0.21& -0.08& -0.08&  4.21& 37.71&  4.11&  6.73&  0.02& -0.08& -0.56&  2.98& 12.86&   239&   582&-0.60& 0.62\\
1448+762&  4.475&  0.31&  0.24&  0.05& -0.62& -0.19& 34.17& 25.08&  0.57&  0.48&  0.01& -0.62& -0.42& 18.85& 24.83&   272&   406&-0.40& 0.59\\
1456+375&  2.920&  0.67&  0.61&  0.20& -0.19& -0.04&  7.45& 36.22&  0.96&  0.95&  0.04& -0.19& -0.18&  5.45& 34.66&   705&   782& 0.08& 1.32\\
1459+480&  0.771&  0.71&  0.65&  0.22& -0.08&  0.05&  1.48& 86.55&  1.55&  1.77&  0.03& -0.08& -0.23&  1.29& 37.74&   405&   526&-0.08& 1.08\\
1504+377& 10.547&  1.34&  1.43&  0.37&  0.12&  0.01& 42.44& 10.66&  5.90& 11.31&  0.05&  0.12& -0.48& 12.46&  8.27&   331&   685&-0.48& 0.68\\
1543+480&  2.265&  0.40&  0.33&  0.08& -0.47& -0.14&  7.89& 46.56&  3.43&  5.80&  0.00& -0.47& -0.89&  2.61& 15.93&   313&   494&-0.15& 1.12\\
1732+389&  5.104& 11.64& 20.08&  8.36&  1.29&  0.38&  6.98&  3.64& 54.51&275.95&  2.21&  1.29& -0.06& 27.50&  0.20&   377&  1228&-0.17& 2.19\\
1734+508&  1.102&  0.59&  0.52&  0.13& -0.28& -0.08&  2.17& 75.71&  1.71&  2.04&  0.01& -0.28& -0.47&  1.50& 34.95&   162&   828&-0.69& 1.04\\
1812+412&  6.689&  0.95&  0.94&  0.30&  0.05&  0.06& 24.61& 16.68&  3.99&  6.65&  0.03&  0.05& -0.43&  7.73& 12.64&   250&   331&-0.33& 0.62\\
1826+796&  0.685&  0.06&  0.03&  0.01& -1.45& -0.38& 13.33&-68.93&  0.09&  0.05&  0.00& -1.45& -0.58&  8.12&-69.12&   138&   597&-0.62& 1.03\\
1843+356&  0.000&  0.14&  0.09&  0.02& -0.97& -0.24&  3.73&  0.00&  0.83&  0.77&  0.00& -0.97& -0.90&  1.02&  0.00&   174&   842&-0.66& 1.06\\
1850+402&  4.390&  0.36&  0.29&  0.06& -0.53& -0.15& 28.42& 25.51&  1.27&  1.38&  0.00& -0.53& -0.61&  8.60& 23.82&   182&   657&-0.47& 1.23\\
1910+375&  1.476&  0.45&  0.38&  0.14& -0.27&  0.02&  3.74& 64.90&  2.05&  2.69&  0.01& -0.27& -0.52&  1.80& 28.78&   191&   374&-0.32& 0.93\\
1943+546&  0.183&  0.01&  0.00&  0.00& -2.30& -0.43& 83.79&-20.74&  0.09&  0.03&  0.00& -2.30& -1.50&  6.01&-20.89&    34&   967&-1.44& 1.03\\
2007+659&  0.645&  5.47&  7.98&  3.88&  0.99&  0.37&  2.86&  2.52& 23.40& 82.48&  0.78&  0.99& -0.07& 11.73&  0.14&   228&   526&-0.52& 0.70\\
2021+614&  0.318&  0.91&  0.89&  0.29& -0.01&  0.03&  1.06&-83.97&  1.90&  2.30&  0.05& -0.01& -0.23&  1.24& 13.19&  1219&  2171&-0.35& 0.79\\
2136+824&  4.464&  1.87&  2.15&  1.29&  0.57&  0.34&  6.53& 21.71& 13.79& 49.29&  0.12&  0.57& -0.29&  7.65&  2.44&    93&   285&-0.74& 0.56\\
2235+731&  3.389&  0.37&  0.30&  0.10& -0.35&  0.01& 17.12& 32.54&  0.66&  0.59&  0.01& -0.35& -0.21&  9.75& 31.82&   214&   319&-0.30& 0.75\\
2255+416&  1.735&  0.75&  0.71&  0.26&  0.00&  0.10&  3.04& 53.38&  6.36& 14.95&  0.02&  0.00& -0.61&  3.50&  4.67&   220&  1132&-0.70& 1.02\\
2310+385&  4.459&  0.06&  0.03&  0.01& -1.34& -0.32&165.62& 25.28&  0.52&  0.39&  0.00& -1.34& -1.12& 20.51& 24.97&   142&   511&-0.53& 1.06\\
2319+444&  7.690&  2.28&  2.73&  0.96&  0.48&  0.17& 14.35& 13.65&  4.95&  8.05&  0.21&  0.48& -0.09&  8.55& 10.54&   499&   552& 0.14& 1.51\\
2351+456&  9.380&  1.31&  1.40&  0.38&  0.17&  0.07& 34.52& 11.94& 10.92& 31.74&  0.03&  0.17& -0.62&  9.53&  5.20&   589&   791&-0.29& 0.69\\
2352+495&  0.362&  0.01&  0.01&  0.00& -2.07& -0.52& 38.59&-39.81&  0.16&  0.07&  0.00& -2.07& -1.45&  3.62&-40.64&   138&  1088&-1.05& 0.70\\
\end{supertabular}
}
\clearpage
\onecolumn
\landscape
\textheight21cm
\textwidth13.5cm
\tabcolsep0.3mm
\vspace*{1.6cm}
\tablecaption{Calculated values for those sources that have been detected by {\it ROSAT}. This table is only available in the online edition of the Journal.} {\smallskip}
\setcounter{table}{4}
\tablehead{\noalign{\smallskip} \hline \noalign{\smallskip}
& &&&&$\alpha=-0.75$&&&& &&&&observed $\alpha$&&& &&&\\ \hline
Source &  $\beta_{\rm app}$ & $\delta_{\rm IC}$ & $\delta_{\rm IC}^{\rm con}$ & $\delta_{\rm EQ}$ & log (T$_{\rm B}$/10$^{11}$) & log (T$_{\rm Bi}^{\rm con}$/10$^{11}$) & $\Gamma_{\rm sl}$ & $\phi$ & $\delta_{\rm IC}$ & $\delta_{\rm IC}^{\rm con}$ & $\delta_{\rm EQ}$ & log (T$_{\rm B}$/10$^{11}$) & log (T$_{\rm Bi}^{\rm con}$ /10$^{11}$) & $\Gamma_{\rm sl}$ & $\phi$ & VLBI flux &VLBI core flux & log(R$_{\rm C}$) & R$_{\rm V}$ \\
\multicolumn{1}{c}{\tiny ~~~ } &
 \multicolumn{1}{c}{\tiny ~[$c$]} &
 \multicolumn{1}{c}{\tiny ~~~}&
 \multicolumn{1}{c}{\tiny ~~~~~} & \multicolumn{1}{c}{\tiny ~~~} &
 \multicolumn{1}{c}{\tiny ~~[K] } &
 \multicolumn{1}{c}{\tiny ~~[K]}&
 \multicolumn{1}{c}{\tiny ~}&
 \multicolumn{1}{c}{\tiny ~~[deg]}&
 \multicolumn{1}{c}{\tiny ~~~}&
 \multicolumn{1}{c}{\tiny ~~~~~} & \multicolumn{1}{c}{\tiny ~~~} &
 \multicolumn{1}{c}{\tiny ~~[K] } &
 \multicolumn{1}{c}{\tiny ~~[K]}&
 \multicolumn{1}{c}{\tiny }&
 \multicolumn{1}{c}{\tiny ~~~~[deg]}&
 \multicolumn{1}{c}{\tiny ~~~~[mJy]}&
 \multicolumn{1}{c}{\tiny ~~~~[mJy]}\\
 (1)&(2)&(3)&(4)&(5)&(6)&(7)&(8)&(9)&(10)&(11)&(12)&(13)&(14)&(15)&(16)&(17)&(18)&(19)&(20)\\
 \noalign{\smallskip} \hline \hline  \noalign{\smallskip}}
 \tabletail{\hline\multicolumn{12}{r}{continued on next page}\\}
 \tablelasttail{\hline}
 \small
 \begin{supertabular}{c|c|ccccccc|ccccccc|cccc}
\label{det}
0014+813&  0.431&  3.49&  4.61&  2.02&  0.81&  0.35&  1.92&  4.33& 12.02& 30.41&  0.30&  0.81& -0.04&  6.06&  0.34&   851&  1093& 0.19& 1.98\\
0035+413&  5.430&  6.42&  9.71&  5.90&  1.16&  0.47&  5.58&  8.85&  6.43&  9.73&  4.73&  1.16&  0.47&  5.59&  8.83&   302&  1014&-0.57& 0.91\\
0110+495&  0.788&  3.27&  4.25&  3.84&  0.95&  0.51&  1.88&  8.70&  4.81&  7.44&  1.82&  0.95&  0.39&  2.57&  3.96&   561&   891&-0.10& 1.25\\
0212+735&  4.291&  1.43&  1.54&  0.58&  0.34&  0.21&  7.52& 23.81&  4.83&  8.51&  0.07&  0.34& -0.20&  4.42& 11.91&  1230&  2066&-0.27& 0.91\\
0219+428&  5.810&  0.47&  0.40&  0.25& -0.04&  0.23& 36.86& 19.41&  1.23&  1.32&  0.03& -0.04& -0.11& 14.73& 18.73&   629&   923&-0.11& 1.15\\
0248+430&  3.329&  0.36&  0.29&  0.13& -0.27&  0.10& 16.89& 33.10&  0.51&  0.43&  0.02& -0.27& -0.03& 12.06& 32.77&   134&  1280&-1.02& 0.91\\
0251+393&  1.089&  0.50&  0.43&  0.37&  0.05&  0.31&  2.44& 78.73&  0.81&  0.76&  0.08&  0.05&  0.13&  1.76& 68.95&   309&   494&-0.12& 1.21\\
0454+844&  0.768&  0.59&  0.52&  0.39&  0.03&  0.23&  1.64&-87.66&  0.85&  0.82&  0.11&  0.03&  0.09&  1.36& 78.21&   229&  1030&-0.79& 0.74\\
0537+531&  5.592&  0.86&  0.83&  0.31&  0.06&  0.12& 19.15& 19.83&  2.40&  3.20&  0.04&  0.06& -0.24&  7.93& 17.25&   478&   747&-0.14& 1.12\\
0546+726&  0.860&  0.25&  0.18&  0.09& -0.42&  0.09&  3.59&-83.46&  0.96&  0.95&  0.01& -0.42& -0.41&  1.38& 68.72&    99&   262&-0.61& 0.65\\
0600+442&  5.562&  0.07&  0.04&  0.02& -1.02& -0.05&225.93& 20.38&  0.63&  0.50&  0.00& -1.02& -0.87& 25.85& 20.14&   150&   461&-0.67& 0.65\\
0609+607&  5.657&  1.43&  1.55&  0.45&  0.23&  0.10& 12.25& 18.90&  4.26&  6.90&  0.05&  0.23& -0.27&  6.00& 12.97&   236&   849&-1.65& 0.08\\
0620+389&  1.201&  1.18&  1.22&  0.52&  0.27&  0.21&  1.62& 52.60&  6.63& 14.84&  0.04&  0.27& -0.36&  3.50&  3.09&   351&   574&-0.36& 0.71\\
0633+734& 16.892&  1.41&  1.52&  0.57&  0.30&  0.17&102.34&  6.73&  8.10& 19.48&  0.05&  0.30& -0.40& 21.72&  5.51&   390&   732&-0.28& 0.98\\
0710+439&  0.894&  0.17&  0.12&  0.05& -0.65& -0.01&  5.30&-84.54&  0.61&  0.51&  0.00& -0.65& -0.48&  1.79& 84.65&   161&  1505&-1.00& 0.92\\
0731+479&  2.005&  2.48&  3.03&  2.50&  0.80&  0.47&  2.25& 23.67&  4.34&  6.73&  0.81&  0.80&  0.29&  2.75& 10.42&   244&   511&-0.34& 0.96\\
0733+597&  0.212&  0.00&  0.00&  0.00& -2.74& -0.48&249.19&-23.94&  0.02&  0.00&  0.00& -2.74& -1.43& 27.26&-23.95&    23&   132&-1.19& 0.37\\
0743+744&  9.740&  0.85&  0.82&  0.28&  0.03&  0.08& 56.72& 11.64&  1.60&  1.84&  0.04&  0.03& -0.14& 30.69& 11.42&   307&   473&-0.19& 0.99\\
0805+410&  3.987&  1.36&  1.45&  0.54&  0.29&  0.18&  6.91& 25.48&  1.61&  1.80&  0.16&  0.29&  0.12&  6.06& 24.53&   902&  1050& 0.08& 1.41\\
0806+573&  1.848&  0.17&  0.11&  0.06& -0.61&  0.05& 13.20& 56.53&  0.50&  0.40&  0.00& -0.61& -0.37&  4.66& 54.20&   315&   400&-0.11& 0.99\\
0814+425&  1.617&  2.29&  2.76&  1.47&  0.59&  0.29&  1.93& 25.22&  4.27&  6.55&  0.47&  0.59&  0.08&  2.56&  9.25&   630&  1699&-0.48& 0.90\\
0820+560&  2.151&  1.87&  2.14&  0.84&  0.45&  0.22&  2.44& 31.18&  5.55& 10.02&  0.13&  0.45& -0.14&  3.28&  7.12&  1056&  1585&-0.06& 1.32\\
0821+621&  1.795&  0.10&  0.06&  0.02& -0.94& -0.08& 21.84& 58.14&  0.32&  0.21&  0.00& -0.94& -0.55&  6.79& 57.09&   562&   583&-0.04& 0.95\\
0831+557&  0.396&  0.05&  0.02&  0.01& -1.36& -0.23& 12.41&-43.28&  0.28&  0.17&  0.00& -1.36& -0.92&  2.23&-45.93&   903&  4092&-0.81& 0.71\\
0833+416&  3.499&  1.39&  1.49&  1.12&  0.50&  0.38&  5.46& 27.99&  3.69&  5.85&  0.19&  0.50&  0.05&  3.64& 15.71&    79&   321&-0.69& 0.83\\
0833+585& 14.048&  0.23&  0.17&  0.04& -0.66& -0.12&      &  8.14&  0.61&  0.53&  0.00& -0.66& -0.49&161.71&  8.13&   441&   642&-0.18& 0.96\\
0836+710&  4.108&  3.95&  5.35&  3.98&  1.05&  0.54&  4.24& 14.64& 20.76& 88.12&  0.57&  1.05&  0.05& 10.81&  1.05&   489&  2408&-0.69& 0.99\\
0850+581&  1.353&  2.86&  3.61&  1.53&  0.66&  0.28&  1.93& 16.70& 10.22& 24.11&  0.26&  0.66& -0.13&  5.25&  1.47&   481&  1023&-0.39& 0.86\\
0859+470&  1.791&  2.04&  2.39&  1.08&  0.53&  0.27&  2.05& 29.38& 16.52& 63.34&  0.11&  0.53& -0.39&  8.39&  0.75&   435&  1151&-0.47& 0.90\\
0859+681&  4.550&  1.60&  1.78&  0.76&  0.39&  0.21&  7.57& 22.23&  3.12&  4.36&  0.15&  0.39& -0.01&  5.04& 17.19&   373&   673&-0.30& 0.90\\
0902+490&  1.168&  1.38&  1.49&  0.46&  0.24&  0.12&  1.55& 45.74&  5.23&  9.52&  0.05&  0.24& -0.32&  2.84&  4.82&   476&   620&-0.06& 1.13\\
0917+449&  2.098&  2.00&  2.33&  1.02&  0.53&  0.28&  2.35& 29.56&  3.60&  5.21&  0.22&  0.53&  0.08&  2.55& 14.40&   777&  1407&-0.12& 1.36\\
0917+624&  0.547&  3.26&  4.24&  1.33&  0.63&  0.20&  1.83&  6.30&  8.30& 16.35&  0.28&  0.63& -0.11&  4.23&  0.92&   822&  1302&-0.21& 0.98\\
0923+392&  2.506&  3.41&  4.48&  1.46&  0.69&  0.24&  2.77& 16.50&  3.22&  4.14&  0.91&  0.69&  0.26&  2.74& 17.79&  9990&  7041& 0.13& 0.94\\
0954+658&  0.159&  0.93&  0.91&  0.55&  0.22&  0.25&  1.02&-70.81&  1.04&  1.04&  0.21&  0.22&  0.21&  1.01& 72.88&   405&   632&-0.54& 0.45\\
1003+830&  0.120&  3.21&  4.16&  3.19&  0.87&  0.44&  1.76&  1.48&  9.51& 21.77&  0.83&  0.87&  0.10&  4.81&  0.15&   191&   479&-0.57& 0.67\\
1010+350& 10.444&  3.28&  4.26&  1.60&  0.68&  0.24& 18.44&  9.97&  5.59&  9.04&  0.47&  0.68&  0.07& 12.65&  8.53&   235&   354&-0.41& 0.59\\
1014+615&  1.984&  1.40&  1.50&  0.34&  0.14&  0.01&  2.47& 39.09&  2.02&  2.41&  0.07&  0.14& -0.12&  2.23& 29.55&   466&   589&-0.13& 0.93\\
1015+359&  6.543&  2.56&  3.16&  1.39&  0.62&  0.28&  9.83& 15.14& 15.21& 49.82&  0.18&  0.62& -0.28&  9.05&  2.74&   487&   701&-0.08& 1.19\\
1020+400& 13.107&  1.19&  1.24&  0.46&  0.22&  0.16& 72.92&  8.65&  7.05& 16.06&  0.04&  0.22& -0.43& 15.78&  6.78&   713&   872&-0.04& 1.11\\
1030+611&  0.564&  1.89&  2.18&  1.32&  0.52&  0.29&  1.29& 21.25&  7.72& 17.22&  0.23&  0.52& -0.17&  3.95&  1.10&   236&   378&-0.39& 0.65\\
1039+811& 10.125&  1.89&  2.17&  1.18&  0.56&  0.33& 28.39& 10.91&  2.65&  3.43&  0.37&  0.56&  0.21& 20.84& 10.57&   825&  1214&-0.14& 1.06\\
1041+536&  2.506&  0.20&  0.14&  0.05& -0.63& -0.04& 18.33& 43.29&  0.74&  0.67&  0.00& -0.63& -0.53&  5.28& 40.68&   191&   400&-0.40& 0.83\\
1058+629&  4.263&  0.27&  0.20&  0.09& -0.42&  0.06& 35.82& 26.31&  0.59&  0.51&  0.01& -0.42& -0.24& 16.46& 25.94&   233&   306&-0.48& 0.44\\
1101+384&  0.187&  0.92&  0.90&  3.70&  0.81&  0.84&  1.02&-72.80&  1.56&  1.83&  1.12&  0.81&  0.66&  1.11& 14.27&   370&   582&-0.29& 0.81\\
1124+455&  2.675&  0.82&  0.79&  0.43&  0.17&  0.24&  5.37& 38.07&  3.64&  6.13&  0.04&  0.17& -0.27&  2.94& 15.43&   139&   359&-0.41& 1.01\\
1144+352&  1.886&  0.15&  0.10&  0.05& -0.76& -0.05& 15.66& 55.65&  0.47&  0.36&  0.00& -0.76& -0.50&  5.09& 53.64&   250&   558&-0.42& 0.84\\
1150+812& 14.008&  1.30&  1.38&  0.43&  0.20&  0.11& 76.49&  8.10&  4.63&  7.98&  0.05&  0.20& -0.32& 23.60&  7.37&   637&  1233&-0.27& 1.04\\
1240+381&  6.421&  0.36&  0.28&  0.06& -0.54& -0.16& 59.49& 17.65&  0.43&  0.35&  0.01& -0.54& -0.23& 49.84& 17.63&   332&   526&-0.36& 0.69\\
1258+507&  2.227&  1.36&  1.45&  0.60&  0.29&  0.18&  2.87& 37.50&  6.33& 13.13&  0.07&  0.29& -0.33&  3.64&  5.78&   193&   345&-0.31& 0.88\\
1305+804&  3.697&  0.03&  0.02&  0.01& -1.43& -0.18&223.38& 30.27&  0.35&  0.21&  0.00& -1.43& -1.09& 21.09& 30.03&    75&   146&-0.70& 0.39\\
1307+562&  4.648&  1.31&  1.39&  0.73&  0.37&  0.27&  9.28& 22.61&  2.16&  2.69&  0.16&  0.37&  0.10&  6.31& 20.19&   219&   323&-0.28& 0.78\\
1309+555&  2.986&  1.29&  1.36&  0.58&  0.27&  0.18&  4.50& 31.97&  1.10&  1.12&  0.30&  0.27&  0.24&  5.05& 33.19&   236&   292&-0.46& 0.43\\
1333+459&  8.175&  0.84&  0.81&  0.21& -0.05&  0.01& 40.83& 13.81&  1.09&  1.11&  0.04& -0.05& -0.08& 31.68& 13.71&   279&   643&-0.33& 1.07\\
1335+552&  2.202&  1.55&  1.71&  0.76&  0.38&  0.22&  2.66& 35.12&  3.50&  5.17&  0.15&  0.38& -0.06&  2.58& 15.32&   144&   626&-0.75& 0.77\\
1415+463&  3.579&  0.19&  0.13&  0.04& -0.67& -0.06& 36.44& 31.14&  0.71&  0.63&  0.00& -0.67& -0.56& 10.11& 30.18&   157&   664&-0.76& 0.73\\
1418+546&  2.044&  0.97&  0.96&  0.70&  0.27&  0.29&  3.16& 44.82&  2.11&  2.67&  0.15&  0.27&  0.01&  2.28& 28.22&  1192&  2198&-0.16& 1.29\\
1435+638&  2.632&  0.38&  0.31&  0.11& -0.30&  0.06& 10.65& 40.93&  3.60&  6.64&  0.01& -0.30& -0.72&  2.90& 15.55&   137&   992&-0.76& 1.25\\
1505+428&  7.899&  0.41&  0.33&  0.13& -0.30&  0.04& 78.30& 14.39&  1.34&  1.49&  0.01& -0.30& -0.40& 24.29& 14.04&   321&   452&-0.10& 1.12\\
1543+517&  1.793&  0.78&  0.73&  0.24& -0.01&  0.08&  3.10& 51.84&  1.89&  2.31&  0.03& -0.01& -0.24&  2.06& 31.76&   271&   622&-0.30& 1.14\\
1547+507&  0.610&  4.00&  5.44&  2.34&  0.84&  0.33&  2.17&  4.54&  9.30& 18.99&  0.54&  0.84&  0.06&  4.72&  0.81&   441&   846&-0.21& 1.17\\
1619+491&  5.633&  1.29&  1.37&  0.63&  0.32&  0.22& 13.33& 19.18&  3.41&  5.13&  0.10&  0.32& -0.11&  6.50& 14.89&   204&   407&-0.36& 0.87\\
1623+578& 10.709&  0.53&  0.46&  0.14& -0.25& -0.01&110.30& 10.64&  1.25&  1.34&  0.02& -0.25& -0.33& 46.91& 10.53&   647&   777& 0.04& 1.32\\
1624+416&  8.249&  0.54&  0.47&  0.12& -0.29& -0.07& 64.24& 13.77&  2.66&  3.83&  0.01& -0.29& -0.63& 14.33& 12.55&   242&  1117&-0.75& 0.82\\
1633+382&  5.108&  3.84&  5.19&  1.52&  0.69&  0.19&  5.45& 14.37&  5.62&  8.81&  0.49&  0.69&  0.07&  5.22& 10.23&   734&  1710&-0.64& 0.54\\
1637+574&  5.639&  2.40&  2.92&  1.75&  0.70&  0.38&  8.03& 17.14&  3.71&  5.32&  0.60&  0.70&  0.23&  6.28& 14.21&   696&  1830&-0.41& 1.01\\
1652+398&  0.221&  1.14&  1.17&  3.28&  0.77&  0.72&  1.03& 52.63&  2.03&  2.58&  1.05&  0.77&  0.53&  1.27&  7.96&   142&   888&-0.98& 0.65\\
1700+685&  1.037&  0.37&  0.29&  0.15& -0.29&  0.08&  3.00& 84.19&  0.62&  0.55&  0.03& -0.29& -0.12&  1.98& 77.41&   161&   227&-0.43& 0.52\\
1716+686&  0.505&  2.11&  2.48&  2.52&  0.78&  0.50&  1.35& 15.32&  2.17&  2.59&  1.74&  0.78&  0.49&  1.38& 14.23&   399&   750&-0.39& 0.76\\
1726+455&  5.686&  1.66&  1.85&  0.61&  0.32&  0.13& 10.89& 18.46&  1.67&  1.88&  0.26&  0.32&  0.13& 10.80& 18.43&  1132&  1314& 0.03& 1.23\\
1738+476&  3.756&  1.45&  1.57&  0.58&  0.29&  0.16&  5.94& 26.30&  4.47&  7.48&  0.09&  0.29& -0.23&  3.93& 12.78&   514&  1080&-0.19& 1.37\\
1738+499&  3.069&  1.77&  2.01&  0.87&  0.44&  0.23&  3.83& 27.99&  6.28& 12.33&  0.12&  0.44& -0.19&  3.97&  7.30&   270&   464&-0.25& 0.97\\
1744+557&  0.046&  0.31&  0.24&  0.22& -0.26&  0.17&  1.78& -5.82&  1.12&  1.18&  0.03& -0.26& -0.30&  1.01& 19.06&   162&   356&-0.57& 0.59\\
1755+578&  3.950&  0.09&  0.06&  0.02& -1.02& -0.15& 87.75& 28.40&  0.85&  0.79&  0.00& -1.02& -0.96& 10.22& 27.28&     8&   480&-1.75& 1.05\\
1803+784&  8.047&  2.55&  3.13&  1.43&  0.64&  0.30& 14.19& 12.91&  4.36&  6.62&  0.44&  0.64&  0.12&  9.72& 11.01&   204&  2853&-1.11& 1.08\\
1809+568&  1.939&  0.29&  0.22&  0.07& -0.51& -0.06&  8.27& 53.73&  0.90&  0.86&  0.00& -0.51& -0.48&  3.10& 47.42&    88&   475&-0.82& 0.82\\
1823+568&  4.497&  0.78&  0.74&  0.42&  0.15&  0.24& 13.91& 24.39&  3.06&  4.71&  0.05&  0.15& -0.22&  5.00& 17.48&  1414&  1278& 0.10& 1.13\\
1849+670&  8.650&  2.39&  2.90&  1.47&  0.63&  0.31& 17.05& 12.27&  5.35&  9.10&  0.36&  0.63&  0.04&  9.76&  9.59&   391&   667&-0.40& 0.67\\
1856+737&  0.695&  0.84&  0.81&  0.57&  0.24&  0.30&  1.30& 82.00&  2.11&  2.71&  0.10&  0.24& -0.02&  1.41& 19.50&   178&   434&-0.49& 0.80\\
1908+484&  4.095&  0.21&  0.15&  0.07& -0.57&  0.00& 42.65& 27.38&  1.08&  1.11&  0.00& -0.57& -0.60&  8.80& 25.82&    71&   145&-0.78& 0.34\\
1924+507&  4.522&  0.55&  0.48&  0.23& -0.07&  0.15& 19.67& 24.60&  5.48& 12.80&  0.01& -0.07& -0.63&  4.70& 10.35&   271&   422&-0.12& 1.19\\
%       &       &      &      &      &      &      &      &      &      &      &      &      &      &      &      &      &      &     &     \\
%       &    & &&&&&&&&&&&&&&&&&\\
%       &    & &&&&&&&&&&&&&&&&&\\
1954+513& 13.003&  1.64&  1.83&  1.08&  0.52&  0.34& 52.73&  8.66&  4.07&  6.53&  0.20&  0.52&  0.04& 22.93&  8.02&   615&  1071&-0.42& 0.67\\
2017+745&  0.963&  5.79&  8.55&  5.42&  1.13&  0.49&  3.06&  3.30& 12.39& 27.95&  1.59&  1.13&  0.25&  6.27&  0.72&   141&   312&-0.55& 0.62\\
2116+818&  1.947&  0.06&  0.03&  0.04& -0.82&  0.20& 38.70& 54.33&  0.21&  0.11&  0.00& -0.82& -0.29& 11.58& 53.95&    61&   143&-0.79& 0.38\\
2214+350&  0.152&  0.64&  0.58&  0.25& -0.06&  0.10&  1.12&-28.02&  2.03&  2.59&  0.03& -0.06& -0.31&  1.27&  5.54&   534&   600& 0.05& 1.26\\
2253+417&  7.415&  0.42&  0.35&  0.17& -0.17&  0.14& 66.26& 15.31&  1.67&  2.03&  0.01& -0.17& -0.35& 17.60& 14.64&   292&  1099&-0.58& 0.98\\
2346+385&  5.617&  0.26&  0.20&  0.08& -0.44&  0.05& 61.70& 20.15&  0.32&  0.25&  0.01& -0.44& -0.03& 50.37& 20.13&   565&   655&-0.05& 1.02\\
2356+385&  2.439&  1.19&  1.23&  0.54&  0.29&  0.23&  3.52& 37.50&  5.90& 12.27&  0.05&  0.29& -0.31&  3.54&  6.99&   207&   382&-0.34& 0.85\\
2356+390&  2.346&  0.71&  0.65&  0.35&  0.08&  0.21&  4.96& 43.17&  2.29&  3.09&  0.04&  0.08& -0.20&  2.57& 25.71&   220&   314&-0.23& 0.85\\
\end{supertabular}
\onecolumn
\landscape
\textheight24cm
\textwidth17.5cm
\oddsidemargin-2cm
\tabcolsep0.3mm
\tablecaption{Calculated values for those sources that have not been detected by {\it ROSAT} and are not shown in the plots of this paper. This table is only available in the online edition of the Journal.} {\smallskip}
\setcounter{table}{5}
\tablehead{\noalign{\smallskip} \hline \noalign{\smallskip}
&&& $\alpha=-0.75$ &&&&& observed $\alpha$ &&&&&& \\ \hline
Source &  $\delta_{\rm IC}$ & $\delta_{\rm IC}^{\rm con}$ & $\delta_{\rm EQ}$ & log (T$_{\rm B}$/10$^
{11}$) & log (T$_{\rm Bi}^{\rm con}$/10$^{11}$) & $ \delta_{\rm IC}$ & $\delta_{\rm IC}^{\rm con}$ & $\delta_{\rm EQ}$ & log (T$_{\rm B}$/10$^{11}$) & log (T$_{\rm Bi}^{\rm con}$/10$^{11}$)& VLBI flux & VLBI core flux & log(R$_{\rm C}$) & R$_{\rm V}$ \\
\multicolumn{1}{c}{\tiny ~~~ } &
 \multicolumn{1}{c}{\tiny ~~~}&
 \multicolumn{1}{c}{\tiny ~~~~~} & \multicolumn{1}{c}{\tiny ~~~} &
 \multicolumn{1}{c}{\tiny ~~[K] } &
 \multicolumn{1}{c}{\tiny ~~[K]}&
 \multicolumn{1}{c}{\tiny ~~~}&
 \multicolumn{1}{c}{\tiny ~~~~~} & \multicolumn{1}{c}{\tiny ~~~} &
 \multicolumn{1}{c}{\tiny ~~[K] } &
 \multicolumn{1}{c}{\tiny ~~[K]}&
 \multicolumn{1}{c}{\tiny ~~~~[mJy]}&
 \multicolumn{1}{c}{\tiny ~~~~[mJy]}&&\\
(1)&(2)&(3)&(4)&(5)&(6)&(7)&(8)&(9)&(10)&(11)&(12)&(13)&(14)&(15)\\
\noalign{\smallskip} \hline \hline  \noalign{\smallskip}}
\tabletail{\hline\multicolumn{12}{r}{continued on next page}\\}
\tablelasttail{\hline}
{\small
\begin{supertabular}{c|ccccc|ccccc|cccc}
\label{nonrest}
0003+380& 0.21& 0.14&  0.05& -0.68& -0.10&  0.76&  0.70&  0.00& -0.68& -0.59&   335&   721&-0.21& 1.31\\
0022+390& 1.23& 1.29&  0.45&  0.21&  0.14&  4.47&  7.67&  0.05&  0.21& -0.30&   130&   600&-0.71& 0.91\\
0145+386& 0.24& 0.18&  0.03& -0.76& -0.24&  0.58&  0.49&  0.00& -0.76& -0.57&   296&   458&-0.10& 1.24\\
0151+474& 7.91&12.53&  5.58&  1.13&  0.37& 13.07& 26.80&  2.58&  1.13&  0.21&   717&   798& 0.15& 1.58\\
0205+722& 1.89& 2.18&  1.42&  0.59&  0.35&  9.11& 23.38&  0.20&  0.59& -0.15&   109&   272&-0.71& 0.49\\
0249+383& 0.31& 0.24&  0.06& -0.53& -0.10&  3.72&  6.92&  0.00& -0.53& -0.96&   216&   361&-0.32& 0.80\\
0307+380& 3.72& 4.99&  2.01&  0.76&  0.27&  2.17&  2.47&  3.43&  0.76&  0.46&   516&   521&-0.17& 0.69\\
0340+362& 0.59& 0.52&  0.19& -0.11&  0.08&  1.11&  1.15&  0.03& -0.11& -0.15&   475&   559& 0.10& 1.49\\
0604+728& 0.20& 0.14&  0.06& -0.58&  0.01&  1.42&  1.66&  0.00& -0.58& -0.69&   109&   427&-0.78& 0.65\\
0615+820& 0.34& 0.26&  0.06& -0.57& -0.17&  1.24&  1.33&  0.00& -0.57& -0.64&   709&   782&-0.15& 0.78\\
0636+680& 1.07& 1.09&  0.21& -0.04& -0.06&  0.77&  0.73&  0.08& -0.04&  0.06&   481&   482&-0.02& 0.97\\
0650+371& 0.97& 0.96&  0.38&  0.14&  0.16&  1.41&  1.54&  0.08&  0.14&  0.02&   334&  1310&-0.47& 1.34\\
0650+453& 0.95& 0.94&  0.33&  0.07&  0.09&  5.36& 10.62&  0.03&  0.07& -0.49&   415&   467&-0.01& 1.11\\
0707+476&&&  0.49&  0.23&&&&  0.06&  0.23&&   272&   854&-0.52& 0.94\\
0730+504& 4.83& 6.85&  3.81&  0.97&  0.39&  5.16&  7.53&  2.67&  0.97&  0.37&   561&   828&-0.20& 0.93\\
0738+491& 4.31& 5.97&  2.18&  0.81&  0.27&  3.31&  4.18&  1.73&  0.81&  0.36&   677&   691& 0.28& 1.96\\
0749+426& 0.78& 0.74&  0.19& -0.10& -0.01&  6.05& 13.14&  0.01& -0.10& -0.70&   274&   424&-0.23& 0.92\\
0803+452& 3.46& 4.56&  2.84&  0.88&  0.42&  7.55& 14.48&  0.69&  0.88&  0.17&    87&   384&-0.68& 0.93\\
0824+355& 1.09& 1.11&  0.16& -0.16& -0.19&  5.59& 10.42&  0.01& -0.16& -0.75&   453&   659&-0.22& 0.88\\
0929+533& 1.04& 1.05&  0.46&  0.16&  0.15&  5.42& 10.77&  0.05&  0.16& -0.40&   133&   270&-0.46& 0.70\\
1053+704& 3.20& 4.14&  1.40&  0.64&  0.21&  7.70& 14.68&  0.27&  0.64& -0.08&   363&   575&-0.27& 0.85\\
1144+402& 3.86& 5.22&  2.13&  0.79&  0.29& 14.63& 40.11&  0.40&  0.79& -0.13&   507&   639&-0.16& 0.86\\
1218+444& 1.65& 1.85&  0.64&  0.32&  0.13&  7.31& 15.57&  0.08&  0.32& -0.36&   390&   541&-0.09& 1.13\\
1239+376& 0.34& 0.27&  0.04& -0.70& -0.30&  2.07&  2.71&  0.00& -0.70& -0.95&   166&   420&-0.43& 0.94\\
1342+663& 1.35& 1.44&  0.29&  0.06& -0.05& 17.39& 67.30&  0.02&  0.06& -0.88&   688&   698& 0.13& 1.37\\
1355+441& 0.03& 0.02&  0.00& -1.71& -0.44&  0.36&  0.23&  0.00& -1.71& -1.36&   195&   537&-0.38& 1.16\\
1421+482& 0.35& 0.28&  0.07& -0.51& -0.13&  0.63&  0.56&  0.01& -0.51& -0.35&   235&   398&-0.36& 0.74\\
1432+422& 1.63& 1.81&  0.62&  0.30&  0.12&  3.30&  4.70&  0.12&  0.30& -0.12&   242&   298&-0.16& 0.84\\
1534+501& 0.15& 0.10&  0.02& -0.96& -0.27&  0.26&  0.19&  0.00& -0.96& -0.49&   180&   315&-0.30& 0.88\\
1645+635& 2.33& 2.81&  0.68&  0.37&  0.06&  4.05&  5.95&  0.14&  0.37& -0.13&   128&   178&-0.54& 0.40\\
1818+356& 0.01& 0.00&  0.00& -2.46& -0.55&  0.09&  0.03&  0.00& -2.46& -1.67&    70&    70&-0.91& 0.12\\
1839+389& 2.56& 3.16&  1.12&  0.55&  0.21&  4.59&  7.08&  0.24&  0.55&  0.01&   219&   202&-0.34& 0.42\\
1851+488& 1.05& 1.07&  0.31&  0.05&  0.03&  2.41&  3.15&  0.05&  0.05& -0.26&   276&   273&-0.10& 0.78\\
1936+714& 0.90& 0.88&  0.20& -0.09& -0.06&  8.58& 22.22&  0.01& -0.09& -0.81&   227&   422&-0.24& 1.08\\
1946+708& 0.01& 0.00&  0.00& -2.20& -0.49&  0.08&  0.03&  0.00& -2.20& -1.36&   107&   629&-0.78& 0.97\\
2054+611& 1.08& 1.10&  0.55&  0.25&  0.22&  2.95&  4.24&  0.08&  0.25& -0.13&   182&   360&-0.36& 0.87\\
2229+695& 0.52& 0.45&  0.12& -0.29& -0.04&  0.48&  0.41&  0.03& -0.29& -0.02&   273&   983&-0.70& 0.72\\
2238+410& 1.21& 1.26&  0.62&  0.29&  0.22&  2.59&  3.48&  0.12&  0.29& -0.05&   272&   293&-0.40& 0.43\\
2259+371& 1.07& 1.09&  0.34&  0.11&  0.08&  6.88& 15.45&  0.03&  0.11& -0.54&    95&   341&-0.63& 0.84\\
2309+454& 1.31& 1.39&  0.42&  0.18&  0.08&  1.66&  1.88&  0.11&  0.18& -0.01&   375&   503&-0.20& 0.84\\
2353+816& 2.02& 2.36&  0.94&  0.47&  0.21&  4.45&  7.06&  0.19&  0.47& -0.05&   318&   484&-0.17& 1.02\\
\end{supertabular}
}
\onecolumn
\landscape
\textheight24cm
\textwidth17.5cm
\oddsidemargin-2cm
\tabcolsep0.3mm
\tablecaption{Calculated values for those sources that have been detected by {\it ROSAT} but are not shown in the plots of this paper. This table is only available in the online edition of the Journal.} {\smallskip}
\setcounter{table}{6}
\tablehead{\noalign{\smallskip} \hline \noalign{\smallskip}
&&& $\alpha=-0.75$ &&&&& observed $\alpha$ &&&&&&\\ \hline
Source &  $\delta_{\rm IC}$&$\delta_{\rm IC}^{\rm con}$&$\delta_{\rm EQ}$&log (T$_{\rm B}$/10$^
{11}$)&log (T$_{\rm Bi}^{\rm con}$/10$^{11}$)& $\delta_{\rm IC}$&$\delta_{\rm IC}^{\rm con}$&$\delta_{\rm EQ}$&log (T$_{\rm B}$/10$^{11}$)&log (T$_{\rm Bi}^{\rm con}$/10$^{11}$)&VLBI flux&VLBI core flux & log(R$_{\rm C}$) & R$_{\rm V}$ \\
\multicolumn{1}{c}{\tiny ~~~ } &
 \multicolumn{1}{c}{\tiny ~~~}&
  \multicolumn{1}{c}{\tiny ~~~~~} & \multicolumn{1}{c}{\tiny ~~~} &
   \multicolumn{1}{c}{\tiny ~~[K] } &
    \multicolumn{1}{c}{\tiny ~~[K]}&
     \multicolumn{1}{c}{\tiny ~~~}&
      \multicolumn{1}{c}{\tiny ~~~~~} & \multicolumn{1}{c}{\tiny ~~~} &
       \multicolumn{1}{c}{\tiny ~~[K] } &
        \multicolumn{1}{c}{\tiny ~~[K]}&
	 \multicolumn{1}{c}{\tiny ~~~~[mJy]}&
	  \multicolumn{1}{c}{\tiny ~~~~[mJy]}\\
	  (1)&(2)&(3)&(4)&(5)&(6)&(7)&(8)&(9)&(10)&(11)&(12)&(13)&(14)&(15)\\
	 \noalign{\smallskip} \hline \hline  \noalign{\smallskip}}
	  \tabletail{\hline\multicolumn{12}{r}{continued on next page}\\}
	  \tablelasttail{\hline}
	  {\small
	  \begin{supertabular}{c|ccccc|ccccc|cccc}
	  \label{detrest}
0010+405& 0.55& 0.48&  0.29& -0.03&  0.19&  4.12&  8.02&  0.02& -0.03& -0.49&   346&   481&-0.48& 0.46\\
0035+367& 0.18& 0.12&  0.06& -0.61&  0.03&  1.82&  2.44&  0.00& -0.61& -0.81&    91&   113&-0.72& 0.23\\
0109+351& 3.55& 4.70&  3.66&  0.94&  0.47&  8.00& 15.97&  1.16&  0.94&  0.21&   261&   379&-0.14& 1.05\\
0153+744& 0.07& 0.04&  0.01& -1.17& -0.21&  0.45&  0.33&  0.00& -1.17& -0.90&    87&  1264&-1.25& 0.82\\
0309+411& 0.17& 0.11&  0.09& -0.48&  0.17&  0.37&  0.27&  0.01& -0.48& -0.14&   334&   462&-0.19& 0.90\\
0316+413& 0.15& 0.10&  0.15& -0.36&  0.34&  0.17&  0.11&  0.04& -0.36&  0.28&  3080& 22410&-1.14& 0.53\\
0402+379& 0.02& 0.01&  0.01& -1.38&  0.03&  0.13&  0.05&  0.00& -1.38& -0.70&    18&   706&-1.72& 0.75\\
0602+673& 1.59& 1.76&  0.62&  0.35&  0.18&  1.79&  2.05&  0.20&  0.35&  0.14&   244&   678&-0.43& 1.03\\
0727+409& 2.65& 3.30&  2.33&  0.80&  0.44&  5.22&  8.74&  0.54&  0.80&  0.22&   135&   417&-0.54& 0.89\\
0746+483& 0.54& 0.47&  0.13& -0.24& -0.01&  1.10&  1.14&  0.01& -0.24& -0.27&   472&   841&-0.26& 0.98\\
0804+499& 1.67& 1.87&  0.72&  0.39&  0.20&  2.99&  4.09&  0.16&  0.39&  0.00&   511&  1494&-0.38& 1.22\\
0812+367& 1.40& 1.50&  0.44&  0.20&  0.08&  4.54&  7.62&  0.06&  0.20& -0.32&   671&   950&-0.17& 0.97\\
0821+394& 1.86& 2.14&  1.03&  0.53&  0.30&  7.99& 18.34&  0.14&  0.53& -0.18&   673&  1063&-0.18& 1.05\\
0847+379& 0.60& 0.54&  0.28& -0.04&  0.15&  4.13&  7.77&  0.02& -0.04& -0.51&   190&   245&-0.30& 0.64\\
0945+408& 1.71& 1.93&  0.59&  0.32&  0.12&  4.52&  7.35&  0.10&  0.32& -0.20&  1439&  1109&-0.04& 0.70\\
0954+556& 0.00& 0.00&  0.00& -3.33& -0.81&  0.01&  0.00&  0.00& -3.33& -1.74&   584&  1394&-0.59& 0.61\\
0955+476& 4.05& 5.53&  1.93&  0.78&  0.27&  8.38& 15.92&  0.48&  0.78&  0.03&  1022&  1073& 0.09& 1.29\\
1038+528& 0.19& 0.13&  0.04& -0.67& -0.07&  0.59&  0.49&  0.00& -0.67& -0.49&   627&   749&-0.05& 1.06\\
1044+719& 2.10& 2.47&  0.69&  0.38&  0.11&  1.56&  1.70&  0.44&  0.38&  0.22&   577&   972&-0.62& 0.40\\
1053+815& 0.62& 0.56&  0.25& -0.05&  0.12&  4.16&  7.77&  0.02& -0.05& -0.53&   522&   564&-0.17& 0.73\\
1058+726& 1.25& 1.31&  0.72&  0.37&  0.29&  6.51& 14.59&  0.07&  0.37& -0.26&    75&   430&-1.10& 0.45\\
1105+437& 2.88& 3.65&  2.07&  0.74&  0.35&  4.72&  7.30&  0.67&  0.74&  0.19&   253&   312&-0.17& 0.83\\
1125+596& 2.04& 2.39&  1.01&  0.50&  0.24&  4.88&  8.11&  0.19&  0.50& -0.05&   332&   330&-0.07& 0.84\\
1128+385&17.68&33.48& 17.06&  1.58&  0.52& 58.91&276.16&  5.69&  1.58&  0.19&   592&  1005&-0.10& 1.35\\
1213+350& 0.02& 0.01&  0.00& -1.72& -0.29&  0.17&  0.08&  0.00& -1.72& -1.12&    31&  1002&-1.57& 0.87\\
1226+373& 1.39& 1.49&  0.52&  0.26&  0.14&  0.91&  0.90&  0.37&  0.26&  0.30&   710&   742&-0.13& 0.78\\
1254+571& 0.06& 0.03&  0.01& -1.29& -0.27&  0.12&  0.06&  0.00& -1.29& -0.52&   187&   188&-0.35& 0.45\\
1300+580& 0.61& 0.54&  0.16& -0.17&  0.01&  0.62&  0.55&  0.04& -0.17&  0.00&   888&   903& 0.07& 1.19\\
1306+360& 6.99&10.77&  5.82&  1.15&  0.43&  8.34& 13.98&  3.79&  1.15&  0.38&   378&   418&-0.06& 0.96\\
1347+539& 0.29& 0.22&  0.08& -0.46& -0.01&  3.32&  5.95&  0.00& -0.46& -0.85&   427&   784&-0.17& 1.24\\
1413+373& 0.83& 0.80&  0.35&  0.09&  0.15&  2.14&  2.74&  0.04&  0.09& -0.18&   104&   390&-0.57& 1.02\\
1417+385& 3.45& 4.54&  1.21&  0.62&  0.16&  7.44& 13.58&  0.27&  0.62& -0.09&   621&   619&-0.15& 0.71\\
1427+543& 1.95& 2.26&  0.73&  0.40&  0.16&  8.44& 18.99&  0.08&  0.40& -0.32&   317&   489&-0.35& 0.68\\
1442+637& 1.83& 2.09&  0.92&  0.46&  0.24& 10.27& 27.70&  0.11&  0.46& -0.32&   301&   502&-0.18& 1.10\\
1526+670& 1.41& 1.52&  0.46&  0.22&  0.10&  4.32&  7.09&  0.05&  0.22& -0.28&   261&   414&-0.20& 0.99\\
1531+722& 0.27& 0.20&  0.06& -0.55& -0.07&  1.88&  2.45&  0.00& -0.55& -0.76&   170&   352&-0.42& 0.78\\
1550+582& 2.20& 2.62&  1.82&  0.69&  0.40&  3.06&  4.11&  0.64&  0.69&  0.29&   215&   287&-0.23& 0.78\\
1622+665& 2.45& 2.99&  2.29&  0.72&  0.39&  2.45&  2.98&  1.70&  0.72&  0.39&   197&   242&-0.42& 0.47\\
1629+495& 2.10& 2.47&  1.72&  0.64&  0.37&  3.92&  5.92&  0.51&  0.64&  0.16&   322&   447&-0.09& 1.14\\
1636+473& 2.37& 2.88&  1.32&  0.59&  0.27&  4.04&  6.00&  0.40&  0.59&  0.09&   340&   750&-0.59& 0.56\\
1638+398& 1.94& 2.25&  0.47&  0.25&  0.01&  2.49&  3.12&  0.13&  0.25& -0.08&  1766&  1787& 0.14& 1.39\\
1638+540& 0.18& 0.12&  0.02& -1.01& -0.38&  0.51&  0.41&  0.00& -1.01& -0.77&   205&   286&-0.26& 0.78\\
1641+399& 2.79& 3.50&  1.64&  0.70&  0.33&  6.50& 11.90&  0.41&  0.70&  0.05&  3420&  7400&-0.39& 0.89\\
1642+690& 0.64& 0.58&  0.22& -0.07&  0.09&  1.83&  2.24&  0.03& -0.07& -0.28&   371&  1870&-0.61& 1.23\\
1645+410& 1.46& 1.59&  0.70&  0.34&  0.20&  2.96&  4.10&  0.15&  0.34& -0.04&   492&   570& 0.10& 1.47\\
1656+477& 1.46& 1.59&  0.70&  0.36&  0.22&  1.93&  2.28&  0.20&  0.36&  0.12&   729&  1554&-0.29& 1.09\\
1656+571& 1.07& 1.08&  0.54&  0.25&  0.23&  2.60&  3.56&  0.08&  0.25& -0.08&   279&   417&-0.48& 0.49\\
1719+357& 0.87& 0.85&  0.83&  0.33&  0.38&  1.88&  2.31&  0.18&  0.33&  0.11&   212&   454&-0.61& 0.52\\
1722+401& 0.96& 0.95&  0.48&  0.20&  0.22&  2.65&  3.68&  0.07&  0.20& -0.14&   220&   391&-0.38& 0.73\\
1739+522& 0.92& 0.90&  0.34&  0.11&  0.14&  8.29& 22.67&  0.02&  0.11& -0.59&  1075&   965&-0.02& 0.85\\
1745+624& 4.29& 5.94&  2.48&  0.87&  0.33& 17.65& 54.24&  0.35&  0.87& -0.11&   329&   472&-0.25& 0.81\\
1749+701& 0.19& 0.14&  0.06& -0.60&  0.00&  2.00&  2.80&  0.00& -0.60& -0.82&   246&  1268&-0.47& 1.74\\
1751+441& 0.33& 0.26&  0.09& -0.40&  0.01&  0.67&  0.60&  0.01& -0.40& -0.26&   717&   953&-0.14& 0.95\\
1758+388& 2.50& 3.06&  1.26&  0.62&  0.28&  4.20&  6.28&  0.31&  0.62&  0.11&   625&   919&-0.06& 1.27\\
1800+440& 2.25& 2.69&  1.53&  0.64&  0.34&  4.00&  5.99&  0.45&  0.64&  0.15&   387&   557&-0.47& 0.48\\
1807+698& 1.12& 1.15&  1.11&  0.40&  0.36&  2.72&  3.81&  0.26&  0.40&  0.06&   579&  1604&-0.58& 0.73\\
1834+612& 1.64& 1.83&  0.71&  0.37&  0.19&  3.09&  4.28&  0.14&  0.37& -0.03&   368&   551&-0.20& 0.93\\
1842+681& 4.41& 6.14&  4.73&  1.04&  0.49&  6.29& 10.35&  2.49&  1.04&  0.38&   401&   851&-0.37& 0.91\\
1928+738& 0.31& 0.24&  0.12& -0.31&  0.12&  0.96&  0.94&  0.01& -0.31& -0.30&  1383&  2835&-0.41& 0.80\\
1950+573& 0.52& 0.45&  0.25& -0.07&  0.17&  1.82&  2.27&  0.03& -0.07& -0.27&   205&   323&-0.37& 0.68\\
2005+642& 0.58& 0.52&  0.13& -0.24& -0.04&  0.40&  0.34&  0.05& -0.24&  0.10&   805&   802& 0.04& 1.08\\
2007+777& 4.10& 5.62&  4.24&  0.98&  0.46&  4.23&  5.87&  3.49&  0.98&  0.45&  1234&  2100&-0.02& 1.64\\
2200+420& 0.13& 0.08&  0.15& -0.40&  0.35&  0.38&  0.26&  0.01& -0.40& -0.08&   838&  1988&-0.63& 0.55\\
\end{supertabular}
}
\end{document}